\documentclass[prl,showpacs,twocolumn,floats,10pt,aps,citeautoscript,longbibliography,superscriptaddress]{revtex4-2}

\usepackage[final]{changes}
\definechangesauthor[name=GK, color=red]{GK}
\definechangesauthor[name=AG, color=red]{AG}

\usepackage{microtype} 
\usepackage{graphicx}
\usepackage{enumerate}
\usepackage{bm}
\usepackage{amsmath}
\usepackage{amsfonts}
\usepackage{dsfont}
\usepackage{color}
\usepackage{upgreek}
\usepackage[normalem]{ulem}

\allowdisplaybreaks 

\usepackage{xspace} 

\renewcommand{\vec}[1]{{\mathbf{#1}}} 
\newcommand{\bk}{{\vec{k}}}

\usepackage{bbm}  

\newcommand{\MoTe}{{MoTe${}_2$}}
\newcommand{\WSe}{{WSe${}_2$}}

\newcommand{\mc}[1]{\ensuremath{\mathcal{#1}}}
\newcommand{\mr}[1]{\ensuremath{\mathrm{#1}}}

\newcommand{\bG}{\mathbf{G}}

\newcommand{\bv}{\mathbf{v}}
\newcommand{\bSigma}{\boldsymbol{\Sigma}}

\newcommand{\Eq}[1]{Eq.~\eqref{#1}}

\definecolor{darkgreen}{rgb}{0,0.5,0}

\newcommand{\pdag}{{\vphantom{dagger}}}  

\RequirePackage[
  hyperindex,colorlinks,bookmarksnumbered,
  plainpages=true,pdfstartview=FitH,hypertexnames=false]{hyperref}
\hypersetup{linkcolor=blue,urlcolor=blue,citecolor=blue} 
\usepackage[hypertexnames=false]{hyperref}
\usepackage[all]{hypcap}

\makeatletter
\def\maketitle{
\@author@finish
\title@column\titleblock@produce
\suppressfloats[t]}
\makeatother

 \newcommand{\maintitle}{Topological Kondo Insulator from Spin Loop Currents}

\begin{document}

\title{\maintitle}
\author{Andreas Gleis}
\affiliation{Department of Physics and Astronomy, Rutgers University, Piscataway, NJ 08854, USA}
\author{Kevin Lucht}
\affiliation{Department of Physics and Astronomy, Rutgers University, Piscataway, NJ 08854, USA}
\author{Po-Jui Chen}
\affiliation{Department of Physics and Astronomy, Rutgers University, Piscataway, NJ 08854, USA}
\author{Daniele Guerci}
\affiliation{Department of Physics, Massachusetts Institute of Technology, Cambridge, MA 02139,USA}
\author{Andrew J Millis}
\affiliation{Department of Physics, Columbia University, New York, NY 10027, USA}
\affiliation{Center for Computational Quantum Physics, Flatiron Institute, 162 5th Avenue, New York, NY 10010}
\author{J. H. Pixley}
\affiliation{Department of Physics and Astronomy, Rutgers University, Piscataway, NJ 08854, USA}
\affiliation{Center for Computational Quantum Physics, Flatiron Institute, 162 5th Avenue, New York, NY 10010}

\date{\today}

\begin{abstract}
We demonstrate that interacting electrons in AB-stacked $\mathrm{MoTe}_2/\mathrm{WSe}_2$ realize a topological Kondo insulator at hole filling $\nu=2$ per moir\'e unit cell.
In the presence of
only local correlations, a symmetry of the moir\'e-scale bandstructure enforces a compensated topological semimetal by tying band inversion to band overlap.
We show that non-local interactions change the physics qualitatively, since they allow intrinsic, quantum-geometry-induced spin loop currents to feed back on the effective bandstructure, which lift the remaining accidental degeneracies and open a full gap in the spectrum, leading to a fully gapped topological Kondo insulator. We establish this using real-frequency dynamical mean-field theory to capture Kondo physics alongside Hartree-Fock for non-local interactions.
The topological Kondo insulator emerges at intermediate displacement fields, where strong correlations manifest through an enhanced spin susceptibility, a suppressed charge susceptibility, and a stronger thermal dependence of the resistivity. Our results are in good agreement with recent  experiments on $\mathrm{MoTe}_2/\mathrm{WSe}_2$ bilayers demonstrating topological to trivial phase transitions controlled by the displacement field.

\end{abstract}

\maketitle

\textit{Introduction.---}
In recent years, bilayer transition metal dichalcogenide~(TMD) moir\'e materials have emerged as highly tunable and versatile platforms for realizing strongly correlated and topological electron systems~\cite{Kennes2021,Mak2022}.
Moir\'e patterns emerge either from twisting in homobilayers, such as twisted \MoTe~\cite{Cai2023,Jia2025,Xu2025} or \WSe~\cite{Ghiotto2021,Wei2024,Xia2024,Guo2025,Xia2025},
or from lattice mismatch in heterobilayers such as \WSe/WS${}_2$~\cite{Regan2020,Li2021b} or \MoTe/\WSe~\cite{Li2021a,Li2021,Pan2022,Devakul2022,Xie2022,Xie2023,Dong2023,Xie2024,Tao2024,Zhao2024,Guerci2023,Zhao2023,Zhao2024a,Zhang2025,Han2025,Zhao2025}.
Monolayer TMDs exhibit Ising spin-orbit coupling that leads to spin-valley locking, which is particularly strong in the valence band~\cite{Kormanyos2015}.
As a result, the moir\'e-scale physics of hole-doped TMD bilayers is governed by spin--valley--locked carriers~\cite{Wu2018,Devakul2021}, effectively realizing a generalized Kane--Mele model~\cite{Kane2005}.

Recently, $\mathrm{MoTe}_2/\mathrm{WSe}_2$ bilayer systems have been proposed as a platform to realize a chiral Kondo lattice model~\cite{Guerci2023}, in which the interlayer exchange coupling between local moments and itinerant carriers exhibits a nontrivial winding.
In this system, first-principles calculations~\cite{Zhang2021} have shown that holes in $\mathrm{MoTe}_2$ are much more strongly affected by the moir\'e potential than those in $\mathrm{WSe}_2$, due to their larger effective mass.
As a result, the $\mathrm{MoTe}_2$ moir\'e band is strongly correlated, exhibiting a pronounced tendency toward a Mott (or charge-transfer) insulating state and the formation of local moments~\cite{Li2021a,Zhao2023}.
When the more itinerant and weakly correlated $\mathrm{WSe}_2$ moir\'e band is tuned to the Fermi level by an applied displacement field, interlayer hybridization induces a chiral antiferromagnetic exchange between localized moments and itinerant carriers, realizing the chiral Kondo lattice. This scenario was proposed in Ref.~\cite{Guerci2023} and  observed experimentally in Refs.~\cite{Zhao2023,Zhao2024a,Zhang2025,Han2025,Zhao2025}.

The chiral Kondo exchange originates from the opposite eigenvalues under three-fold rotations around $z$ ($C_3$) of the two lowest energy bands at the high-symmetry points where band inversion takes place~\cite{Zhang2021,Guerci2024,Xie2024}.
The resulting mismatch in angular momentum enforces nodes in the interlayer hybridization, enabling a displacement-field-induced band inversion and, consequently, the emergence of nontrivial band topology~\cite{Zhao2024}.
When strong correlations are taken into account, \MoTe/\WSe\ bilayers may therefore realize a tunable two-dimensional topological Kondo insulator (TKI) platform~\cite{Dzero2010,Dzero2012} at hole filling $\nu = 2$, providing a controlled setting to elucidate several long-standing mysteries in candidate three-dimensional TKIs such as SmB${}_6$~\cite{Menth1969,Li2014,Phelan2014,Tan2015,Li2020} and YbB${}_{12}$~\cite{Liu2018,Sato2019,Chen2025}.

Theoretically, the emergence of a TKI phase in \MoTe/\WSe\ bilayers has been investigated in Refs.~\cite{Guerci2024,Mendez2024,Xie2025}.
 These studies predict a compensated topological semimetal (CTSM) in a clean material,  although interlayer strain inhomogeneity can induce a TKI~\cite{Guerci2024}.
 The compensated semimetal nature of the phase is a consequence of a symmetry of the band-structure of the moir\'{e}-scale model that ties band inversion to band overlap ([cf.\ Eqs.~\eqref{eq:C6_epsf}--\eqref{eq:Ep_Em}] below). It is an open question whether in the presence of general interactions the band symmetries remain robust, or whether a gapped TKI phase can occur intrinsically in a clean system.

In this letter, we show that due to the presence of non-local interactions the aforementioned symmetry is not protected, which will have general consequences for stacked and twisted TMDs.
Here, we focus on \MoTe/\WSe bilayers, where the quantum geometry arises from interlayer hybridization and induces symmetry-allowed spin loop currents (SLCs).
Through Fock exchange from non-local interactions, these SLCs qualitatively alter the effective bandstructure, which leads to an intrinsic TKI.
Starting from a realistic microscopic Hamiltonian, we obtain quantitative results using real-frequency dynamical mean-field theory (DMFT)~\cite{Georges1996,Kotliar2006}
together with Hartree-Fock (HF) to capture SLC-induced feedback from non-local interactions, which is crucial to obtain a TKI.

At hole density $\nu = 2$, we investigate the evolution of the gap under displacement-field tuning.
This evolution is governed by the interplay between electronic correlations and proximity to a topological phase transition, enabling controlled tuning from the chiral Kondo regime into a topological band-insulating regime as mixed-valence charge fluctuations in the local-moment layer are enhanced. As a result, DMFT is essential to accurately capture this crossover from strong to weak coupling.
Furthermore, our theory predicts a strong temperature sensitivity of the TKI, where thermal fluctuations destroy the SLC-induced topological gap, with direct consequences for electronic transport.

\textit{Model.---}
Our work is based on a tight-binding parametrization of the two highest-energy valence bands of AB-stacked \MoTe/\WSe~\cite{Devakul2022,Guerci2024}.
These form a honeycomb lattice with lattice vectors $\vec{a}_{1/2} = a_M(\pm \sqrt{3}/2,1/2)^T$ and moir\'e lattice constant $a_M \simeq 4.65 \, \mr{nm}$.
The two sites in the moir\'e unit cell reside in the $\mathrm{MoTe}_2$ and $\mathrm{WSe}_2$ layers, respectively; throughout the remainder of this work, we refer to them as $f$ and $c$ orbitals.
The tight-binding Hamiltonian is
\begin{align}
\label{eq:Hamiltonian}
H &= \sum_{\vec{k}\sigma} \xi^{c}_{\vec{k}\sigma} c^{\dagger}_{\vec{k}\sigma} c^{\pdag}_{\vec{k}\sigma} + U^c_{\mr{loc}} \sum_{i} n^{c}_{i\uparrow} n^{c}_{i\downarrow}
\\ \nonumber
&+ \sum_{\vec{k}\sigma} \xi^{f}_{\vec{k}} f^{\dagger}_{\vec{k}\sigma} f^{\pdag}_{\vec{k}\sigma} + U^f_{\mr{loc}} \sum_{i} n^{f}_{i\uparrow} n^{f}_{i\downarrow} + U^f_{nn} \sum_{\langle\!\langle i,j \rangle\!\rangle} n^{f}_i n^{f}_j
\\ \nonumber
&+ \sum_{\vec{k}\sigma} (V_{\vec{k}} c^{\dagger}_{\vec{k}\sigma} f^{\pdag}_{\vec{k}\sigma} + \mr{h.c.}) + U^{cf}_n \sum_{\langle i,j \rangle} n^{f}_{i} n^{c}_j \, ,
\end{align}
where $c_{\vec{k}\sigma}$ and $f_{\vec{k}\sigma}$ annihilate a momentum $\vec{k}$, spin $\sigma$ hole in the $c$ and $f$ band, respectively.
Further, $\xi^{f}_{\vec{k}\sigma} = \epsilon_{f\vec{k}} - \frac{\Delta}{2} - \mu$ and $\xi^{c}_{\vec{k}\sigma} = \epsilon_{c\vec{k}\sigma} + \frac{\Delta}{2} - \mu$,
with chemical potential $\mu$, displacement field $\Delta$ and dispersions $\epsilon_{c\vec{k}\sigma} = -2t_c \sum_{n} \cos(\vec{k} \boldsymbol{\gamma}_n + \phi^{c}_{\sigma})$
and $\epsilon_{f\vec{k}} = -2t_f \sum_{n} \cos(\vec{k} \boldsymbol{\gamma}_n)$, where $\boldsymbol{\gamma}_n = C^{n-1}_{3z} \vec{a}_2$, $\phi^{c}_{\sigma} = 2\pi s_{\sigma}/3$, and $s_{\uparrow} = -s_{\downarrow} = 1$.
The $\vec{k}$-dependent interlayer hybridization is $V_{\vec{k}} = - t_{\perp} \sum_{n} \mr{e}^{\mr{i}\vec{k}\vec{u}_n}$, with $\vec{u}_1 = a_M (1,0)^T/\sqrt{3}$ and $\vec{u}_n = C^{n-1}_{3} \vec{u}_1$.
Below the $f$ and $c$ hole densities are denoted by $\nu_f$ and $\nu_c$, respectively, and the total hole density by $\nu = \nu_f + \nu_c$.

We consider local interactions  $U^{f}_{\mr{loc}}$ and  $U^{c}_{\mr{loc}}$ in the $f$ and $c$ layers, a nearest-neighbor density interaction $U^{cf}_n$ and a next-nearest-neighbor density interaction in the $f$ layer, $U^{f}_{nn}$.
As we will show, $U^{f}_{nn}$ is
responsible for opening a full topological gap at $\nu = 2$.
Other interactions are smaller and not considered here for simplicity.
We choose $t_c = 8.3 \, \mr{meV}$, $t_f = 4 \, \mr{meV}$, and $t_{\perp} = 2 \, \mr{meV}$ for the hoppings, and
 $U^{f}_{\mr{loc}} = 90 \, \mr{meV}$, $U^{c}_{\mr{loc}} = 65 \, \mr{meV}$, $U^{cf}_{n} = 45 \, \mr{meV}$, and $U^{f}_{nn} = 20 \, \mr{meV}$ for the interactions.
 These choices are in reasonable agreement with ab-initio estimates~\cite{Guerci2024}.

 \textit{Method.---}
We treat $U^f_{\mr{loc}}$ within DMFT to capture the dynamics of local-moment formation and Kondo screening in the $f$-band,
while both the non-local and the $c$-band interactions are treated within HF, to keep computations manageable.
We use the QSpace-based~\cite{Weichselbaum2012,Weichselbaum2020,Weichselbaum2024} MuNRG implementation~\cite{Lee2016,Lee2017,Lee2021} of the Numerical Renormalization Group~(NRG)~\cite{Wilson1975,Weichselbaum2007,Bulla2008}
to solve the impurity model,
which provides real-frequency dynamics at arbitrary temperatures; see the Supplemental Material~(SM)~\cite{supplement} 
\nocite{Zitko2009,Kugler2022,BonbienManchon2020,Bastin1971,SmrckaStreda1977,Munoz2025}
for details.

The resulting retarded self-energy has the form
\begin{align}
\label{eq:SEf_HF}
\boldsymbol{\Sigma}_{\vec{k}\sigma}(\omega) =
\begin{pmatrix}
\Sigma^{f}_{\mr{loc}}(\omega) + \Sigma^{f}_{\vec{k}\sigma} & \Sigma^{fc}_{\vec{k}} \\
\Sigma^{cf}_{\vec{k}} & \Sigma^{c}
\end{pmatrix} \, ,
\end{align}
where $\Sigma^{f}_{\mr{loc}}(\omega)$ includes both the DMFT contribution and the local Hartree terms from $U^{cf}_{n}$ and $U^{f}_{nn}$,
while all other HF terms are static and given by
\begin{align}
\label{eq:SEf_nn}
\Sigma^{f}_{\vec{k}\sigma} &=
-2U^{f}_{nn} \chi^{ff} \sum_{n=1}^{3} \cos(\boldsymbol{\gamma}_n \vec{k} + s_{\sigma}\phi^{f})
\\ \nonumber
\Sigma^{fc}_{\vec{k}} &= -U^{cf}_{n}\chi^{fc} \sum_{n=1}^{3} \mr{e}^{\mr{i}\vec{k}\vec{u}_n} \, , \; \Sigma^{c} = \frac{U_{\mr{loc}}^{c}}{2} \nu_c + 3 U^{cf}_{n} \nu_f \, .
\end{align}
Here, we have defined the bond expectation values
\begin{align}
\label{eq:chifc_chiff}
\chi^{fc} = \langle f^{\dagger}_{i\sigma} c^{\pdag}_{j\sigma} \rangle \, , \quad \chi^{ff} \mr{e}^{\mr{i} s_{\sigma} \nu_{\ell\ell'} \phi^{f}} = \langle f^{\dagger}_{\ell\sigma} f^{\pdag}_{\ell'\sigma} \rangle \, ,
\end{align}
where $(i,j)$ and $(\ell,\ell')$ are nearest and next-nearest neighbors, respectively, $\nu_{\ell\ell'} = \pm 1$ if the bond points along $\pm \boldsymbol{\gamma}_n$, $n \in \{1,2,3\}$, and $\chi^{ff} = |\langle f^{\dagger}_{\ell\sigma} f^{\pdag}_{\ell'\sigma} \rangle|$.
Importantly, a spin and bond direction dependent phase $s_{\sigma} \nu_{\ell\ell'} \phi^{f}$ on $\langle f^{\dagger}_{\ell\sigma} f^{\pdag}_{\ell'\sigma} \rangle$ does not break the $C_{3z}$ or time reversal symmetries of the model.
Therefore, $\phi^{f}$ will generically be non-zero, which implies the presence of SLCs, as discussed below.

 \begin{figure}[tb!]
\includegraphics[width = \linewidth]{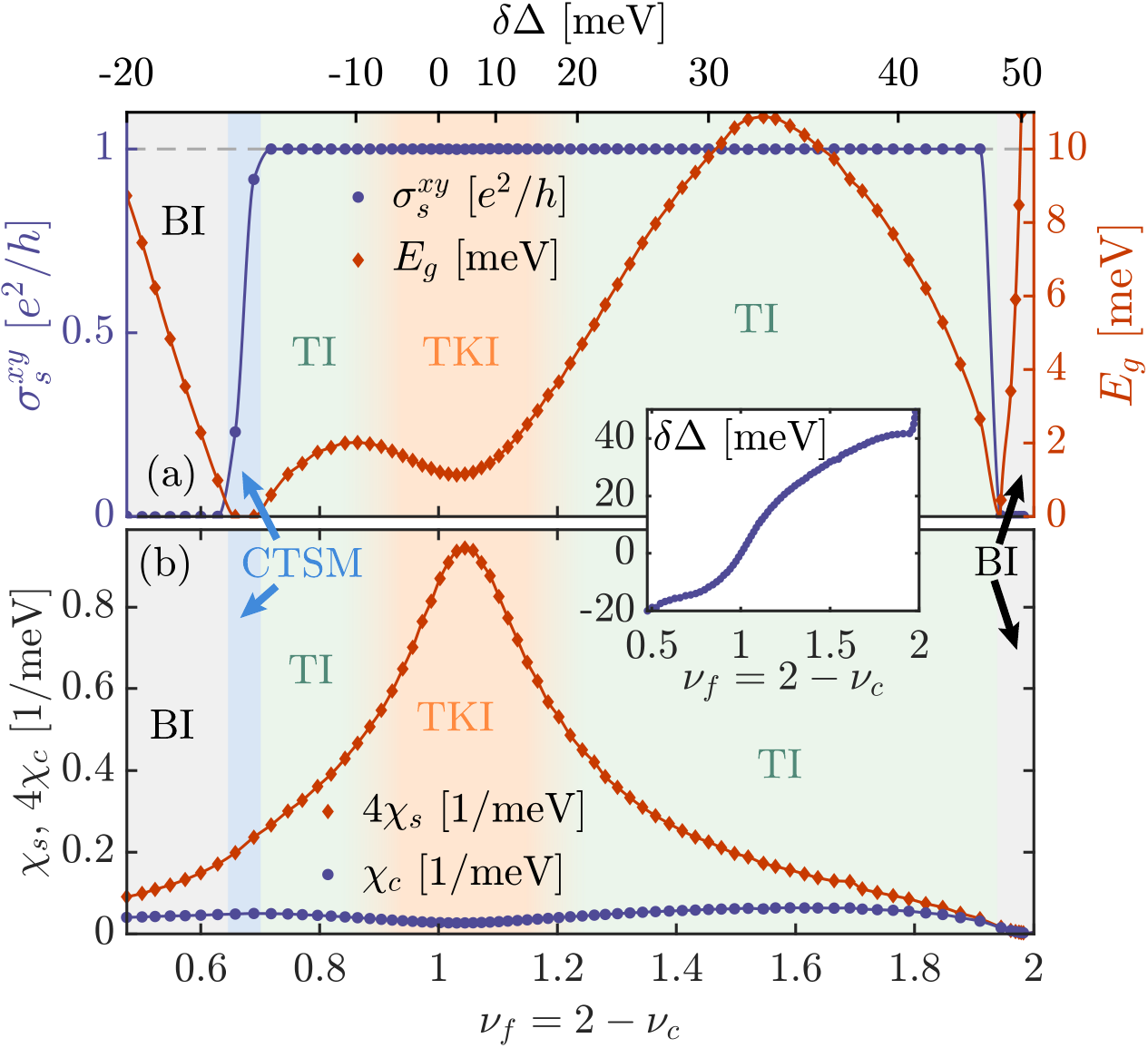}
\caption{Phase diagram of \MoTe/\WSe\ at hole filling $\nu = \nu_f + \nu_c = 2$ and $T=0$ with varying displacement field $\Delta$.
The top abscissa shows the displacement field $\delta \Delta = \Delta(\nu_f) - \Delta(\nu_f = 1)$ relative to $\nu_f = 1$, while all other abscissas show the equivalent $f$-band filling $\nu_f$.
The relation between $\delta\Delta$ and $\nu_f$ is shown in the inset.
In (a), we show the gap $E_g$ and the anomalous spin-Hall conductance $\sigma^{xy}_s$, while (b) shows
the local $f$-band spin and charge susceptibilities $\chi_s$ and $\chi_c$, respectively.
The latter are an indication of the degree of correlations, since $\chi_c = 4 \chi_s$ in the non-interacting limit and $\chi_c \ll 4 \chi_s$ in the Kondo limit.
Based on those quantities, we distinguish four different regimes. A topological Kondo insulator (TKI, orange) with quantized $\sigma^{xy}_s = e^2/h$ and strong correlations ($\chi_c \ll 4 \chi_s$) around $\nu_f = 1$,
a more weakly correlated topological insulator (TI, green) with $\sigma^{xy}_s = e^2/h$ further away from $\nu_f = 1$, a trivial band insulator (BI, gray) with $\sigma^{xy}_s = 0$ far away from $\nu_f = 1$, and
a compensated topological semimetal (CTSM, blue) with $E_g = 0$ and non-zero but non-quantized $\sigma^{xy}_s$, separating TI from BI in the $\nu_f < 1$ region.
TI and TKI correspond to the same phase and are only qualitatively distinguished by the degree of their correlations. We use $4 \chi_s/\chi_c > 20$ for the TKI and $4 \chi_s/\chi_c < 10$ for the TI, with a crossover region in between.
\label{fig:Fig3_nf_dependence}}
\end{figure}

\textit{Phase diagram at $\nu = 2$.---}
The resulting displacement field driven phase diagram at fixed $\nu = \nu_f + \nu_c= 2$ and $T=0$ is shown in Fig.~\ref{fig:Fig3_nf_dependence}.
Figure~\ref{fig:Fig3_nf_dependence}(a) shows the the gap $E_g$ and anomalous spin-Hall conductance $\sigma^{xy}_{s} = \tfrac{1}{2}(\sigma^{xy}_{\uparrow} -\sigma^{xy}_{\downarrow})$,
which indicate whether the system is insulating or semi-metallic and whether it is topological, respectively.
Figure~\ref{fig:Fig3_nf_dependence}(b) shows the local $f$-orbital spin and charge susceptibilities $\chi_s$ and $\chi_c$, respectively.
These are indicators for the degree of correlations, since $\chi_c = 4\chi_s$ in the non-interacting limit, while $\chi_c \ll 4\chi_s$ in the Kondo limit where local moments form in the $f$ orbitals.

As we vary the displacement field, we find four different regimes at $\nu = 2$.
Most importantly, around $\nu_f \simeq \nu_c \simeq 1$, we find an intrinsic TKI regime that does not require disorder or symmetry breaking, which appears to be the most consistent with recent experiments~\cite{Han2025}.
The TKI exhibits an (indirect) gap of $E_g \simeq 1 \, \mr{meV}$, a quantized spin-Hall conductance $\sigma_s^{xy} = e^2/h$,
and a local $f$-band spin susceptilility clearly exceeding the charge susceptibility, $\chi_c \ll 4\chi_s$, which places it well within the Kondo regime.
As we tune away from $\nu_f \simeq 1$, we find that correlations weaken and the gap increases while the spin-Hall conductance remains quantized, resulting in a more weakly correlated topological insulator~(TI).
Note that the TI and TKI belong to the same phase and are only qualitatively distinguished by the degree of their correlations.
For both the TKI and TI regime, it is absolutely crucial that (i) \MoTe/\WSe\ intrinsically supports SLCs and (ii) that the next-nearest neighbor interaction in the $f$-band, $U^f_{nn}$, is non-zero.
This allows SLCs to feed back on the effective bandstructure, which ultimately leads to an intrinsic full topological gap, as will be discussed in more detail below.

Tuning further away from $\nu_f \simeq \nu_c \simeq 1$ deep into the TI regime, the charge gap $E_g$ reaches a maximum and subsequently narrows, even as correlations continue to weaken and $\sigma^{xy}_s$ remains quantized.
This reduction of the gap signals an impending topological phase transition. As $|\nu_f - 1| \to 1$, the system must ultimately evolve into a weakly-correlated trivial band insulator~(BI) with $\sigma^{xy}_s = 0$.
Because the TI and BI states are topologically distinct, the gap must close between them.
For $\nu_f > 1$, this transition occurs at a single critical point near $\nu_f \simeq 1.94$.
In contrast, for $\nu_f < 1$, the transition occurs via an intermediate CTSM region ($0.65 \lesssim \nu_f \lesssim 0.7$) featuring a non-zero but non-quantized anomalous spin-Hall conductance.

\textit{Importance of non-local interactions.---}
Above, we have shown that \MoTe/\WSe\ supports an intrinsic (symmetric and disorder-free) TKI around $\nu_f \simeq \nu_c \simeq 1$, in stark contrast to previous findings in Refs.~\cite{Guerci2024,Xie2025}, which intrinsically found a CTSM.
These previous findings are due to a symmetry of the
non-interacting bandstructure of \MoTe/\WSe, which is not a symmetry of the full interacting Hamiltonian.
We will see below that non-local interactions in the presence of quantum geometry explicitly break this symmetry and enable an intrinsic TKI.

This symmetry
arises from two basic properties of the non-interacting bandstructure, (i) a 6-fold rotation symmetry of the continuum model moir\'e bands of \MoTe\ and \WSe\ around their band minima~\cite{Rademaker2022},
and (ii) the presence of nodes in the interlayer hybridization $V_{\vec{k}}$ and their location in $\vec{k}$-space due to the different $C_{3z}$ eigenvalues of the low-energy \MoTe\ and \WSe\ moir\'e bands~\cite{Zhang2021}.
Both properties are present in the continuum model and respected by the tight-binding fit.
Property (i) relates the $f$-band dispersion $\epsilon_{f\vec{k}}$ (minimum at $\boldsymbol{\gamma} = (0,0)^T$) at $\boldsymbol{\kappa} = \tfrac{4\pi}{3 a_M} (\sqrt{3}/2,-1/2)^T$ and $\boldsymbol{\kappa}' = \tfrac{4\pi}{3 a_M} (\sqrt{3}/2,1/2)^T$,
\begin{align}
\label{eq:C6_epsf}
\epsilon_{f\boldsymbol{\kappa}} = \epsilon_{f\boldsymbol{\kappa}'} \, .
\end{align}
It holds to a very good approximation at the moir\'e scale, but is weakly broken beyond the quadratic single-valley band approximation.
Property (ii) is exact and, in our gauge choice with real interlayer hopping, leads to hybridization nodes at $\boldsymbol{\kappa}$ and $\boldsymbol{\kappa}'$,
\begin{align}
\label{eq:hyb_nodes}
V_{\boldsymbol{\kappa}} = V_{\boldsymbol{\kappa}'} = 0 \, .
\end{align}
Therefore, the nodes of $V_{\vec{k}}$ align with the maxima of $\epsilon_{f\vec{k}}$.

Tuning $\Delta$ such that the $f$ and $c$ bands overlap triggers band inversion, with hybridized bands featuring spin-contrasting Chern numbers.
Denoting the upper and lower band dispersions by $E^{+}_{\vec{k}\sigma}$ and $E^{-}_{\vec{k}\sigma}$, respectively,
band inversion and \Eq{eq:hyb_nodes} necessarily lead to
\begin{align}
E^{+}_{\boldsymbol{\kappa}\sigma} &= \epsilon_{f\boldsymbol{\kappa}} & E^{-}_{\boldsymbol{\kappa}'\sigma} &= \epsilon_{f\boldsymbol{\kappa}'} \, ,
\end{align}
or vice versa. Combining this with Eq.~\eqref{eq:C6_epsf}, we find
\begin{align}
\label{eq:Ep_Em}
E^{+}_{\boldsymbol{\kappa}\sigma} = E^{-}_{\boldsymbol{\kappa}'\sigma} \, ,
\end{align}
i.e.\ the bands overlap if they are inverted, i.e. topological.

This aligns with the findings in Refs.~\cite{Guerci2024,Xie2025}, which generically found a CTSM at $\nu_f = \nu_c = 1$.
Ref.~\cite{Mendez2024} implemented a Hamiltonian which strongly violates Eq.~\eqref{eq:C6_epsf} and is therefore not compatible with the continuum models of Refs.~\cite{Zhang2021,Rademaker2022}.
In Ref.~\cite{Guerci2024}, Eq.~\eqref{eq:hyb_nodes} was lifted on average by introducing disorder to open a full band gap.

Since Eq.~\eqref{eq:Ep_Em} is accidental, i.e.\ not enforced by the $C_{3z}$ or time reversal symmetries of $H$,
it can be lifted, without symmetry-breaking, by interaction effects that effectively break the accidental 6-fold symmetry of $\epsilon_{f\vec{k}}$ in Eq.~\eqref{eq:C6_epsf}.
A physical quantity that does not respect this accidental symmetry is SLCs in the $f$ band.
Their presence is described by the spin current operator for bond $(\ell,\ell')$,
\begin{align}
\hat{J}^{s}_{\ell\ell'} = \mr{i} t_f \nu_{\ell\ell'} \sum_{\sigma} s_{\sigma} \left[ f^{\dagger}_{\ell\sigma} f^{\pdag}_{\ell'\sigma} - \mr{h.c.}\right] \, ,
\end{align}
with bond orientation dependent direction convention via $\nu_{\ell\ell'}$.
If, in Eq.~\eqref{eq:chifc_chiff}, $0 < \phi^{f} < \pi$, its expectation value
\begin{align}
\label{eq:SLC_EV}
J^{s} = \langle \hat{J}^{s}_{\ell\ell'} \rangle = 4 t_f \chi^{ff} \sin \phi^{f}
\end{align}
is non-zero and bond-independent, implying SLCs that run around triangles in the $f$ sublattice~\footnote{The interlayer hybridization leads to spin loop currents \textit{both} in the $f$ and the $c$ layer.
However, as a result of the bandwidth mismatch, they are much weaker in the latter, compared to the kinetic energy.
The reason is that the kinetic energy due to an $f$-$f$ bond is $-4 t_f \chi^{ff} \cos \phi^{f}$, i.e.\ it is minimized for $\phi^{f} = 0$, and likewise for $c$.
Since $|t_f| \ll |t_c|$, a phase mismatch between hopping amplitude and bond expectation value is energetically much more costly in the $c$ band than in the $f$ band.
As a result, such phase mismatches and therefore spin loop currents are most prominent in the $f$ band}.

\begin{figure}[tb!]
\includegraphics[width = \linewidth]{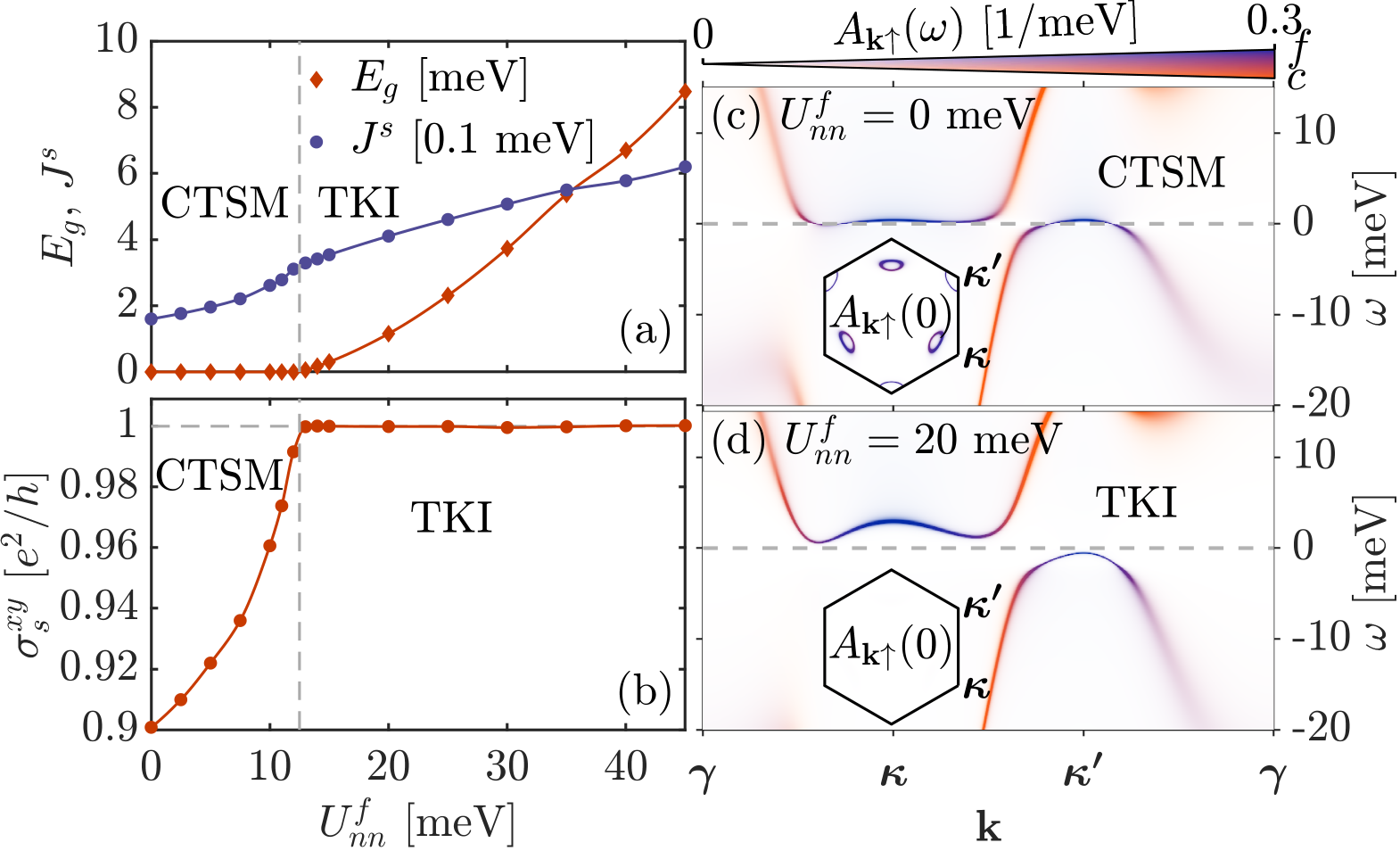}
\caption{(a) spin loop currents (blue) and gap (orange) versus $U^f_{nn}$. (b) spin Hall conductance versus $U^{f}_{nn}$.
(c,d) Total spectral function $A_{\bk \uparrow}(\omega) = A_{f \bk \uparrow}(\omega) + A_{c \bk \uparrow}(\omega)$, in (c) the CTSM and (d) the TKI phase.
The insets show $A_{\bk \uparrow}(0)$ (Fermi surface). All results are obtained at $T=0$. 
\label{fig:Fig1_Uff_dependence}}
\end{figure}

Feedback of the SLCs on the effective bandstructure requires a non-zero non-local $f$-band interaction $U^{f}_{nn}$
and, on the HF level, occurs via Eq.~\eqref{eq:SEf_nn}, which lifts property~\eqref{eq:C6_epsf} and thereby enables an intrinsic TKI.
For concreteness, consider the interacting retarded Green's function
\begin{align}
\label{eq:Ginv}
\mathbf{G}^{-1}_{\vec{k}\sigma}(\omega) = \omega  -
\begin{pmatrix}
\xi_{f\vec{k}} & V_{\vec{k}} \\
V_{\vec{k}}^{\ast} &\xi_{c\vec{k}\sigma}
\end{pmatrix}
- \boldsymbol{\Sigma}_{\vec{k}\sigma}(\omega) \, .
\end{align}
Topology is encoded in the effective Hamiltonian~\cite{Wang2012}
\begin{align}
\label{eq:Heff}
\mathbf{H}^{\mr{eff}}_{\vec{k}\sigma} = -\tfrac{1}{2}\bigl(\mathbf{G}^{-1}_{\vec{k}\sigma}(0) + [\mathbf{G}^{-1}_{\vec{k}\sigma}(0)]^{\dagger}\bigr) \, .
\end{align}
Its \textit{effective} $f$-band dispersion
\begin{align}
\label{eq:efk_effective}
\epsilon^{\mr{eff}}_{f\vec{k}\sigma} = \epsilon_{f\vec{k}} + \Sigma^{f}_{\mr{loc}}(0)  + \Sigma^{f}_{\vec{k}\sigma} \, , \quad \epsilon^{\mr{eff}}_{f\boldsymbol{\kappa}\sigma} \neq \epsilon^{\mr{eff}}_{f\boldsymbol{\kappa}'\sigma}
\end{align}
does not fulfill Eq.~\eqref{eq:C6_epsf} if $U^{f}_{nn} \neq 0$ and $\phi^{f} \neq 0$.

To show how a non-zero $U^{f}_{nn}$ induces a TKI
we have performed DMFT+HF calculations at filling $\nu = \nu_f + \nu_c = 2$ and $\nu_f \simeq \nu_c \simeq 1$ with varying $U^{f}_{nn}$,
while the remaining parameters are fixed as specified below Eq.~\eqref{eq:Hamiltonian}.
Figure~\ref{fig:Fig1_Uff_dependence}(a,b) shows the resulting $J^{s}$, gap $E_g$, and spin-Hall conductance $\sigma^{xy}_{s} = \tfrac{1}{2}(\sigma^{xy}_{\uparrow} -\sigma^{xy}_{\downarrow})$.
At $U^{f}_{nn} = 0$, $J^{s}$ is non-zero, while $E_g = 0$ and $\sigma^{xy}_{s} \simeq 0.9 e^2/h$ is not quantized, i.e.\ the system is a CTSM, consistent with Refs.~\cite{Guerci2024,Xie2025}.
However, from the corresponding spectral function, shown in Fig.~\ref{fig:Fig1_Uff_dependence}(c), we can infer that the system is very close to a TKI.
Spectral features in the vicinity of the Fermi level are extremely flat due to strong correlations, and the bands only overlap in a narrow energy window.
Further, the inset of Fig.~\ref{fig:Fig1_Uff_dependence}(c) shows that the electron and hole pockets occupy only a tiny fraction of the Brillouin zone.

Once $U^{f}_{nn}$ is increased to non-zero values, $J^{s}$ moderately increases while $\sigma^{xy}_{s}$ rises steeply until $\sigma^{xy}_{s} = e^2/h$ is reached around $U^f_{nn} = 13 \, \mr{meV}$, where a full topological gap opens.
This gap increases with increasing $U^f_{nn}$ and remains topological for $U^f_{nn} \geq 13 \, \mr{meV}$.
Realistic values for $U^{f}_{nn}$ are in the 20-30 meV range, where $E_g \simeq \textrm{1-4 meV}$, thus it is reasonable that AB-stacked \MoTe/\WSe\ may be a SLC-induced TKI.
However, even though our Hamiltonian parameters are reasonable, there are uncertainties, also due to dynamical screening from remote bands or even phonons~\cite{Lau2025}.
Since AB-stacked \MoTe/\WSe\ is at the boundary between CTSM and TKI, a definitive prediction is therefore not possible.

Figure~\ref{fig:Fig1_Uff_dependence}(d) shows the spectral function at $U^{f}_{nn} = 20 \,  \mr{meV}$, our value of choice for the results shown in Figs.~\ref{fig:Fig3_nf_dependence} and~\ref{fig:Fig2_T_dependence}.
Compared to $U^{f}_{nn} = 0$, changes in the spectral function are only visible at low energies, where we find that non-zero $U^{f}_{nn}$ leads to more dispersive and coherent features, and a full gap.

\begin{figure}[tb!]
\includegraphics[width = \linewidth]{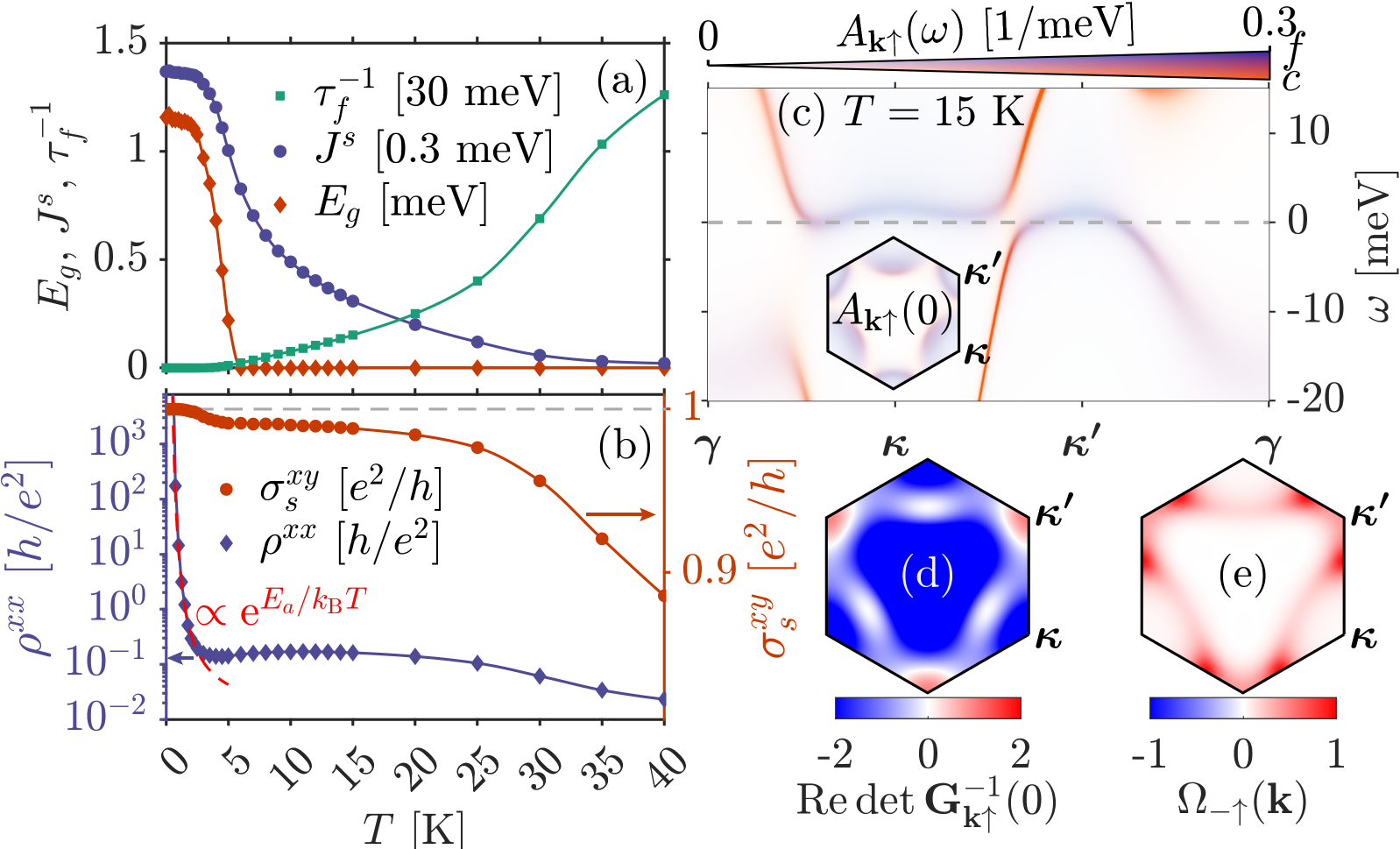}
\caption{(a) the gap $E_g$, the SLCs $J^s$, and the $f$ band scattering rate $\tau^{-1}_f =  -\mr{Im} \, \Sigma^{f}_{\mr{loc}}(\omega = 0)$ versus temperature.
(b) Resistivity (blue) and spin Hall conductance (orange) versus temperature.
The red dashed line: Arrhenius fit for $T \leq 2 \, \mr{K}$.
(c) Total spectral function $A_{\bk \uparrow}(\omega) = A_{f \bk \uparrow}(\omega) + A_{c \bk \uparrow}(\omega)$ at $T=15$ K; the inset shows $A_{\bk \uparrow}(0)$. (d) 
real part of $\mr{det} \, \mathbf{G}^{-1}_{\vec{k}\uparrow}(0)$ and (e) valence-band Berry curvature of $\mathbf{H}^{\mr{eff}}_{\bk\uparrow}$ [c.f.\ Eq.~\eqref{eq:Heff}].
\label{fig:Fig2_T_dependence}}
\end{figure}

\textit{Temperature dependence.---}
Last, we study the temperature dependence of the SLC-induced TKI, focusing on the
gap and transport coefficients, which are experimentally accessible.
In Fig.~\ref{fig:Fig2_T_dependence}(a), we show the temperature dependence of the gap $E_g$, alongside $J^s$, $\chi^{ff}$ [c.f.\ Eq.~\eqref{eq:chifc_chiff}],
and the imaginary part of $\Sigma^{f}_{\mr{loc}}(\omega = 0)$, which corresponds to the $f$-band scattering rate at the Fermi level.
Figure~\ref{fig:Fig2_T_dependence}(b) shows the corresponding resistivity $\rho^{xx}(T)$ and $\sigma^{xy}_{s}(T)$, see the SM~\cite{supplement} for computational details.
At $T=0$, the gap is $E_g = 1.16 \, \mr{meV}$, which corresponds to $13.5 \, \mr{K}$, while $\sigma^{xy}_{s}(0) = e^2/h$ is quantized, and $\rho^{xx}(T\to 0) \to \infty$.

When the system is heated, both the gap and the SLCs are almost temperature independent for $T \lesssim 2 \, \mr{K}$.
In the same temperature region, $\sigma^{xy}_{s}(T)$ is quantized and the resistivity has Arrhenius behavior, $\rho^{xx}(T) \propto \mr{e}^{E_a/k_{\mr{B}} T}$, i.e.\ it behaves as a thermally activated rigid-gap insulator.
We find an activation energy of $2 E_a = 1.22 \, \mr{meV}$, in reasonable agreement with $E_g = 1.16 \, \mr{meV}$.

When heating beyond $T \sim 2 \, \mr{K}$, thermal fluctuations start to significantly reduce the SLCs with increasing temperature, which causes the gap to collapse over a relatively small temperature window $2 \, \mr{K} \lesssim T \lesssim 5 \, \mr{K}$.
In that temperature window, $\chi^{ff}$ decreases less drastically than $J^{s}$, i.e.\ the $J^s$ reduction is driven largely by a reduction of $\phi^{f}$.
Once the gap collapses, the resistivity deviates from its rigid-gap $\propto \mr{e}^{E_a/k_{\mr{B}} T}$ behavior above 2\,K, becomes almost temperature independent between 2 and 30\,K, and slowly decreases further towards even higher temperatures.

Surprisingly, even though the gap rapidly collapses for $T \gtrsim 2 \, \mr{K}$, the resistivity is \textit{larger} than what would be expected from an extrapolation of the low-temperature $\propto \mr{e}^{E_a/k_{\mr{B}} T}$ dependence, even though there should be more mobile charge carriers.
This counterintuitive behavior is due to (i) a strong increase of the $f$-band scattering rate $\tau^{-1}_{f} = -\mr{Im} \, \Sigma^{f}_{\mr{loc}}(0)$ and (ii) a decrease of SLCs, which decreases the effective $f$-band velocity $v^{\alpha}_{f\vec{k}\sigma} = \partial_{k^{\alpha}} \epsilon_{f\vec{k}\sigma}^{\mr{eff}}$.
Therefore, increasing temperature strongly decreases the mean free path $l_f\sim v_f \tau_f$ in the $f$-band, which contributes most of the activated charge carriers at the Fermi level.
The dominant $f$ content at the Fermi level is apparent from the spectral function at $T= 15\,\mr{K}$, shown in Fig.~\ref{fig:Fig2_T_dependence}(c).
By contrasting this with $T=0$ [Fig.~\ref{fig:Fig1_Uff_dependence}(d)], the decreased coherence and dispersiveness at $T= 15\,\mr{K}$ is also clearly visible.
This decrease in the mean free path counteracts charge carrier activation, leading to the resistivity plateau.

The spectral function, together with the valence-band Berry curvature $\Omega_{-\uparrow}(\vec{k})$ of $\mathbf{H}^{\mr{eff}}_{\vec{k}\uparrow}$ [Fig.~\ref{fig:Fig2_T_dependence}(e)] and the real part of the determinant of $\mathbf{G}_{\vec{k}\uparrow}^{-1}(0)$ [Fig.~\ref{fig:Fig2_T_dependence}(d)] also help to understand the behavior of $\sigma^{xy}_{s}(T)$. The latter slightly deviates from its quantized $T \to 0$ value once the gap collapses,
becomes almost constant for $5\, \mr{K} \lesssim T \lesssim 30 \, \mr{K}$, and shows significant temperature dependence only beyond 30\,K.
This is due to the distribution of Berry curvature in the Brillouin zone, which is concentrated in regions somewhat separated from the unoccupied region around $\boldsymbol{\kappa}'$ where $\mr{Re} \, \mr{det}\, \mathbf{G}_{\vec{k}\uparrow}^{-1}(0) > 0$ [red in Fig.~\ref{fig:Fig2_T_dependence}(d)].
The spectral weight in the region where $\Omega_{-\uparrow}(\vec{k})$ is large is well below the Fermi level, with significant thermal activation occurring only beyond 30\,K.

\textit{Conclusion and Outlook.---}
We have studied a tight-binding model relevant to AB-stacked MoTe$_2$/WSe$_2$ using real-frequency DMFT plus HF. Our calculations show that the interplay between quantum geometry and non-local interactions qualitatively changes the physics and enables an SLC-induced TKI in an experimentally relevant parameter regime. At $\nu=2$, displacement-field tuning first drives a crossover from a strongly correlated TKI near $\nu_f \simeq 1$ to a more weakly correlated TI as mixed-valence fluctuations grow, and then a topological transition to a trivial BI: for $\nu_f>1$ via a single gap closing near $\nu_f \simeq 1.94$, and for $\nu_f<1$ via an intermediate CTSM window, $0.65 \lesssim \nu_f \lesssim 0.7$. 
Further, we find that the gap of the TKI at $\nu_f \simeq 1$ collapses at temperatures much lower than the gap scale, with observable consequences for $\rho^{xx}(T)$.

Interestingly, similar behavior has been found in recent experiments on AA-stacked \MoTe/\WSe~\cite{Han2025}.
In its TKI regime, it has a topological gap slightly larger than $1\, \mr{meV}$, which closes around $2\, \mr{K}$ as indicated by the edge contribution to the resistance.
Tuning away from the TKI regime by displacement-field reveals an increase of the gap, followed by a gap closing, a metallic region, and a subsequent opening of a trivial gap, which mirrors our findings for $\nu_f < 1$.
It is an interesting future direction to conduct a similar study on AA-stacked \MoTe/\WSe.

Another interesting direction for future work on \MoTe/\WSe\ would be to identify a parameter regime where Kondo breakdown~\cite{Si2001,Coleman2001,Senthil2003,Senthil2004,Gleis2024,Gleis2025} physics occurs.
One particularly interesting issue would be its interplay with topology, which arises from a $p$-wave interlayer hybridization in this system.

\textit{Acknowledgements.---}
We thank Andrei Bernevig, Piers Coleman,  Kin-Fai Mak, Juan Felipe Mendez-Valderrama, and Jie Shan for useful discussions.
AG is supported by the Abrahams Postdoctoral Fellowship
of the Center for Materials Theory at Rutgers University.
K.P.L. and J.H.P. are partially supported by NSF Career Grant No.~DMR-1941569, and
J.H.P. has been supported in part by the US Department of Energy Office of Science, Basic Energy Sciences, under Award No.~DE-SC0026023 (work involved the conceptualization and construction of the project, as well as the interpretation of results and the presentation). PJ was supported by Office of Basic
Energy Sciences, Material Sciences and Engineering Division, U.S. Department of Energy (DOE) under Contract
DE-FG02-99ER45790.
The work of AJM was supported in part by the National Science Foundation (NSF) Materials
Research Science and Engineering Centers (MRSEC) program
through Columbia University under the Precision-Assembled
Quantum Materials (PAQM) Grant No. DMR-2011738. 
The Flatiron Institute is a division of the Simons Foundation.

\bibliography{references}

\begin{thebibliography}{74}%
\makeatletter
\providecommand \@ifxundefined [1]{%
 \@ifx{#1\undefined}
}%
\providecommand \@ifnum [1]{%
 \ifnum #1\expandafter \@firstoftwo
 \else \expandafter \@secondoftwo
 \fi
}%
\providecommand \@ifx [1]{%
 \ifx #1\expandafter \@firstoftwo
 \else \expandafter \@secondoftwo
 \fi
}%
\providecommand \natexlab [1]{#1}%
\providecommand \enquote  [1]{``#1''}%
\providecommand \bibnamefont  [1]{#1}%
\providecommand \bibfnamefont [1]{#1}%
\providecommand \citenamefont [1]{#1}%
\providecommand \href@noop [0]{\@secondoftwo}%
\providecommand \href [0]{\begingroup \@sanitize@url \@href}%
\providecommand \@href[1]{\@@startlink{#1}\@@href}%
\providecommand \@@href[1]{\endgroup#1\@@endlink}%
\providecommand \@sanitize@url [0]{\catcode `\\12\catcode `\$12\catcode
  `\&12\catcode `\#12\catcode `\^12\catcode `\_12\catcode `\%12\relax}%
\providecommand \@@startlink[1]{}%
\providecommand \@@endlink[0]{}%
\providecommand \url  [0]{\begingroup\@sanitize@url \@url }%
\providecommand \@url [1]{\endgroup\@href {#1}{\urlprefix }}%
\providecommand \urlprefix  [0]{URL }%
\providecommand \Eprint [0]{\href }%
\providecommand \doibase [0]{https://doi.org/}%
\providecommand \selectlanguage [0]{\@gobble}%
\providecommand \bibinfo  [0]{\@secondoftwo}%
\providecommand \bibfield  [0]{\@secondoftwo}%
\providecommand \translation [1]{[#1]}%
\providecommand \BibitemOpen [0]{}%
\providecommand \bibitemStop [0]{}%
\providecommand \bibitemNoStop [0]{.\EOS\space}%
\providecommand \EOS [0]{\spacefactor3000\relax}%
\providecommand \BibitemShut  [1]{\csname bibitem#1\endcsname}%
\let\auto@bib@innerbib\@empty
\bibitem [{\citenamefont {Kennes}\ \emph {et~al.}(2021)\citenamefont {Kennes},
  \citenamefont {Claassen}, \citenamefont {Xian}, \citenamefont {Georges},
  \citenamefont {Millis}, \citenamefont {Hone}, \citenamefont {Dean},
  \citenamefont {Basov}, \citenamefont {Pasupathy},\ and\ \citenamefont
  {Rubio}}]{Kennes2021}%
  \BibitemOpen
  \bibfield  {author} {\bibinfo {author} {\bibfnamefont {D.~M.}\ \bibnamefont
  {Kennes}}, \bibinfo {author} {\bibfnamefont {M.}~\bibnamefont {Claassen}},
  \bibinfo {author} {\bibfnamefont {L.}~\bibnamefont {Xian}}, \bibinfo {author}
  {\bibfnamefont {A.}~\bibnamefont {Georges}}, \bibinfo {author} {\bibfnamefont
  {A.~J.}\ \bibnamefont {Millis}}, \bibinfo {author} {\bibfnamefont
  {J.}~\bibnamefont {Hone}}, \bibinfo {author} {\bibfnamefont {C.~R.}\
  \bibnamefont {Dean}}, \bibinfo {author} {\bibfnamefont {D.~N.}\ \bibnamefont
  {Basov}}, \bibinfo {author} {\bibfnamefont {A.~N.}\ \bibnamefont
  {Pasupathy}},\ and\ \bibinfo {author} {\bibfnamefont {A.}~\bibnamefont
  {Rubio}},\ }\bibfield  {title} {\bibinfo {title} {Moiré heterostructures as
  a condensed-matter quantum simulator},\ }\href
  {https://doi.org/10.1038/s41567-020-01154-3} {\bibfield  {journal} {\bibinfo
  {journal} {Nature Physics}\ }\textbf {\bibinfo {volume} {17}},\ \bibinfo
  {pages} {155} (\bibinfo {year} {2021})}\BibitemShut {NoStop}%
\bibitem [{\citenamefont {Mak}\ and\ \citenamefont {Shan}(2022)}]{Mak2022}%
  \BibitemOpen
  \bibfield  {author} {\bibinfo {author} {\bibfnamefont {K.~F.}\ \bibnamefont
  {Mak}}\ and\ \bibinfo {author} {\bibfnamefont {J.}~\bibnamefont {Shan}},\
  }\bibfield  {title} {\bibinfo {title} {Semiconductor moiré materials},\
  }\href {https://doi.org/10.1038/s41565-022-01165-6} {\bibfield  {journal}
  {\bibinfo  {journal} {Nature Nanotechnology}\ }\textbf {\bibinfo {volume}
  {17}},\ \bibinfo {pages} {686} (\bibinfo {year} {2022})}\BibitemShut
  {NoStop}%
\bibitem [{\citenamefont {Cai}\ \emph {et~al.}(2023)\citenamefont {Cai},
  \citenamefont {Anderson}, \citenamefont {Wang}, \citenamefont {Zhang},
  \citenamefont {Liu}, \citenamefont {Holtzmann}, \citenamefont {Zhang},
  \citenamefont {Fan}, \citenamefont {Taniguchi}, \citenamefont {Watanabe},
  \citenamefont {Ran}, \citenamefont {Cao}, \citenamefont {Fu}, \citenamefont
  {Xiao}, \citenamefont {Yao},\ and\ \citenamefont {Xu}}]{Cai2023}%
  \BibitemOpen
  \bibfield  {author} {\bibinfo {author} {\bibfnamefont {J.}~\bibnamefont
  {Cai}}, \bibinfo {author} {\bibfnamefont {E.}~\bibnamefont {Anderson}},
  \bibinfo {author} {\bibfnamefont {C.}~\bibnamefont {Wang}}, \bibinfo {author}
  {\bibfnamefont {X.}~\bibnamefont {Zhang}}, \bibinfo {author} {\bibfnamefont
  {X.}~\bibnamefont {Liu}}, \bibinfo {author} {\bibfnamefont {W.}~\bibnamefont
  {Holtzmann}}, \bibinfo {author} {\bibfnamefont {Y.}~\bibnamefont {Zhang}},
  \bibinfo {author} {\bibfnamefont {F.}~\bibnamefont {Fan}}, \bibinfo {author}
  {\bibfnamefont {T.}~\bibnamefont {Taniguchi}}, \bibinfo {author}
  {\bibfnamefont {K.}~\bibnamefont {Watanabe}}, \bibinfo {author}
  {\bibfnamefont {Y.}~\bibnamefont {Ran}}, \bibinfo {author} {\bibfnamefont
  {T.}~\bibnamefont {Cao}}, \bibinfo {author} {\bibfnamefont {L.}~\bibnamefont
  {Fu}}, \bibinfo {author} {\bibfnamefont {D.}~\bibnamefont {Xiao}}, \bibinfo
  {author} {\bibfnamefont {W.}~\bibnamefont {Yao}},\ and\ \bibinfo {author}
  {\bibfnamefont {X.}~\bibnamefont {Xu}},\ }\bibfield  {title} {\bibinfo
  {title} {Signatures of fractional quantum anomalous hall states in twisted
  mote2},\ }\href {https://doi.org/10.1038/s41586-023-06289-w} {\bibfield
  {journal} {\bibinfo  {journal} {Nature}\ }\textbf {\bibinfo {volume} {622}},\
  \bibinfo {pages} {63} (\bibinfo {year} {2023})}\BibitemShut {NoStop}%
\bibitem [{\citenamefont {Jia}\ \emph {et~al.}(2025)\citenamefont {Jia},
  \citenamefont {Song}, \citenamefont {Zheng}, \citenamefont {Cheng},
  \citenamefont {Uzan}, \citenamefont {Yu}, \citenamefont {Tang}, \citenamefont
  {Pollak}, \citenamefont {Yuan}, \citenamefont {Onyszczak}, \citenamefont
  {Watanabe}, \citenamefont {Taniguchi}, \citenamefont {Lei}, \citenamefont
  {Yao}, \citenamefont {Schoop}, \citenamefont {Ong},\ and\ \citenamefont
  {Wu}}]{Jia2025}%
  \BibitemOpen
  \bibfield  {author} {\bibinfo {author} {\bibfnamefont {Y.}~\bibnamefont
  {Jia}}, \bibinfo {author} {\bibfnamefont {T.}~\bibnamefont {Song}}, \bibinfo
  {author} {\bibfnamefont {Z.~J.}\ \bibnamefont {Zheng}}, \bibinfo {author}
  {\bibfnamefont {G.}~\bibnamefont {Cheng}}, \bibinfo {author} {\bibfnamefont
  {A.~J.}\ \bibnamefont {Uzan}}, \bibinfo {author} {\bibfnamefont
  {G.}~\bibnamefont {Yu}}, \bibinfo {author} {\bibfnamefont {Y.}~\bibnamefont
  {Tang}}, \bibinfo {author} {\bibfnamefont {C.~J.}\ \bibnamefont {Pollak}},
  \bibinfo {author} {\bibfnamefont {F.}~\bibnamefont {Yuan}}, \bibinfo {author}
  {\bibfnamefont {M.}~\bibnamefont {Onyszczak}}, \bibinfo {author}
  {\bibfnamefont {K.}~\bibnamefont {Watanabe}}, \bibinfo {author}
  {\bibfnamefont {T.}~\bibnamefont {Taniguchi}}, \bibinfo {author}
  {\bibfnamefont {S.}~\bibnamefont {Lei}}, \bibinfo {author} {\bibfnamefont
  {N.}~\bibnamefont {Yao}}, \bibinfo {author} {\bibfnamefont {L.~M.}\
  \bibnamefont {Schoop}}, \bibinfo {author} {\bibfnamefont {N.~P.}\
  \bibnamefont {Ong}},\ and\ \bibinfo {author} {\bibfnamefont {S.}~\bibnamefont
  {Wu}},\ }\bibfield  {title} {\bibinfo {title} {Anomalous superconductivity in
  twisted mote 2 nanojunctions},\ }\bibfield  {journal} {\bibinfo  {journal}
  {Science Advances}\ }\textbf {\bibinfo {volume} {11}},\ \href
  {https://doi.org/10.1126/sciadv.adq5712} {10.1126/sciadv.adq5712} (\bibinfo
  {year} {2025})\BibitemShut {NoStop}%
\bibitem [{\citenamefont {Xu}\ \emph {et~al.}(2025)\citenamefont {Xu},
  \citenamefont {Sun}, \citenamefont {Li}, \citenamefont {Zheng}, \citenamefont
  {Xu}, \citenamefont {Gao}, \citenamefont {Jia}, \citenamefont {Watanabe},
  \citenamefont {Taniguchi}, \citenamefont {Tong}, \citenamefont {Lu},
  \citenamefont {Jia}, \citenamefont {Shi}, \citenamefont {Jiang},
  \citenamefont {Zhang}, \citenamefont {Zhang}, \citenamefont {Lei},
  \citenamefont {Liu},\ and\ \citenamefont {Li}}]{Xu2025}%
  \BibitemOpen
  \bibfield  {author} {\bibinfo {author} {\bibfnamefont {F.}~\bibnamefont
  {Xu}}, \bibinfo {author} {\bibfnamefont {Z.}~\bibnamefont {Sun}}, \bibinfo
  {author} {\bibfnamefont {J.}~\bibnamefont {Li}}, \bibinfo {author}
  {\bibfnamefont {C.}~\bibnamefont {Zheng}}, \bibinfo {author} {\bibfnamefont
  {C.}~\bibnamefont {Xu}}, \bibinfo {author} {\bibfnamefont {J.}~\bibnamefont
  {Gao}}, \bibinfo {author} {\bibfnamefont {T.}~\bibnamefont {Jia}}, \bibinfo
  {author} {\bibfnamefont {K.}~\bibnamefont {Watanabe}}, \bibinfo {author}
  {\bibfnamefont {T.}~\bibnamefont {Taniguchi}}, \bibinfo {author}
  {\bibfnamefont {B.}~\bibnamefont {Tong}}, \bibinfo {author} {\bibfnamefont
  {L.}~\bibnamefont {Lu}}, \bibinfo {author} {\bibfnamefont {J.}~\bibnamefont
  {Jia}}, \bibinfo {author} {\bibfnamefont {Z.}~\bibnamefont {Shi}}, \bibinfo
  {author} {\bibfnamefont {S.}~\bibnamefont {Jiang}}, \bibinfo {author}
  {\bibfnamefont {Y.}~\bibnamefont {Zhang}}, \bibinfo {author} {\bibfnamefont
  {Y.}~\bibnamefont {Zhang}}, \bibinfo {author} {\bibfnamefont
  {S.}~\bibnamefont {Lei}}, \bibinfo {author} {\bibfnamefont {X.}~\bibnamefont
  {Liu}},\ and\ \bibinfo {author} {\bibfnamefont {T.}~\bibnamefont {Li}},\
  }\href {https://doi.org/10.48550/ARXIV.2504.06972} {\bibinfo {title}
  {Signatures of unconventional superconductivity near reentrant and fractional
  quantum anomalous hall insulators}} (\bibinfo {year} {2025})\BibitemShut
  {NoStop}%
\bibitem [{\citenamefont {Ghiotto}\ \emph {et~al.}(2021)\citenamefont
  {Ghiotto}, \citenamefont {Shih}, \citenamefont {Pereira}, \citenamefont
  {Rhodes}, \citenamefont {Kim}, \citenamefont {Zang}, \citenamefont {Millis},
  \citenamefont {Watanabe}, \citenamefont {Taniguchi}, \citenamefont {Hone},
  \citenamefont {Wang}, \citenamefont {Dean},\ and\ \citenamefont
  {Pasupathy}}]{Ghiotto2021}%
  \BibitemOpen
  \bibfield  {author} {\bibinfo {author} {\bibfnamefont {A.}~\bibnamefont
  {Ghiotto}}, \bibinfo {author} {\bibfnamefont {E.-M.}\ \bibnamefont {Shih}},
  \bibinfo {author} {\bibfnamefont {G.~S. S.~G.}\ \bibnamefont {Pereira}},
  \bibinfo {author} {\bibfnamefont {D.~A.}\ \bibnamefont {Rhodes}}, \bibinfo
  {author} {\bibfnamefont {B.}~\bibnamefont {Kim}}, \bibinfo {author}
  {\bibfnamefont {J.}~\bibnamefont {Zang}}, \bibinfo {author} {\bibfnamefont
  {A.~J.}\ \bibnamefont {Millis}}, \bibinfo {author} {\bibfnamefont
  {K.}~\bibnamefont {Watanabe}}, \bibinfo {author} {\bibfnamefont
  {T.}~\bibnamefont {Taniguchi}}, \bibinfo {author} {\bibfnamefont {J.~C.}\
  \bibnamefont {Hone}}, \bibinfo {author} {\bibfnamefont {L.}~\bibnamefont
  {Wang}}, \bibinfo {author} {\bibfnamefont {C.~R.}\ \bibnamefont {Dean}},\
  and\ \bibinfo {author} {\bibfnamefont {A.~N.}\ \bibnamefont {Pasupathy}},\
  }\bibfield  {title} {\bibinfo {title} {Quantum criticality in twisted
  transition metal dichalcogenides},\ }\href
  {https://doi.org/10.1038/s41586-021-03815-6} {\bibfield  {journal} {\bibinfo
  {journal} {Nature}\ }\textbf {\bibinfo {volume} {597}},\ \bibinfo {pages}
  {345} (\bibinfo {year} {2021})}\BibitemShut {NoStop}%
\bibitem [{\citenamefont {Wei}\ \emph {et~al.}(2024)\citenamefont {Wei},
  \citenamefont {Xu}, \citenamefont {He}, \citenamefont {Li}, \citenamefont
  {Huang}, \citenamefont {Zhu}, \citenamefont {Watanabe}, \citenamefont
  {Taniguchi}, \citenamefont {Claassen}, \citenamefont {Rhodes}, \citenamefont
  {Kennes}, \citenamefont {Xian}, \citenamefont {Rubio},\ and\ \citenamefont
  {Wang}}]{Wei2024}%
  \BibitemOpen
  \bibfield  {author} {\bibinfo {author} {\bibfnamefont {L.}~\bibnamefont
  {Wei}}, \bibinfo {author} {\bibfnamefont {Q.}~\bibnamefont {Xu}}, \bibinfo
  {author} {\bibfnamefont {Y.}~\bibnamefont {He}}, \bibinfo {author}
  {\bibfnamefont {Q.}~\bibnamefont {Li}}, \bibinfo {author} {\bibfnamefont
  {Y.}~\bibnamefont {Huang}}, \bibinfo {author} {\bibfnamefont
  {W.}~\bibnamefont {Zhu}}, \bibinfo {author} {\bibfnamefont {K.}~\bibnamefont
  {Watanabe}}, \bibinfo {author} {\bibfnamefont {T.}~\bibnamefont {Taniguchi}},
  \bibinfo {author} {\bibfnamefont {M.}~\bibnamefont {Claassen}}, \bibinfo
  {author} {\bibfnamefont {D.~A.}\ \bibnamefont {Rhodes}}, \bibinfo {author}
  {\bibfnamefont {D.~M.}\ \bibnamefont {Kennes}}, \bibinfo {author}
  {\bibfnamefont {L.}~\bibnamefont {Xian}}, \bibinfo {author} {\bibfnamefont
  {A.}~\bibnamefont {Rubio}},\ and\ \bibinfo {author} {\bibfnamefont
  {L.}~\bibnamefont {Wang}},\ }\bibfield  {title} {\bibinfo {title} {Linear
  resistivity at van hove singularities in twisted bilayer wse 2},\ }\bibfield
  {journal} {\bibinfo  {journal} {Proceedings of the National Academy of
  Sciences}\ }\textbf {\bibinfo {volume} {121}},\ \href
  {https://doi.org/10.1073/pnas.2321665121} {10.1073/pnas.2321665121} (\bibinfo
  {year} {2024})\BibitemShut {NoStop}%
\bibitem [{\citenamefont {Xia}\ \emph {et~al.}(2024)\citenamefont {Xia},
  \citenamefont {Han}, \citenamefont {Watanabe}, \citenamefont {Taniguchi},
  \citenamefont {Shan},\ and\ \citenamefont {Mak}}]{Xia2024}%
  \BibitemOpen
  \bibfield  {author} {\bibinfo {author} {\bibfnamefont {Y.}~\bibnamefont
  {Xia}}, \bibinfo {author} {\bibfnamefont {Z.}~\bibnamefont {Han}}, \bibinfo
  {author} {\bibfnamefont {K.}~\bibnamefont {Watanabe}}, \bibinfo {author}
  {\bibfnamefont {T.}~\bibnamefont {Taniguchi}}, \bibinfo {author}
  {\bibfnamefont {J.}~\bibnamefont {Shan}},\ and\ \bibinfo {author}
  {\bibfnamefont {K.~F.}\ \bibnamefont {Mak}},\ }\bibfield  {title} {\bibinfo
  {title} {Superconductivity in twisted bilayer wse2},\ }\href
  {https://doi.org/10.1038/s41586-024-08116-2} {\bibfield  {journal} {\bibinfo
  {journal} {Nature}\ }\textbf {\bibinfo {volume} {637}},\ \bibinfo {pages}
  {833} (\bibinfo {year} {2024})}\BibitemShut {NoStop}%
\bibitem [{\citenamefont {Guo}\ \emph {et~al.}(2025)\citenamefont {Guo},
  \citenamefont {Pack}, \citenamefont {Swann}, \citenamefont {Holtzman},
  \citenamefont {Cothrine}, \citenamefont {Watanabe}, \citenamefont
  {Taniguchi}, \citenamefont {Mandrus}, \citenamefont {Barmak}, \citenamefont
  {Hone}, \citenamefont {Millis}, \citenamefont {Pasupathy},\ and\
  \citenamefont {Dean}}]{Guo2025}%
  \BibitemOpen
  \bibfield  {author} {\bibinfo {author} {\bibfnamefont {Y.}~\bibnamefont
  {Guo}}, \bibinfo {author} {\bibfnamefont {J.}~\bibnamefont {Pack}}, \bibinfo
  {author} {\bibfnamefont {J.}~\bibnamefont {Swann}}, \bibinfo {author}
  {\bibfnamefont {L.}~\bibnamefont {Holtzman}}, \bibinfo {author}
  {\bibfnamefont {M.}~\bibnamefont {Cothrine}}, \bibinfo {author}
  {\bibfnamefont {K.}~\bibnamefont {Watanabe}}, \bibinfo {author}
  {\bibfnamefont {T.}~\bibnamefont {Taniguchi}}, \bibinfo {author}
  {\bibfnamefont {D.~G.}\ \bibnamefont {Mandrus}}, \bibinfo {author}
  {\bibfnamefont {K.}~\bibnamefont {Barmak}}, \bibinfo {author} {\bibfnamefont
  {J.}~\bibnamefont {Hone}}, \bibinfo {author} {\bibfnamefont {A.~J.}\
  \bibnamefont {Millis}}, \bibinfo {author} {\bibfnamefont {A.}~\bibnamefont
  {Pasupathy}},\ and\ \bibinfo {author} {\bibfnamefont {C.~R.}\ \bibnamefont
  {Dean}},\ }\bibfield  {title} {\bibinfo {title} {Superconductivity in 5.0°
  twisted bilayer wse2},\ }\href {https://doi.org/10.1038/s41586-024-08381-1}
  {\bibfield  {journal} {\bibinfo  {journal} {Nature}\ }\textbf {\bibinfo
  {volume} {637}},\ \bibinfo {pages} {839} (\bibinfo {year}
  {2025})}\BibitemShut {NoStop}%
\bibitem [{\citenamefont {Xia}\ \emph {et~al.}(2025)\citenamefont {Xia},
  \citenamefont {Han}, \citenamefont {Zhu}, \citenamefont {Zhang},
  \citenamefont {Knüppel}, \citenamefont {Watanabe}, \citenamefont
  {Taniguchi}, \citenamefont {Mak},\ and\ \citenamefont {Shan}}]{Xia2025}%
  \BibitemOpen
  \bibfield  {author} {\bibinfo {author} {\bibfnamefont {Y.}~\bibnamefont
  {Xia}}, \bibinfo {author} {\bibfnamefont {Z.}~\bibnamefont {Han}}, \bibinfo
  {author} {\bibfnamefont {J.}~\bibnamefont {Zhu}}, \bibinfo {author}
  {\bibfnamefont {Y.}~\bibnamefont {Zhang}}, \bibinfo {author} {\bibfnamefont
  {P.}~\bibnamefont {Knüppel}}, \bibinfo {author} {\bibfnamefont
  {K.}~\bibnamefont {Watanabe}}, \bibinfo {author} {\bibfnamefont
  {T.}~\bibnamefont {Taniguchi}}, \bibinfo {author} {\bibfnamefont {K.~F.}\
  \bibnamefont {Mak}},\ and\ \bibinfo {author} {\bibfnamefont {J.}~\bibnamefont
  {Shan}},\ }\href {https://doi.org/10.48550/ARXIV.2508.02662} {\bibinfo
  {title} {Simulating high-temperature superconductivity in moiré wse2}}
  (\bibinfo {year} {2025})\BibitemShut {NoStop}%
\bibitem [{\citenamefont {Regan}\ \emph {et~al.}(2020)\citenamefont {Regan},
  \citenamefont {Wang}, \citenamefont {Jin}, \citenamefont {Bakti~Utama},
  \citenamefont {Gao}, \citenamefont {Wei}, \citenamefont {Zhao}, \citenamefont
  {Zhao}, \citenamefont {Zhang}, \citenamefont {Yumigeta}, \citenamefont
  {Blei}, \citenamefont {Carlström}, \citenamefont {Watanabe}, \citenamefont
  {Taniguchi}, \citenamefont {Tongay}, \citenamefont {Crommie}, \citenamefont
  {Zettl},\ and\ \citenamefont {Wang}}]{Regan2020}%
  \BibitemOpen
  \bibfield  {author} {\bibinfo {author} {\bibfnamefont {E.~C.}\ \bibnamefont
  {Regan}}, \bibinfo {author} {\bibfnamefont {D.}~\bibnamefont {Wang}},
  \bibinfo {author} {\bibfnamefont {C.}~\bibnamefont {Jin}}, \bibinfo {author}
  {\bibfnamefont {M.~I.}\ \bibnamefont {Bakti~Utama}}, \bibinfo {author}
  {\bibfnamefont {B.}~\bibnamefont {Gao}}, \bibinfo {author} {\bibfnamefont
  {X.}~\bibnamefont {Wei}}, \bibinfo {author} {\bibfnamefont {S.}~\bibnamefont
  {Zhao}}, \bibinfo {author} {\bibfnamefont {W.}~\bibnamefont {Zhao}}, \bibinfo
  {author} {\bibfnamefont {Z.}~\bibnamefont {Zhang}}, \bibinfo {author}
  {\bibfnamefont {K.}~\bibnamefont {Yumigeta}}, \bibinfo {author}
  {\bibfnamefont {M.}~\bibnamefont {Blei}}, \bibinfo {author} {\bibfnamefont
  {J.~D.}\ \bibnamefont {Carlström}}, \bibinfo {author} {\bibfnamefont
  {K.}~\bibnamefont {Watanabe}}, \bibinfo {author} {\bibfnamefont
  {T.}~\bibnamefont {Taniguchi}}, \bibinfo {author} {\bibfnamefont
  {S.}~\bibnamefont {Tongay}}, \bibinfo {author} {\bibfnamefont
  {M.}~\bibnamefont {Crommie}}, \bibinfo {author} {\bibfnamefont
  {A.}~\bibnamefont {Zettl}},\ and\ \bibinfo {author} {\bibfnamefont
  {F.}~\bibnamefont {Wang}},\ }\bibfield  {title} {\bibinfo {title} {Mott and
  generalized wigner crystal states in wse2/ws2 moir\'e superlattices},\ }\href
  {https://doi.org/10.1038/s41586-020-2092-4} {\bibfield  {journal} {\bibinfo
  {journal} {Nature}\ }\textbf {\bibinfo {volume} {579}},\ \bibinfo {pages}
  {359} (\bibinfo {year} {2020})}\BibitemShut {NoStop}%
\bibitem [{\citenamefont {Li}\ \emph {et~al.}(2021{\natexlab{a}})\citenamefont
  {Li}, \citenamefont {Li}, \citenamefont {Regan}, \citenamefont {Wang},
  \citenamefont {Zhao}, \citenamefont {Kahn}, \citenamefont {Yumigeta},
  \citenamefont {Blei}, \citenamefont {Taniguchi}, \citenamefont {Watanabe},
  \citenamefont {Tongay}, \citenamefont {Zettl}, \citenamefont {Crommie},\ and\
  \citenamefont {Wang}}]{Li2021b}%
  \BibitemOpen
  \bibfield  {author} {\bibinfo {author} {\bibfnamefont {H.}~\bibnamefont
  {Li}}, \bibinfo {author} {\bibfnamefont {S.}~\bibnamefont {Li}}, \bibinfo
  {author} {\bibfnamefont {E.~C.}\ \bibnamefont {Regan}}, \bibinfo {author}
  {\bibfnamefont {D.}~\bibnamefont {Wang}}, \bibinfo {author} {\bibfnamefont
  {W.}~\bibnamefont {Zhao}}, \bibinfo {author} {\bibfnamefont {S.}~\bibnamefont
  {Kahn}}, \bibinfo {author} {\bibfnamefont {K.}~\bibnamefont {Yumigeta}},
  \bibinfo {author} {\bibfnamefont {M.}~\bibnamefont {Blei}}, \bibinfo {author}
  {\bibfnamefont {T.}~\bibnamefont {Taniguchi}}, \bibinfo {author}
  {\bibfnamefont {K.}~\bibnamefont {Watanabe}}, \bibinfo {author}
  {\bibfnamefont {S.}~\bibnamefont {Tongay}}, \bibinfo {author} {\bibfnamefont
  {A.}~\bibnamefont {Zettl}}, \bibinfo {author} {\bibfnamefont {M.~F.}\
  \bibnamefont {Crommie}},\ and\ \bibinfo {author} {\bibfnamefont
  {F.}~\bibnamefont {Wang}},\ }\bibfield  {title} {\bibinfo {title} {Imaging
  two-dimensional generalized wigner crystals},\ }\href
  {https://doi.org/10.1038/s41586-021-03874-9} {\bibfield  {journal} {\bibinfo
  {journal} {Nature}\ }\textbf {\bibinfo {volume} {597}},\ \bibinfo {pages}
  {650} (\bibinfo {year} {2021}{\natexlab{a}})}\BibitemShut {NoStop}%
\bibitem [{\citenamefont {Li}\ \emph {et~al.}(2021{\natexlab{b}})\citenamefont
  {Li}, \citenamefont {Jiang}, \citenamefont {Li}, \citenamefont {Zhang},
  \citenamefont {Kang}, \citenamefont {Zhu}, \citenamefont {Watanabe},
  \citenamefont {Taniguchi}, \citenamefont {Chowdhury}, \citenamefont {Fu},
  \citenamefont {Shan},\ and\ \citenamefont {Mak}}]{Li2021a}%
  \BibitemOpen
  \bibfield  {author} {\bibinfo {author} {\bibfnamefont {T.}~\bibnamefont
  {Li}}, \bibinfo {author} {\bibfnamefont {S.}~\bibnamefont {Jiang}}, \bibinfo
  {author} {\bibfnamefont {L.}~\bibnamefont {Li}}, \bibinfo {author}
  {\bibfnamefont {Y.}~\bibnamefont {Zhang}}, \bibinfo {author} {\bibfnamefont
  {K.}~\bibnamefont {Kang}}, \bibinfo {author} {\bibfnamefont {J.}~\bibnamefont
  {Zhu}}, \bibinfo {author} {\bibfnamefont {K.}~\bibnamefont {Watanabe}},
  \bibinfo {author} {\bibfnamefont {T.}~\bibnamefont {Taniguchi}}, \bibinfo
  {author} {\bibfnamefont {D.}~\bibnamefont {Chowdhury}}, \bibinfo {author}
  {\bibfnamefont {L.}~\bibnamefont {Fu}}, \bibinfo {author} {\bibfnamefont
  {J.}~\bibnamefont {Shan}},\ and\ \bibinfo {author} {\bibfnamefont {K.~F.}\
  \bibnamefont {Mak}},\ }\bibfield  {title} {\bibinfo {title} {Continuous mott
  transition in semiconductor moir\'e superlattices},\ }\href
  {https://doi.org/10.1038/s41586-021-03853-0} {\bibfield  {journal} {\bibinfo
  {journal} {Nature}\ }\textbf {\bibinfo {volume} {597}},\ \bibinfo {pages}
  {350} (\bibinfo {year} {2021}{\natexlab{b}})}\BibitemShut {NoStop}%
\bibitem [{\citenamefont {Li}\ \emph {et~al.}(2021{\natexlab{c}})\citenamefont
  {Li}, \citenamefont {Jiang}, \citenamefont {Shen}, \citenamefont {Zhang},
  \citenamefont {Li}, \citenamefont {Tao}, \citenamefont {Devakul},
  \citenamefont {Watanabe}, \citenamefont {Taniguchi}, \citenamefont {Fu},
  \citenamefont {Shan},\ and\ \citenamefont {Mak}}]{Li2021}%
  \BibitemOpen
  \bibfield  {author} {\bibinfo {author} {\bibfnamefont {T.}~\bibnamefont
  {Li}}, \bibinfo {author} {\bibfnamefont {S.}~\bibnamefont {Jiang}}, \bibinfo
  {author} {\bibfnamefont {B.}~\bibnamefont {Shen}}, \bibinfo {author}
  {\bibfnamefont {Y.}~\bibnamefont {Zhang}}, \bibinfo {author} {\bibfnamefont
  {L.}~\bibnamefont {Li}}, \bibinfo {author} {\bibfnamefont {Z.}~\bibnamefont
  {Tao}}, \bibinfo {author} {\bibfnamefont {T.}~\bibnamefont {Devakul}},
  \bibinfo {author} {\bibfnamefont {K.}~\bibnamefont {Watanabe}}, \bibinfo
  {author} {\bibfnamefont {T.}~\bibnamefont {Taniguchi}}, \bibinfo {author}
  {\bibfnamefont {L.}~\bibnamefont {Fu}}, \bibinfo {author} {\bibfnamefont
  {J.}~\bibnamefont {Shan}},\ and\ \bibinfo {author} {\bibfnamefont {K.~F.}\
  \bibnamefont {Mak}},\ }\bibfield  {title} {\bibinfo {title} {Quantum
  anomalous hall effect from intertwined moiré bands},\ }\href
  {https://doi.org/10.1038/s41586-021-04171-1} {\bibfield  {journal} {\bibinfo
  {journal} {Nature}\ }\textbf {\bibinfo {volume} {600}},\ \bibinfo {pages}
  {641} (\bibinfo {year} {2021}{\natexlab{c}})}\BibitemShut {NoStop}%
\bibitem [{\citenamefont {Pan}\ \emph {et~al.}(2022)\citenamefont {Pan},
  \citenamefont {Xie}, \citenamefont {Wu},\ and\ \citenamefont
  {Das~Sarma}}]{Pan2022}%
  \BibitemOpen
  \bibfield  {author} {\bibinfo {author} {\bibfnamefont {H.}~\bibnamefont
  {Pan}}, \bibinfo {author} {\bibfnamefont {M.}~\bibnamefont {Xie}}, \bibinfo
  {author} {\bibfnamefont {F.}~\bibnamefont {Wu}},\ and\ \bibinfo {author}
  {\bibfnamefont {S.}~\bibnamefont {Das~Sarma}},\ }\bibfield  {title} {\bibinfo
  {title} {Topological phases in ab-stacked mote2/wse2 : Z2 topological
  insulators, chern insulators, and topological charge density waves},\ }\href
  {https://doi.org/10.1103/physrevlett.129.056804} {\bibfield  {journal}
  {\bibinfo  {journal} {Physical Review Letters}\ }\textbf {\bibinfo {volume}
  {129}},\ \bibinfo {pages} {056804} (\bibinfo {year} {2022})}\BibitemShut
  {NoStop}%
\bibitem [{\citenamefont {Devakul}\ and\ \citenamefont
  {Fu}(2022)}]{Devakul2022}%
  \BibitemOpen
  \bibfield  {author} {\bibinfo {author} {\bibfnamefont {T.}~\bibnamefont
  {Devakul}}\ and\ \bibinfo {author} {\bibfnamefont {L.}~\bibnamefont {Fu}},\
  }\bibfield  {title} {\bibinfo {title} {Quantum anomalous hall effect from
  inverted charge transfer gap},\ }\href
  {https://doi.org/10.1103/physrevx.12.021031} {\bibfield  {journal} {\bibinfo
  {journal} {Physical Review X}\ }\textbf {\bibinfo {volume} {12}},\ \bibinfo
  {pages} {021031} (\bibinfo {year} {2022})}\BibitemShut {NoStop}%
\bibitem [{\citenamefont {Xie}\ \emph {et~al.}(2022)\citenamefont {Xie},
  \citenamefont {Zhang}, \citenamefont {Hu}, \citenamefont {Mak},\ and\
  \citenamefont {Law}}]{Xie2022}%
  \BibitemOpen
  \bibfield  {author} {\bibinfo {author} {\bibfnamefont {Y.-M.}\ \bibnamefont
  {Xie}}, \bibinfo {author} {\bibfnamefont {C.-P.}\ \bibnamefont {Zhang}},
  \bibinfo {author} {\bibfnamefont {J.-X.}\ \bibnamefont {Hu}}, \bibinfo
  {author} {\bibfnamefont {K.~F.}\ \bibnamefont {Mak}},\ and\ \bibinfo {author}
  {\bibfnamefont {K.}~\bibnamefont {Law}},\ }\bibfield  {title} {\bibinfo
  {title} {Valley-polarized quantum anomalous hall state in moir\'e mote2/wse2
  heterobilayers},\ }\href {https://doi.org/10.1103/physrevlett.128.026402}
  {\bibfield  {journal} {\bibinfo  {journal} {Physical Review Letters}\
  }\textbf {\bibinfo {volume} {128}},\ \bibinfo {pages} {026402} (\bibinfo
  {year} {2022})}\BibitemShut {NoStop}%
\bibitem [{\citenamefont {Xie}\ \emph {et~al.}(2023)\citenamefont {Xie},
  \citenamefont {Pan}, \citenamefont {Wu},\ and\ \citenamefont
  {Das~Sarma}}]{Xie2023}%
  \BibitemOpen
  \bibfield  {author} {\bibinfo {author} {\bibfnamefont {M.}~\bibnamefont
  {Xie}}, \bibinfo {author} {\bibfnamefont {H.}~\bibnamefont {Pan}}, \bibinfo
  {author} {\bibfnamefont {F.}~\bibnamefont {Wu}},\ and\ \bibinfo {author}
  {\bibfnamefont {S.}~\bibnamefont {Das~Sarma}},\ }\bibfield  {title} {\bibinfo
  {title} {Nematic excitonic insulator in transition metal dichalcogenide
  moir\'e heterobilayers},\ }\href
  {https://doi.org/10.1103/physrevlett.131.046402} {\bibfield  {journal}
  {\bibinfo  {journal} {Physical Review Letters}\ }\textbf {\bibinfo {volume}
  {131}},\ \bibinfo {pages} {046402} (\bibinfo {year} {2023})}\BibitemShut
  {NoStop}%
\bibitem [{\citenamefont {Dong}\ and\ \citenamefont {Zhang}(2023)}]{Dong2023}%
  \BibitemOpen
  \bibfield  {author} {\bibinfo {author} {\bibfnamefont {Z.}~\bibnamefont
  {Dong}}\ and\ \bibinfo {author} {\bibfnamefont {Y.-H.}\ \bibnamefont
  {Zhang}},\ }\bibfield  {title} {\bibinfo {title} {Excitonic chern insulator
  and kinetic ferromagnetism in a mote2/wse2 moir\'e bilayer},\ }\href
  {https://doi.org/10.1103/physrevb.107.l081101} {\bibfield  {journal}
  {\bibinfo  {journal} {Physical Review B}\ }\textbf {\bibinfo {volume}
  {107}},\ \bibinfo {pages} {l081101} (\bibinfo {year} {2023})}\BibitemShut
  {NoStop}%
\bibitem [{\citenamefont {Xie}\ \emph {et~al.}(2024)\citenamefont {Xie},
  \citenamefont {Zhang},\ and\ \citenamefont {Law}}]{Xie2024}%
  \BibitemOpen
  \bibfield  {author} {\bibinfo {author} {\bibfnamefont {Y.-M.}\ \bibnamefont
  {Xie}}, \bibinfo {author} {\bibfnamefont {C.-P.}\ \bibnamefont {Zhang}},\
  and\ \bibinfo {author} {\bibfnamefont {K.~T.}\ \bibnamefont {Law}},\
  }\bibfield  {title} {\bibinfo {title} {Topological px+ipy intervalley
  coherent state in moir\'e mote2/wse2 heterobilayers},\ }\href
  {https://doi.org/10.1103/physrevb.110.045115} {\bibfield  {journal} {\bibinfo
   {journal} {Physical Review B}\ }\textbf {\bibinfo {volume} {110}},\ \bibinfo
  {pages} {045115} (\bibinfo {year} {2024})}\BibitemShut {NoStop}%
\bibitem [{\citenamefont {Tao}\ \emph {et~al.}(2024)\citenamefont {Tao},
  \citenamefont {Shen}, \citenamefont {Jiang}, \citenamefont {Li},
  \citenamefont {Li}, \citenamefont {Ma}, \citenamefont {Zhao}, \citenamefont
  {Hu}, \citenamefont {Pistunova}, \citenamefont {Watanabe}, \citenamefont
  {Taniguchi}, \citenamefont {Heinz}, \citenamefont {Mak},\ and\ \citenamefont
  {Shan}}]{Tao2024}%
  \BibitemOpen
  \bibfield  {author} {\bibinfo {author} {\bibfnamefont {Z.}~\bibnamefont
  {Tao}}, \bibinfo {author} {\bibfnamefont {B.}~\bibnamefont {Shen}}, \bibinfo
  {author} {\bibfnamefont {S.}~\bibnamefont {Jiang}}, \bibinfo {author}
  {\bibfnamefont {T.}~\bibnamefont {Li}}, \bibinfo {author} {\bibfnamefont
  {L.}~\bibnamefont {Li}}, \bibinfo {author} {\bibfnamefont {L.}~\bibnamefont
  {Ma}}, \bibinfo {author} {\bibfnamefont {W.}~\bibnamefont {Zhao}}, \bibinfo
  {author} {\bibfnamefont {J.}~\bibnamefont {Hu}}, \bibinfo {author}
  {\bibfnamefont {K.}~\bibnamefont {Pistunova}}, \bibinfo {author}
  {\bibfnamefont {K.}~\bibnamefont {Watanabe}}, \bibinfo {author}
  {\bibfnamefont {T.}~\bibnamefont {Taniguchi}}, \bibinfo {author}
  {\bibfnamefont {T.~F.}\ \bibnamefont {Heinz}}, \bibinfo {author}
  {\bibfnamefont {K.~F.}\ \bibnamefont {Mak}},\ and\ \bibinfo {author}
  {\bibfnamefont {J.}~\bibnamefont {Shan}},\ }\bibfield  {title} {\bibinfo
  {title} {Valley-coherent quantum anomalous hall state in ab-stacked
  mote2/wse2 bilayers},\ }\href {https://doi.org/10.1103/physrevx.14.011004}
  {\bibfield  {journal} {\bibinfo  {journal} {Physical Review X}\ }\textbf
  {\bibinfo {volume} {14}},\ \bibinfo {pages} {011004} (\bibinfo {year}
  {2024})}\BibitemShut {NoStop}%
\bibitem [{\citenamefont {Zhao}\ \emph
  {et~al.}(2024{\natexlab{a}})\citenamefont {Zhao}, \citenamefont {Kang},
  \citenamefont {Zhang}, \citenamefont {Knüppel}, \citenamefont {Tao},
  \citenamefont {Li}, \citenamefont {Tschirhart}, \citenamefont {Redekop},
  \citenamefont {Watanabe}, \citenamefont {Taniguchi}, \citenamefont {Young},
  \citenamefont {Shan},\ and\ \citenamefont {Mak}}]{Zhao2024}%
  \BibitemOpen
  \bibfield  {author} {\bibinfo {author} {\bibfnamefont {W.}~\bibnamefont
  {Zhao}}, \bibinfo {author} {\bibfnamefont {K.}~\bibnamefont {Kang}}, \bibinfo
  {author} {\bibfnamefont {Y.}~\bibnamefont {Zhang}}, \bibinfo {author}
  {\bibfnamefont {P.}~\bibnamefont {Knüppel}}, \bibinfo {author}
  {\bibfnamefont {Z.}~\bibnamefont {Tao}}, \bibinfo {author} {\bibfnamefont
  {L.}~\bibnamefont {Li}}, \bibinfo {author} {\bibfnamefont {C.~L.}\
  \bibnamefont {Tschirhart}}, \bibinfo {author} {\bibfnamefont
  {E.}~\bibnamefont {Redekop}}, \bibinfo {author} {\bibfnamefont
  {K.}~\bibnamefont {Watanabe}}, \bibinfo {author} {\bibfnamefont
  {T.}~\bibnamefont {Taniguchi}}, \bibinfo {author} {\bibfnamefont {A.~F.}\
  \bibnamefont {Young}}, \bibinfo {author} {\bibfnamefont {J.}~\bibnamefont
  {Shan}},\ and\ \bibinfo {author} {\bibfnamefont {K.~F.}\ \bibnamefont
  {Mak}},\ }\bibfield  {title} {\bibinfo {title} {Realization of the haldane
  chern insulator in a moir\'e lattice},\ }\href
  {https://doi.org/10.1038/s41567-023-02284-0} {\bibfield  {journal} {\bibinfo
  {journal} {Nature Physics}\ }\textbf {\bibinfo {volume} {20}},\ \bibinfo
  {pages} {275} (\bibinfo {year} {2024}{\natexlab{a}})}\BibitemShut {NoStop}%
\bibitem [{\citenamefont {Guerci}\ \emph {et~al.}(2023)\citenamefont {Guerci},
  \citenamefont {Wang}, \citenamefont {Zang}, \citenamefont {Cano},
  \citenamefont {Pixley},\ and\ \citenamefont {Millis}}]{Guerci2023}%
  \BibitemOpen
  \bibfield  {author} {\bibinfo {author} {\bibfnamefont {D.}~\bibnamefont
  {Guerci}}, \bibinfo {author} {\bibfnamefont {J.}~\bibnamefont {Wang}},
  \bibinfo {author} {\bibfnamefont {J.}~\bibnamefont {Zang}}, \bibinfo {author}
  {\bibfnamefont {J.}~\bibnamefont {Cano}}, \bibinfo {author} {\bibfnamefont
  {J.~H.}\ \bibnamefont {Pixley}},\ and\ \bibinfo {author} {\bibfnamefont
  {A.}~\bibnamefont {Millis}},\ }\bibfield  {title} {\bibinfo {title} {Chiral
  kondo lattice in doped mote${}_2$/wse${}_2$ bilayers},\ }\href
  {https://doi.org/10.1126/sciadv.ade7701} {\bibfield  {journal} {\bibinfo
  {journal} {Science Advances}\ }\textbf {\bibinfo {volume} {9}},\ \bibinfo
  {pages} {eade7701} (\bibinfo {year} {2023})},\ \Eprint
  {https://arxiv.org/abs/https://www.science.org/doi/pdf/10.1126/sciadv.ade7701}
  {https://www.science.org/doi/pdf/10.1126/sciadv.ade7701} \BibitemShut
  {NoStop}%
\bibitem [{\citenamefont {Zhao}\ \emph {et~al.}(2023)\citenamefont {Zhao},
  \citenamefont {Shen}, \citenamefont {Tao}, \citenamefont {Han}, \citenamefont
  {Kang}, \citenamefont {Watanabe}, \citenamefont {Taniguchi}, \citenamefont
  {Mak},\ and\ \citenamefont {Shan}}]{Zhao2023}%
  \BibitemOpen
  \bibfield  {author} {\bibinfo {author} {\bibfnamefont {W.}~\bibnamefont
  {Zhao}}, \bibinfo {author} {\bibfnamefont {B.}~\bibnamefont {Shen}}, \bibinfo
  {author} {\bibfnamefont {Z.}~\bibnamefont {Tao}}, \bibinfo {author}
  {\bibfnamefont {Z.}~\bibnamefont {Han}}, \bibinfo {author} {\bibfnamefont
  {K.}~\bibnamefont {Kang}}, \bibinfo {author} {\bibfnamefont {K.}~\bibnamefont
  {Watanabe}}, \bibinfo {author} {\bibfnamefont {T.}~\bibnamefont {Taniguchi}},
  \bibinfo {author} {\bibfnamefont {K.~F.}\ \bibnamefont {Mak}},\ and\ \bibinfo
  {author} {\bibfnamefont {J.}~\bibnamefont {Shan}},\ }\bibfield  {title}
  {\bibinfo {title} {Gate-tunable heavy fermions in a moiré kondo lattice},\
  }\href {https://doi.org/10.1038/s41586-023-05800-7} {\bibfield  {journal}
  {\bibinfo  {journal} {Nature}\ }\textbf {\bibinfo {volume} {616}},\ \bibinfo
  {pages} {61} (\bibinfo {year} {2023})}\BibitemShut {NoStop}%
\bibitem [{\citenamefont {Zhao}\ \emph
  {et~al.}(2024{\natexlab{b}})\citenamefont {Zhao}, \citenamefont {Shen},
  \citenamefont {Tao}, \citenamefont {Kim}, \citenamefont {Knüppel},
  \citenamefont {Han}, \citenamefont {Zhang}, \citenamefont {Watanabe},
  \citenamefont {Taniguchi}, \citenamefont {Chowdhury}, \citenamefont {Shan},\
  and\ \citenamefont {Mak}}]{Zhao2024a}%
  \BibitemOpen
  \bibfield  {author} {\bibinfo {author} {\bibfnamefont {W.}~\bibnamefont
  {Zhao}}, \bibinfo {author} {\bibfnamefont {B.}~\bibnamefont {Shen}}, \bibinfo
  {author} {\bibfnamefont {Z.}~\bibnamefont {Tao}}, \bibinfo {author}
  {\bibfnamefont {S.}~\bibnamefont {Kim}}, \bibinfo {author} {\bibfnamefont
  {P.}~\bibnamefont {Knüppel}}, \bibinfo {author} {\bibfnamefont
  {Z.}~\bibnamefont {Han}}, \bibinfo {author} {\bibfnamefont {Y.}~\bibnamefont
  {Zhang}}, \bibinfo {author} {\bibfnamefont {K.}~\bibnamefont {Watanabe}},
  \bibinfo {author} {\bibfnamefont {T.}~\bibnamefont {Taniguchi}}, \bibinfo
  {author} {\bibfnamefont {D.}~\bibnamefont {Chowdhury}}, \bibinfo {author}
  {\bibfnamefont {J.}~\bibnamefont {Shan}},\ and\ \bibinfo {author}
  {\bibfnamefont {K.~F.}\ \bibnamefont {Mak}},\ }\bibfield  {title} {\bibinfo
  {title} {Emergence of ferromagnetism at the onset of moiré kondo
  breakdown},\ }\href {https://doi.org/10.1038/s41567-024-02636-4} {\bibfield
  {journal} {\bibinfo  {journal} {Nature Physics}\ }\textbf {\bibinfo {volume}
  {20}},\ \bibinfo {pages} {1772} (\bibinfo {year}
  {2024}{\natexlab{b}})}\BibitemShut {NoStop}%
\bibitem [{\citenamefont {Zhang}\ \emph {et~al.}(2025)\citenamefont {Zhang},
  \citenamefont {Zhao}, \citenamefont {Han}, \citenamefont {Watanabe},
  \citenamefont {Taniguchi}, \citenamefont {Shan},\ and\ \citenamefont
  {Mak}}]{Zhang2025}%
  \BibitemOpen
  \bibfield  {author} {\bibinfo {author} {\bibfnamefont {Y.}~\bibnamefont
  {Zhang}}, \bibinfo {author} {\bibfnamefont {W.}~\bibnamefont {Zhao}},
  \bibinfo {author} {\bibfnamefont {Z.}~\bibnamefont {Han}}, \bibinfo {author}
  {\bibfnamefont {K.}~\bibnamefont {Watanabe}}, \bibinfo {author}
  {\bibfnamefont {T.}~\bibnamefont {Taniguchi}}, \bibinfo {author}
  {\bibfnamefont {J.}~\bibnamefont {Shan}},\ and\ \bibinfo {author}
  {\bibfnamefont {K.~F.}\ \bibnamefont {Mak}},\ }\href
  {https://doi.org/10.48550/ARXIV.2510.26958} {\bibinfo {title}
  {Thermoelectricity of moiré heavy fermions in mote2/wse2 bilayers}}
  (\bibinfo {year} {2025})\BibitemShut {NoStop}%
\bibitem [{\citenamefont {Han}\ \emph {et~al.}(2025)\citenamefont {Han},
  \citenamefont {Xia}, \citenamefont {Xia}, \citenamefont {Zhao}, \citenamefont
  {Zhang}, \citenamefont {Watanabe}, \citenamefont {Taniguchi}, \citenamefont
  {Shan},\ and\ \citenamefont {Mak}}]{Han2025}%
  \BibitemOpen
  \bibfield  {author} {\bibinfo {author} {\bibfnamefont {Z.}~\bibnamefont
  {Han}}, \bibinfo {author} {\bibfnamefont {Y.}~\bibnamefont {Xia}}, \bibinfo
  {author} {\bibfnamefont {Z.}~\bibnamefont {Xia}}, \bibinfo {author}
  {\bibfnamefont {W.}~\bibnamefont {Zhao}}, \bibinfo {author} {\bibfnamefont
  {Y.}~\bibnamefont {Zhang}}, \bibinfo {author} {\bibfnamefont
  {K.}~\bibnamefont {Watanabe}}, \bibinfo {author} {\bibfnamefont
  {T.}~\bibnamefont {Taniguchi}}, \bibinfo {author} {\bibfnamefont
  {J.}~\bibnamefont {Shan}},\ and\ \bibinfo {author} {\bibfnamefont {K.~F.}\
  \bibnamefont {Mak}},\ }\href {https://doi.org/10.48550/ARXIV.2507.03287}
  {\bibinfo {title} {Evidence of topological kondo insulating state in
  mote2/wse2 moiré bilayers}} (\bibinfo {year} {2025})\BibitemShut {NoStop}%
\bibitem [{\citenamefont {Zhao}\ \emph {et~al.}(2025)\citenamefont {Zhao},
  \citenamefont {Tao}, \citenamefont {Zhang}, \citenamefont {Shen},
  \citenamefont {Han}, \citenamefont {Knüppel}, \citenamefont {Zeng},
  \citenamefont {Xia}, \citenamefont {Watanabe}, \citenamefont {Taniguchi},
  \citenamefont {Shan},\ and\ \citenamefont {Mak}}]{Zhao2025}%
  \BibitemOpen
  \bibfield  {author} {\bibinfo {author} {\bibfnamefont {W.}~\bibnamefont
  {Zhao}}, \bibinfo {author} {\bibfnamefont {Z.}~\bibnamefont {Tao}}, \bibinfo
  {author} {\bibfnamefont {Y.}~\bibnamefont {Zhang}}, \bibinfo {author}
  {\bibfnamefont {B.}~\bibnamefont {Shen}}, \bibinfo {author} {\bibfnamefont
  {Z.}~\bibnamefont {Han}}, \bibinfo {author} {\bibfnamefont {P.}~\bibnamefont
  {Knüppel}}, \bibinfo {author} {\bibfnamefont {Y.}~\bibnamefont {Zeng}},
  \bibinfo {author} {\bibfnamefont {Z.}~\bibnamefont {Xia}}, \bibinfo {author}
  {\bibfnamefont {K.}~\bibnamefont {Watanabe}}, \bibinfo {author}
  {\bibfnamefont {T.}~\bibnamefont {Taniguchi}}, \bibinfo {author}
  {\bibfnamefont {J.}~\bibnamefont {Shan}},\ and\ \bibinfo {author}
  {\bibfnamefont {K.~F.}\ \bibnamefont {Mak}},\ }\href
  {https://doi.org/10.48550/ARXIV.2506.14063} {\bibinfo {title} {Emergence of
  chern metal in a moiré kondo lattice}} (\bibinfo {year} {2025})\BibitemShut
  {NoStop}%
\bibitem [{\citenamefont {Korm\'anyos}\ \emph {et~al.}(2015)\citenamefont
  {Korm\'anyos}, \citenamefont {Burkard}, \citenamefont {Gmitra}, \citenamefont
  {Fabian}, \citenamefont {Z\'olyomi}, \citenamefont {Drummond},\ and\
  \citenamefont {Fal’ko}}]{Kormanyos2015}%
  \BibitemOpen
  \bibfield  {author} {\bibinfo {author} {\bibfnamefont {A.}~\bibnamefont
  {Korm\'anyos}}, \bibinfo {author} {\bibfnamefont {G.}~\bibnamefont
  {Burkard}}, \bibinfo {author} {\bibfnamefont {M.}~\bibnamefont {Gmitra}},
  \bibinfo {author} {\bibfnamefont {J.}~\bibnamefont {Fabian}}, \bibinfo
  {author} {\bibfnamefont {V.}~\bibnamefont {Z\'olyomi}}, \bibinfo {author}
  {\bibfnamefont {N.~D.}\ \bibnamefont {Drummond}},\ and\ \bibinfo {author}
  {\bibfnamefont {V.}~\bibnamefont {Fal’ko}},\ }\bibfield  {title} {\bibinfo
  {title} {k · p theory for two-dimensional transition metal dichalcogenide
  semiconductors},\ }\href {https://doi.org/10.1088/2053-1583/2/2/022001}
  {\bibfield  {journal} {\bibinfo  {journal} {2D Materials}\ }\textbf {\bibinfo
  {volume} {2}},\ \bibinfo {pages} {022001} (\bibinfo {year}
  {2015})}\BibitemShut {NoStop}%
\bibitem [{\citenamefont {Wu}\ \emph {et~al.}(2018)\citenamefont {Wu},
  \citenamefont {Lovorn}, \citenamefont {Tutuc},\ and\ \citenamefont
  {MacDonald}}]{Wu2018}%
  \BibitemOpen
  \bibfield  {author} {\bibinfo {author} {\bibfnamefont {F.}~\bibnamefont
  {Wu}}, \bibinfo {author} {\bibfnamefont {T.}~\bibnamefont {Lovorn}}, \bibinfo
  {author} {\bibfnamefont {E.}~\bibnamefont {Tutuc}},\ and\ \bibinfo {author}
  {\bibfnamefont {A.}~\bibnamefont {MacDonald}},\ }\bibfield  {title} {\bibinfo
  {title} {Hubbard model physics in transition metal dichalcogenide moiré
  bands},\ }\href {https://doi.org/10.1103/physrevlett.121.026402} {\bibfield
  {journal} {\bibinfo  {journal} {Physical Review Letters}\ }\textbf {\bibinfo
  {volume} {121}},\ \bibinfo {pages} {026402} (\bibinfo {year}
  {2018})}\BibitemShut {NoStop}%
\bibitem [{\citenamefont {Devakul}\ \emph {et~al.}(2021)\citenamefont
  {Devakul}, \citenamefont {Crépel}, \citenamefont {Zhang},\ and\
  \citenamefont {Fu}}]{Devakul2021}%
  \BibitemOpen
  \bibfield  {author} {\bibinfo {author} {\bibfnamefont {T.}~\bibnamefont
  {Devakul}}, \bibinfo {author} {\bibfnamefont {V.}~\bibnamefont {Crépel}},
  \bibinfo {author} {\bibfnamefont {Y.}~\bibnamefont {Zhang}},\ and\ \bibinfo
  {author} {\bibfnamefont {L.}~\bibnamefont {Fu}},\ }\bibfield  {title}
  {\bibinfo {title} {Magic in twisted transition metal dichalcogenide
  bilayers},\ }\bibfield  {journal} {\bibinfo  {journal} {Nature
  Communications}\ }\textbf {\bibinfo {volume} {12}},\ \href
  {https://doi.org/10.1038/s41467-021-27042-9} {10.1038/s41467-021-27042-9}
  (\bibinfo {year} {2021})\BibitemShut {NoStop}%
\bibitem [{\citenamefont {Kane}\ and\ \citenamefont {Mele}(2005)}]{Kane2005}%
  \BibitemOpen
  \bibfield  {author} {\bibinfo {author} {\bibfnamefont {C.~L.}\ \bibnamefont
  {Kane}}\ and\ \bibinfo {author} {\bibfnamefont {E.~J.}\ \bibnamefont
  {Mele}},\ }\bibfield  {title} {\bibinfo {title} {Quantum spin hall effect in
  graphene},\ }\href {https://doi.org/10.1103/PhysRevLett.95.226801} {\bibfield
   {journal} {\bibinfo  {journal} {Phys. Rev. Lett.}\ }\textbf {\bibinfo
  {volume} {95}},\ \bibinfo {pages} {226801} (\bibinfo {year}
  {2005})}\BibitemShut {NoStop}%
\bibitem [{\citenamefont {Zhang}\ \emph {et~al.}(2021)\citenamefont {Zhang},
  \citenamefont {Devakul},\ and\ \citenamefont {Fu}}]{Zhang2021}%
  \BibitemOpen
  \bibfield  {author} {\bibinfo {author} {\bibfnamefont {Y.}~\bibnamefont
  {Zhang}}, \bibinfo {author} {\bibfnamefont {T.}~\bibnamefont {Devakul}},\
  and\ \bibinfo {author} {\bibfnamefont {L.}~\bibnamefont {Fu}},\ }\bibfield
  {title} {\bibinfo {title} {Spin-textured chern bands in ab-stacked transition
  metal dichalcogenide bilayers},\ }\href
  {https://doi.org/10.1073/pnas.2112673118} {\bibfield  {journal} {\bibinfo
  {journal} {Proceedings of the National Academy of Sciences}\ }\textbf
  {\bibinfo {volume} {118}},\ \bibinfo {pages} {e2112673118} (\bibinfo {year}
  {2021})}\BibitemShut {NoStop}%
\bibitem [{\citenamefont {Guerci}\ \emph {et~al.}(2024)\citenamefont {Guerci},
  \citenamefont {Lucht}, \citenamefont {Cr\'epel}, \citenamefont {Cano},
  \citenamefont {Pixley},\ and\ \citenamefont {Millis}}]{Guerci2024}%
  \BibitemOpen
  \bibfield  {author} {\bibinfo {author} {\bibfnamefont {D.}~\bibnamefont
  {Guerci}}, \bibinfo {author} {\bibfnamefont {K.~P.}\ \bibnamefont {Lucht}},
  \bibinfo {author} {\bibfnamefont {V.}~\bibnamefont {Cr\'epel}}, \bibinfo
  {author} {\bibfnamefont {J.}~\bibnamefont {Cano}}, \bibinfo {author}
  {\bibfnamefont {J.~H.}\ \bibnamefont {Pixley}},\ and\ \bibinfo {author}
  {\bibfnamefont {A.}~\bibnamefont {Millis}},\ }\bibfield  {title} {\bibinfo
  {title} {Topological kondo semimetal and insulator in ab-stacked
  heterobilayer transition metal dichalcogenides},\ }\href
  {https://doi.org/10.1103/PhysRevB.110.165128} {\bibfield  {journal} {\bibinfo
   {journal} {Phys. Rev. B}\ }\textbf {\bibinfo {volume} {110}},\ \bibinfo
  {pages} {165128} (\bibinfo {year} {2024})}\BibitemShut {NoStop}%
\bibitem [{\citenamefont {Dzero}\ \emph {et~al.}(2010)\citenamefont {Dzero},
  \citenamefont {Sun}, \citenamefont {Galitski},\ and\ \citenamefont
  {Coleman}}]{Dzero2010}%
  \BibitemOpen
  \bibfield  {author} {\bibinfo {author} {\bibfnamefont {M.}~\bibnamefont
  {Dzero}}, \bibinfo {author} {\bibfnamefont {K.}~\bibnamefont {Sun}}, \bibinfo
  {author} {\bibfnamefont {V.}~\bibnamefont {Galitski}},\ and\ \bibinfo
  {author} {\bibfnamefont {P.}~\bibnamefont {Coleman}},\ }\bibfield  {title}
  {\bibinfo {title} {Topological kondo insulators},\ }\href
  {https://doi.org/10.1103/physrevlett.104.106408} {\bibfield  {journal}
  {\bibinfo  {journal} {Physical Review Letters}\ }\textbf {\bibinfo {volume}
  {104}},\ \bibinfo {pages} {106408} (\bibinfo {year} {2010})}\BibitemShut
  {NoStop}%
\bibitem [{\citenamefont {Dzero}\ \emph {et~al.}(2012)\citenamefont {Dzero},
  \citenamefont {Sun}, \citenamefont {Coleman},\ and\ \citenamefont
  {Galitski}}]{Dzero2012}%
  \BibitemOpen
  \bibfield  {author} {\bibinfo {author} {\bibfnamefont {M.}~\bibnamefont
  {Dzero}}, \bibinfo {author} {\bibfnamefont {K.}~\bibnamefont {Sun}}, \bibinfo
  {author} {\bibfnamefont {P.}~\bibnamefont {Coleman}},\ and\ \bibinfo {author}
  {\bibfnamefont {V.}~\bibnamefont {Galitski}},\ }\bibfield  {title} {\bibinfo
  {title} {Theory of topological kondo insulators},\ }\href
  {https://doi.org/10.1103/physrevb.85.045130} {\bibfield  {journal} {\bibinfo
  {journal} {Physical Review B}\ }\textbf {\bibinfo {volume} {85}},\ \bibinfo
  {pages} {045130} (\bibinfo {year} {2012})}\BibitemShut {NoStop}%
\bibitem [{\citenamefont {Menth}\ \emph {et~al.}(1969)\citenamefont {Menth},
  \citenamefont {Buehler},\ and\ \citenamefont {Geballe}}]{Menth1969}%
  \BibitemOpen
  \bibfield  {author} {\bibinfo {author} {\bibfnamefont {A.}~\bibnamefont
  {Menth}}, \bibinfo {author} {\bibfnamefont {E.}~\bibnamefont {Buehler}},\
  and\ \bibinfo {author} {\bibfnamefont {T.~H.}\ \bibnamefont {Geballe}},\
  }\bibfield  {title} {\bibinfo {title} {Magnetic and semiconducting properties
  of smb6},\ }\href {https://doi.org/10.1103/physrevlett.22.295} {\bibfield
  {journal} {\bibinfo  {journal} {Physical Review Letters}\ }\textbf {\bibinfo
  {volume} {22}},\ \bibinfo {pages} {295} (\bibinfo {year} {1969})}\BibitemShut
  {NoStop}%
\bibitem [{\citenamefont {Li}\ \emph {et~al.}(2014)\citenamefont {Li},
  \citenamefont {Xiang}, \citenamefont {Yu}, \citenamefont {Asaba},
  \citenamefont {Lawson}, \citenamefont {Cai}, \citenamefont {Tinsman},
  \citenamefont {Berkley}, \citenamefont {Wolgast}, \citenamefont {Eo},
  \citenamefont {Kim}, \citenamefont {Kurdak}, \citenamefont {Allen},
  \citenamefont {Sun}, \citenamefont {Chen}, \citenamefont {Wang},
  \citenamefont {Fisk},\ and\ \citenamefont {Li}}]{Li2014}%
  \BibitemOpen
  \bibfield  {author} {\bibinfo {author} {\bibfnamefont {G.}~\bibnamefont
  {Li}}, \bibinfo {author} {\bibfnamefont {Z.}~\bibnamefont {Xiang}}, \bibinfo
  {author} {\bibfnamefont {F.}~\bibnamefont {Yu}}, \bibinfo {author}
  {\bibfnamefont {T.}~\bibnamefont {Asaba}}, \bibinfo {author} {\bibfnamefont
  {B.}~\bibnamefont {Lawson}}, \bibinfo {author} {\bibfnamefont
  {P.}~\bibnamefont {Cai}}, \bibinfo {author} {\bibfnamefont {C.}~\bibnamefont
  {Tinsman}}, \bibinfo {author} {\bibfnamefont {A.}~\bibnamefont {Berkley}},
  \bibinfo {author} {\bibfnamefont {S.}~\bibnamefont {Wolgast}}, \bibinfo
  {author} {\bibfnamefont {Y.~S.}\ \bibnamefont {Eo}}, \bibinfo {author}
  {\bibfnamefont {D.-J.}\ \bibnamefont {Kim}}, \bibinfo {author} {\bibfnamefont
  {C.}~\bibnamefont {Kurdak}}, \bibinfo {author} {\bibfnamefont {J.~W.}\
  \bibnamefont {Allen}}, \bibinfo {author} {\bibfnamefont {K.}~\bibnamefont
  {Sun}}, \bibinfo {author} {\bibfnamefont {X.~H.}\ \bibnamefont {Chen}},
  \bibinfo {author} {\bibfnamefont {Y.~Y.}\ \bibnamefont {Wang}}, \bibinfo
  {author} {\bibfnamefont {Z.}~\bibnamefont {Fisk}},\ and\ \bibinfo {author}
  {\bibfnamefont {L.}~\bibnamefont {Li}},\ }\bibfield  {title} {\bibinfo
  {title} {Two-dimensional fermi surfaces in kondo insulator smb 6},\ }\href
  {https://doi.org/10.1126/science.1250366} {\bibfield  {journal} {\bibinfo
  {journal} {Science}\ }\textbf {\bibinfo {volume} {346}},\ \bibinfo {pages}
  {1208} (\bibinfo {year} {2014})}\BibitemShut {NoStop}%
\bibitem [{\citenamefont {Phelan}\ \emph {et~al.}(2014)\citenamefont {Phelan},
  \citenamefont {Koohpayeh}, \citenamefont {Cottingham}, \citenamefont
  {Freeland}, \citenamefont {Leiner}, \citenamefont {Broholm},\ and\
  \citenamefont {McQueen}}]{Phelan2014}%
  \BibitemOpen
  \bibfield  {author} {\bibinfo {author} {\bibfnamefont {W.}~\bibnamefont
  {Phelan}}, \bibinfo {author} {\bibfnamefont {S.}~\bibnamefont {Koohpayeh}},
  \bibinfo {author} {\bibfnamefont {P.}~\bibnamefont {Cottingham}}, \bibinfo
  {author} {\bibfnamefont {J.}~\bibnamefont {Freeland}}, \bibinfo {author}
  {\bibfnamefont {J.}~\bibnamefont {Leiner}}, \bibinfo {author} {\bibfnamefont
  {C.}~\bibnamefont {Broholm}},\ and\ \bibinfo {author} {\bibfnamefont
  {T.}~\bibnamefont {McQueen}},\ }\bibfield  {title} {\bibinfo {title}
  {Correlation between bulk thermodynamic measurements and the
  low-temperature-resistance plateau insmb6},\ }\href
  {https://doi.org/10.1103/physrevx.4.031012} {\bibfield  {journal} {\bibinfo
  {journal} {Physical Review X}\ }\textbf {\bibinfo {volume} {4}},\ \bibinfo
  {pages} {031012} (\bibinfo {year} {2014})}\BibitemShut {NoStop}%
\bibitem [{\citenamefont {Tan}\ \emph {et~al.}(2015)\citenamefont {Tan},
  \citenamefont {Hsu}, \citenamefont {Zeng}, \citenamefont {Hatnean},
  \citenamefont {Harrison}, \citenamefont {Zhu}, \citenamefont {Hartstein},
  \citenamefont {Kiourlappou}, \citenamefont {Srivastava}, \citenamefont
  {Johannes}, \citenamefont {Murphy}, \citenamefont {Park}, \citenamefont
  {Balicas}, \citenamefont {Lonzarich}, \citenamefont {Balakrishnan},\ and\
  \citenamefont {Sebastian}}]{Tan2015}%
  \BibitemOpen
  \bibfield  {author} {\bibinfo {author} {\bibfnamefont {B.~S.}\ \bibnamefont
  {Tan}}, \bibinfo {author} {\bibfnamefont {Y.-T.}\ \bibnamefont {Hsu}},
  \bibinfo {author} {\bibfnamefont {B.}~\bibnamefont {Zeng}}, \bibinfo {author}
  {\bibfnamefont {M.~C.}\ \bibnamefont {Hatnean}}, \bibinfo {author}
  {\bibfnamefont {N.}~\bibnamefont {Harrison}}, \bibinfo {author}
  {\bibfnamefont {Z.}~\bibnamefont {Zhu}}, \bibinfo {author} {\bibfnamefont
  {M.}~\bibnamefont {Hartstein}}, \bibinfo {author} {\bibfnamefont
  {M.}~\bibnamefont {Kiourlappou}}, \bibinfo {author} {\bibfnamefont
  {A.}~\bibnamefont {Srivastava}}, \bibinfo {author} {\bibfnamefont {M.~D.}\
  \bibnamefont {Johannes}}, \bibinfo {author} {\bibfnamefont {T.~P.}\
  \bibnamefont {Murphy}}, \bibinfo {author} {\bibfnamefont {J.-H.}\
  \bibnamefont {Park}}, \bibinfo {author} {\bibfnamefont {L.}~\bibnamefont
  {Balicas}}, \bibinfo {author} {\bibfnamefont {G.~G.}\ \bibnamefont
  {Lonzarich}}, \bibinfo {author} {\bibfnamefont {G.}~\bibnamefont
  {Balakrishnan}},\ and\ \bibinfo {author} {\bibfnamefont {S.~E.}\ \bibnamefont
  {Sebastian}},\ }\bibfield  {title} {\bibinfo {title} {Unconventional fermi
  surface in an insulating state},\ }\href
  {https://doi.org/10.1126/science.aaa7974} {\bibfield  {journal} {\bibinfo
  {journal} {Science}\ }\textbf {\bibinfo {volume} {349}},\ \bibinfo {pages}
  {287} (\bibinfo {year} {2015})}\BibitemShut {NoStop}%
\bibitem [{\citenamefont {Li}\ \emph {et~al.}(2020)\citenamefont {Li},
  \citenamefont {Sun}, \citenamefont {Kurdak},\ and\ \citenamefont
  {Allen}}]{Li2020}%
  \BibitemOpen
  \bibfield  {author} {\bibinfo {author} {\bibfnamefont {L.}~\bibnamefont
  {Li}}, \bibinfo {author} {\bibfnamefont {K.}~\bibnamefont {Sun}}, \bibinfo
  {author} {\bibfnamefont {C.}~\bibnamefont {Kurdak}},\ and\ \bibinfo {author}
  {\bibfnamefont {J.~W.}\ \bibnamefont {Allen}},\ }\bibfield  {title} {\bibinfo
  {title} {Emergent mystery in the kondo insulator samarium hexaboride},\
  }\href {https://doi.org/10.1038/s42254-020-0210-8} {\bibfield  {journal}
  {\bibinfo  {journal} {Nature Reviews Physics}\ }\textbf {\bibinfo {volume}
  {2}},\ \bibinfo {pages} {463} (\bibinfo {year} {2020})}\BibitemShut {NoStop}%
\bibitem [{\citenamefont {Liu}\ \emph {et~al.}(2018)\citenamefont {Liu},
  \citenamefont {Hartstein}, \citenamefont {Wallace}, \citenamefont {Davies},
  \citenamefont {Hatnean}, \citenamefont {Johannes}, \citenamefont
  {Shitsevalova}, \citenamefont {Balakrishnan},\ and\ \citenamefont
  {Sebastian}}]{Liu2018}%
  \BibitemOpen
  \bibfield  {author} {\bibinfo {author} {\bibfnamefont {H.}~\bibnamefont
  {Liu}}, \bibinfo {author} {\bibfnamefont {M.}~\bibnamefont {Hartstein}},
  \bibinfo {author} {\bibfnamefont {G.~J.}\ \bibnamefont {Wallace}}, \bibinfo
  {author} {\bibfnamefont {A.~J.}\ \bibnamefont {Davies}}, \bibinfo {author}
  {\bibfnamefont {M.~C.}\ \bibnamefont {Hatnean}}, \bibinfo {author}
  {\bibfnamefont {M.~D.}\ \bibnamefont {Johannes}}, \bibinfo {author}
  {\bibfnamefont {N.}~\bibnamefont {Shitsevalova}}, \bibinfo {author}
  {\bibfnamefont {G.}~\bibnamefont {Balakrishnan}},\ and\ \bibinfo {author}
  {\bibfnamefont {S.~E.}\ \bibnamefont {Sebastian}},\ }\bibfield  {title}
  {\bibinfo {title} {Fermi surfaces in kondo insulators},\ }\href
  {https://doi.org/10.1088/1361-648x/aaa522} {\bibfield  {journal} {\bibinfo
  {journal} {Journal of Physics: Condensed Matter}\ }\textbf {\bibinfo {volume}
  {30}},\ \bibinfo {pages} {16LT01} (\bibinfo {year} {2018})}\BibitemShut
  {NoStop}%
\bibitem [{\citenamefont {Sato}\ \emph {et~al.}(2019)\citenamefont {Sato},
  \citenamefont {Xiang}, \citenamefont {Kasahara}, \citenamefont {Taniguchi},
  \citenamefont {Kasahara}, \citenamefont {Chen}, \citenamefont {Asaba},
  \citenamefont {Tinsman}, \citenamefont {Murayama}, \citenamefont {Tanaka},
  \citenamefont {Mizukami}, \citenamefont {Shibauchi}, \citenamefont {Iga},
  \citenamefont {Singleton}, \citenamefont {Li},\ and\ \citenamefont
  {Matsuda}}]{Sato2019}%
  \BibitemOpen
  \bibfield  {author} {\bibinfo {author} {\bibfnamefont {Y.}~\bibnamefont
  {Sato}}, \bibinfo {author} {\bibfnamefont {Z.}~\bibnamefont {Xiang}},
  \bibinfo {author} {\bibfnamefont {Y.}~\bibnamefont {Kasahara}}, \bibinfo
  {author} {\bibfnamefont {T.}~\bibnamefont {Taniguchi}}, \bibinfo {author}
  {\bibfnamefont {S.}~\bibnamefont {Kasahara}}, \bibinfo {author}
  {\bibfnamefont {L.}~\bibnamefont {Chen}}, \bibinfo {author} {\bibfnamefont
  {T.}~\bibnamefont {Asaba}}, \bibinfo {author} {\bibfnamefont
  {C.}~\bibnamefont {Tinsman}}, \bibinfo {author} {\bibfnamefont
  {H.}~\bibnamefont {Murayama}}, \bibinfo {author} {\bibfnamefont
  {O.}~\bibnamefont {Tanaka}}, \bibinfo {author} {\bibfnamefont
  {Y.}~\bibnamefont {Mizukami}}, \bibinfo {author} {\bibfnamefont
  {T.}~\bibnamefont {Shibauchi}}, \bibinfo {author} {\bibfnamefont
  {F.}~\bibnamefont {Iga}}, \bibinfo {author} {\bibfnamefont {J.}~\bibnamefont
  {Singleton}}, \bibinfo {author} {\bibfnamefont {L.}~\bibnamefont {Li}},\ and\
  \bibinfo {author} {\bibfnamefont {Y.}~\bibnamefont {Matsuda}},\ }\bibfield
  {title} {\bibinfo {title} {Unconventional thermal metallic state of
  charge-neutral fermions in an insulator},\ }\href
  {https://doi.org/10.1038/s41567-019-0552-2} {\bibfield  {journal} {\bibinfo
  {journal} {Nature Physics}\ }\textbf {\bibinfo {volume} {15}},\ \bibinfo
  {pages} {954} (\bibinfo {year} {2019})}\BibitemShut {NoStop}%
\bibitem [{\citenamefont {Chen}\ \emph {et~al.}(2025)\citenamefont {Chen},
  \citenamefont {Zhu}, \citenamefont {Ratkovski}, \citenamefont {Zheng},
  \citenamefont {Zhang}, \citenamefont {Chan}, \citenamefont {Jenkins},
  \citenamefont {Blawat}, \citenamefont {Asaba}, \citenamefont {Iga},
  \citenamefont {Varma}, \citenamefont {Matsuda}, \citenamefont {Singleton},
  \citenamefont {Bangura},\ and\ \citenamefont {Li}}]{Chen2025}%
  \BibitemOpen
  \bibfield  {author} {\bibinfo {author} {\bibfnamefont {K.-W.}\ \bibnamefont
  {Chen}}, \bibinfo {author} {\bibfnamefont {Y.}~\bibnamefont {Zhu}}, \bibinfo
  {author} {\bibfnamefont {D.}~\bibnamefont {Ratkovski}}, \bibinfo {author}
  {\bibfnamefont {G.}~\bibnamefont {Zheng}}, \bibinfo {author} {\bibfnamefont
  {D.}~\bibnamefont {Zhang}}, \bibinfo {author} {\bibfnamefont
  {A.}~\bibnamefont {Chan}}, \bibinfo {author} {\bibfnamefont {K.}~\bibnamefont
  {Jenkins}}, \bibinfo {author} {\bibfnamefont {J.}~\bibnamefont {Blawat}},
  \bibinfo {author} {\bibfnamefont {T.}~\bibnamefont {Asaba}}, \bibinfo
  {author} {\bibfnamefont {F.}~\bibnamefont {Iga}}, \bibinfo {author}
  {\bibfnamefont {C.~M.}\ \bibnamefont {Varma}}, \bibinfo {author}
  {\bibfnamefont {Y.}~\bibnamefont {Matsuda}}, \bibinfo {author} {\bibfnamefont
  {J.}~\bibnamefont {Singleton}}, \bibinfo {author} {\bibfnamefont {A.~F.}\
  \bibnamefont {Bangura}},\ and\ \bibinfo {author} {\bibfnamefont
  {L.}~\bibnamefont {Li}},\ }\bibfield  {title} {\bibinfo {title} {Quantum
  oscillations in the heat capacity of kondo insulator ybb12},\ }\bibfield
  {journal} {\bibinfo  {journal} {Physical Review Letters}\ }\textbf {\bibinfo
  {volume} {135}},\ \href {https://doi.org/10.1103/ms3x-pjsk}
  {10.1103/ms3x-pjsk} (\bibinfo {year} {2025})\BibitemShut {NoStop}%
\bibitem [{\citenamefont {Mendez-Valderrama}\ \emph {et~al.}(2024)\citenamefont
  {Mendez-Valderrama}, \citenamefont {Kim},\ and\ \citenamefont
  {Chowdhury}}]{Mendez2024}%
  \BibitemOpen
  \bibfield  {author} {\bibinfo {author} {\bibfnamefont {J.~F.}\ \bibnamefont
  {Mendez-Valderrama}}, \bibinfo {author} {\bibfnamefont {S.}~\bibnamefont
  {Kim}},\ and\ \bibinfo {author} {\bibfnamefont {D.}~\bibnamefont
  {Chowdhury}},\ }\bibfield  {title} {\bibinfo {title} {Correlated topological
  mixed-valence insulators in moir\'e heterobilayers},\ }\href
  {https://doi.org/10.1103/PhysRevB.110.L201105} {\bibfield  {journal}
  {\bibinfo  {journal} {Phys. Rev. B}\ }\textbf {\bibinfo {volume} {110}},\
  \bibinfo {pages} {L201105} (\bibinfo {year} {2024})}\BibitemShut {NoStop}%
\bibitem [{\citenamefont {Xie}\ \emph {et~al.}(2025)\citenamefont {Xie},
  \citenamefont {Chen}, \citenamefont {Fang},\ and\ \citenamefont
  {Si}}]{Xie2025}%
  \BibitemOpen
  \bibfield  {author} {\bibinfo {author} {\bibfnamefont {F.}~\bibnamefont
  {Xie}}, \bibinfo {author} {\bibfnamefont {L.}~\bibnamefont {Chen}}, \bibinfo
  {author} {\bibfnamefont {Y.}~\bibnamefont {Fang}},\ and\ \bibinfo {author}
  {\bibfnamefont {Q.}~\bibnamefont {Si}},\ }\bibfield  {title} {\bibinfo
  {title} {Topological kondo semimetals emulated in heterobilayer transition
  metal dichalcogenides},\ }\href {https://doi.org/10.1103/fwbg-kdb9}
  {\bibfield  {journal} {\bibinfo  {journal} {Phys. Rev. Res.}\ }\textbf
  {\bibinfo {volume} {7}},\ \bibinfo {pages} {033093} (\bibinfo {year}
  {2025})}\BibitemShut {NoStop}%
\bibitem [{\citenamefont {Georges}\ \emph {et~al.}(1996)\citenamefont
  {Georges}, \citenamefont {Kotliar}, \citenamefont {Krauth},\ and\
  \citenamefont {Rozenberg}}]{Georges1996}%
  \BibitemOpen
  \bibfield  {author} {\bibinfo {author} {\bibfnamefont {A.}~\bibnamefont
  {Georges}}, \bibinfo {author} {\bibfnamefont {G.}~\bibnamefont {Kotliar}},
  \bibinfo {author} {\bibfnamefont {W.}~\bibnamefont {Krauth}},\ and\ \bibinfo
  {author} {\bibfnamefont {M.~J.}\ \bibnamefont {Rozenberg}},\ }\bibfield
  {title} {\bibinfo {title} {Dynamical mean-field theory of strongly correlated
  fermion systems and the limit of infinite dimensions},\ }\href
  {https://doi.org/10.1103/revmodphys.68.13} {\bibfield  {journal} {\bibinfo
  {journal} {Reviews of Modern Physics}\ }\textbf {\bibinfo {volume} {68}},\
  \bibinfo {pages} {13} (\bibinfo {year} {1996})}\BibitemShut {NoStop}%
\bibitem [{\citenamefont {Kotliar}\ \emph {et~al.}(2006)\citenamefont
  {Kotliar}, \citenamefont {Savrasov}, \citenamefont {Haule}, \citenamefont
  {Oudovenko}, \citenamefont {Parcollet},\ and\ \citenamefont
  {Marianetti}}]{Kotliar2006}%
  \BibitemOpen
  \bibfield  {author} {\bibinfo {author} {\bibfnamefont {G.}~\bibnamefont
  {Kotliar}}, \bibinfo {author} {\bibfnamefont {S.~Y.}\ \bibnamefont
  {Savrasov}}, \bibinfo {author} {\bibfnamefont {K.}~\bibnamefont {Haule}},
  \bibinfo {author} {\bibfnamefont {V.~S.}\ \bibnamefont {Oudovenko}}, \bibinfo
  {author} {\bibfnamefont {O.}~\bibnamefont {Parcollet}},\ and\ \bibinfo
  {author} {\bibfnamefont {C.~A.}\ \bibnamefont {Marianetti}},\ }\bibfield
  {title} {\bibinfo {title} {Electronic structure calculations with dynamical
  mean-field theory},\ }\href {https://doi.org/10.1103/revmodphys.78.865}
  {\bibfield  {journal} {\bibinfo  {journal} {Reviews of Modern Physics}\
  }\textbf {\bibinfo {volume} {78}},\ \bibinfo {pages} {865} (\bibinfo {year}
  {2006})}\BibitemShut {NoStop}%
\bibitem [{\citenamefont {Weichselbaum}(2012)}]{Weichselbaum2012}%
  \BibitemOpen
  \bibfield  {author} {\bibinfo {author} {\bibfnamefont {A.}~\bibnamefont
  {Weichselbaum}},\ }\bibfield  {title} {\bibinfo {title} {Non-abelian
  symmetries in tensor networks: A quantum symmetry space approach},\ }\href
  {https://doi.org/10.1016/j.aop.2012.07.009} {\bibfield  {journal} {\bibinfo
  {journal} {Annals of Physics}\ }\textbf {\bibinfo {volume} {327}},\ \bibinfo
  {pages} {2972} (\bibinfo {year} {2012})}\BibitemShut {NoStop}%
\bibitem [{\citenamefont {Weichselbaum}(2020)}]{Weichselbaum2020}%
  \BibitemOpen
  \bibfield  {author} {\bibinfo {author} {\bibfnamefont {A.}~\bibnamefont
  {Weichselbaum}},\ }\bibfield  {title} {\bibinfo {title} {X-symbols for
  non-abelian symmetries in tensor networks},\ }\href
  {https://doi.org/10.1103/physrevresearch.2.023385} {\bibfield  {journal}
  {\bibinfo  {journal} {Physical Review Research}\ }\textbf {\bibinfo {volume}
  {2}},\ \bibinfo {pages} {023385} (\bibinfo {year} {2020})}\BibitemShut
  {NoStop}%
\bibitem [{\citenamefont {Weichselbaum}(2024)}]{Weichselbaum2024}%
  \BibitemOpen
  \bibfield  {author} {\bibinfo {author} {\bibfnamefont {A.}~\bibnamefont
  {Weichselbaum}},\ }\bibfield  {title} {\bibinfo {title} {Codebase release 4.0
  for qspace},\ }\bibfield  {journal} {\bibinfo  {journal} {SciPost Physics
  Codebases}\ }\href {https://doi.org/10.21468/scipostphyscodeb.40-r4.0}
  {10.21468/scipostphyscodeb.40-r4.0} (\bibinfo {year} {2024})\BibitemShut
  {NoStop}%
\bibitem [{\citenamefont {Lee}\ and\ \citenamefont
  {Weichselbaum}(2016)}]{Lee2016}%
  \BibitemOpen
  \bibfield  {author} {\bibinfo {author} {\bibfnamefont {S.-S.~B.}\
  \bibnamefont {Lee}}\ and\ \bibinfo {author} {\bibfnamefont {A.}~\bibnamefont
  {Weichselbaum}},\ }\bibfield  {title} {\bibinfo {title} {Adaptive broadening
  to improve spectral resolution in the numerical renormalization group},\
  }\href {https://doi.org/10.1103/physrevb.94.235127} {\bibfield  {journal}
  {\bibinfo  {journal} {Physical Review B}\ }\textbf {\bibinfo {volume} {94}},\
  \bibinfo {pages} {235127} (\bibinfo {year} {2016})}\BibitemShut {NoStop}%
\bibitem [{\citenamefont {Lee}\ \emph {et~al.}(2017)\citenamefont {Lee},
  \citenamefont {von Delft},\ and\ \citenamefont {Weichselbaum}}]{Lee2017}%
  \BibitemOpen
  \bibfield  {author} {\bibinfo {author} {\bibfnamefont {S.-S.~B.}\
  \bibnamefont {Lee}}, \bibinfo {author} {\bibfnamefont {J.}~\bibnamefont {von
  Delft}},\ and\ \bibinfo {author} {\bibfnamefont {A.}~\bibnamefont
  {Weichselbaum}},\ }\bibfield  {title} {\bibinfo {title} {Doublon-holon origin
  of the subpeaks at the hubbard band edges},\ }\href
  {https://doi.org/10.1103/physrevlett.119.236402} {\bibfield  {journal}
  {\bibinfo  {journal} {Physical Review Letters}\ }\textbf {\bibinfo {volume}
  {119}},\ \bibinfo {pages} {236402} (\bibinfo {year} {2017})}\BibitemShut
  {NoStop}%
\bibitem [{\citenamefont {Kugler}\ \emph {et~al.}(2021)\citenamefont {Kugler},
  \citenamefont {Lee},\ and\ \citenamefont {von Delft}}]{Lee2021}%
  \BibitemOpen
  \bibfield  {author} {\bibinfo {author} {\bibfnamefont {F.~B.}\ \bibnamefont
  {Kugler}}, \bibinfo {author} {\bibfnamefont {S.-S.~B.}\ \bibnamefont {Lee}},\
  and\ \bibinfo {author} {\bibfnamefont {J.}~\bibnamefont {von Delft}},\
  }\bibfield  {title} {\bibinfo {title} {Multipoint correlation functions:
  Spectral representation and numerical evaluation},\ }\href
  {https://doi.org/10.1103/PhysRevX.11.041006} {\bibfield  {journal} {\bibinfo
  {journal} {Phys. Rev. X}\ }\textbf {\bibinfo {volume} {11}},\ \bibinfo
  {pages} {041006} (\bibinfo {year} {2021})}\BibitemShut {NoStop}%
\bibitem [{\citenamefont {Wilson}(1975)}]{Wilson1975}%
  \BibitemOpen
  \bibfield  {author} {\bibinfo {author} {\bibfnamefont {K.~G.}\ \bibnamefont
  {Wilson}},\ }\bibfield  {title} {\bibinfo {title} {The renormalization group:
  Critical phenomena and the kondo problem},\ }\href
  {https://doi.org/10.1103/revmodphys.47.773} {\bibfield  {journal} {\bibinfo
  {journal} {Reviews of Modern Physics}\ }\textbf {\bibinfo {volume} {47}},\
  \bibinfo {pages} {773} (\bibinfo {year} {1975})}\BibitemShut {NoStop}%
\bibitem [{\citenamefont {Weichselbaum}\ and\ \citenamefont {von
  Delft}(2007)}]{Weichselbaum2007}%
  \BibitemOpen
  \bibfield  {author} {\bibinfo {author} {\bibfnamefont {A.}~\bibnamefont
  {Weichselbaum}}\ and\ \bibinfo {author} {\bibfnamefont {J.}~\bibnamefont {von
  Delft}},\ }\bibfield  {title} {\bibinfo {title} {Sum-rule conserving spectral
  functions from the numerical renormalization group},\ }\href
  {https://doi.org/10.1103/physrevlett.99.076402} {\bibfield  {journal}
  {\bibinfo  {journal} {Physical Review Letters}\ }\textbf {\bibinfo {volume}
  {99}},\ \bibinfo {pages} {076402} (\bibinfo {year} {2007})}\BibitemShut
  {NoStop}%
\bibitem [{\citenamefont {Bulla}\ \emph {et~al.}(2008)\citenamefont {Bulla},
  \citenamefont {Costi},\ and\ \citenamefont {Pruschke}}]{Bulla2008}%
  \BibitemOpen
  \bibfield  {author} {\bibinfo {author} {\bibfnamefont {R.}~\bibnamefont
  {Bulla}}, \bibinfo {author} {\bibfnamefont {T.~A.}\ \bibnamefont {Costi}},\
  and\ \bibinfo {author} {\bibfnamefont {T.}~\bibnamefont {Pruschke}},\
  }\bibfield  {title} {\bibinfo {title} {Numerical renormalization group method
  for quantum impurity systems},\ }\href
  {https://doi.org/10.1103/revmodphys.80.395} {\bibfield  {journal} {\bibinfo
  {journal} {Reviews of Modern Physics}\ }\textbf {\bibinfo {volume} {80}},\
  \bibinfo {pages} {395} (\bibinfo {year} {2008})}\BibitemShut {NoStop}%
\bibitem [{sup()}]{supplement}%
  \BibitemOpen
  \href@noop {} {\bibinfo  {journal} {See Supplemental Material at [url] for
  technical details of our dynamical mean-field theory plus Hartree-Fock
  approach and the numerical evaluation of the conductivity tensor. The
  Supplemental Material includes
  Refs.~\cite{Zitko2009,Kugler2022,BonbienManchon2020,Bastin1971,SmrckaStreda1977,Munoz2025}}\
  }\BibitemShut {NoStop}%
\bibitem [{\citenamefont {\v{Z}itko}(2009)}]{Zitko2009}%
  \BibitemOpen
\bibfield  {journal} {  }\bibfield  {author} {\bibinfo {author} {\bibfnamefont
  {R.}~\bibnamefont {\v{Z}itko}},\ }\bibfield  {title} {\bibinfo {title}
  {Convergence acceleration and stabilization of dynamical mean-field theory
  calculations},\ }\href {https://doi.org/10.1103/PhysRevB.80.125125}
  {\bibfield  {journal} {\bibinfo  {journal} {Phys. Rev. B}\ }\textbf {\bibinfo
  {volume} {80}},\ \bibinfo {pages} {125125} (\bibinfo {year}
  {2009})}\BibitemShut {NoStop}%
\bibitem [{\citenamefont {Kugler}(2022)}]{Kugler2022}%
  \BibitemOpen
  \bibfield  {author} {\bibinfo {author} {\bibfnamefont {F.~B.}\ \bibnamefont
  {Kugler}},\ }\bibfield  {title} {\bibinfo {title} {Improved estimator for
  numerical renormalization group calculations of the self-energy},\ }\href
  {https://doi.org/10.1103/PhysRevB.105.245132} {\bibfield  {journal} {\bibinfo
   {journal} {Phys. Rev. B}\ }\textbf {\bibinfo {volume} {105}},\ \bibinfo
  {pages} {245132} (\bibinfo {year} {2022})}\BibitemShut {NoStop}%
\bibitem [{\citenamefont {Bonbien}\ and\ \citenamefont
  {Manchon}(2020)}]{BonbienManchon2020}%
  \BibitemOpen
  \bibfield  {author} {\bibinfo {author} {\bibfnamefont {V.}~\bibnamefont
  {Bonbien}}\ and\ \bibinfo {author} {\bibfnamefont {A.}~\bibnamefont
  {Manchon}},\ }\bibfield  {title} {\bibinfo {title} {Symmetrized decomposition
  of the kubo--bastin formula},\ }\href
  {https://doi.org/10.1103/PhysRevB.102.085113} {\bibfield  {journal} {\bibinfo
   {journal} {Physical Review B}\ }\textbf {\bibinfo {volume} {102}},\ \bibinfo
  {pages} {085113} (\bibinfo {year} {2020})}\BibitemShut {NoStop}%
\bibitem [{\citenamefont {Bastin}\ \emph {et~al.}(1971)\citenamefont {Bastin},
  \citenamefont {Lewiner}, \citenamefont {Betbeder-Matibet},\ and\
  \citenamefont {Nozieres}}]{Bastin1971}%
  \BibitemOpen
  \bibfield  {author} {\bibinfo {author} {\bibfnamefont {A.}~\bibnamefont
  {Bastin}}, \bibinfo {author} {\bibfnamefont {C.}~\bibnamefont {Lewiner}},
  \bibinfo {author} {\bibfnamefont {O.}~\bibnamefont {Betbeder-Matibet}},\ and\
  \bibinfo {author} {\bibfnamefont {P.}~\bibnamefont {Nozieres}},\ }\bibfield
  {title} {\bibinfo {title} {Quantum oscillations of the hall effect of a
  fermion gas with random impurity scattering},\ }\href
  {https://doi.org/10.1016/S0022-3697(71)80147-6} {\bibfield  {journal}
  {\bibinfo  {journal} {Journal of Physics and Chemistry of Solids}\ }\textbf
  {\bibinfo {volume} {32}},\ \bibinfo {pages} {1811} (\bibinfo {year}
  {1971})}\BibitemShut {NoStop}%
\bibitem [{\citenamefont {Smrcka}\ and\ \citenamefont
  {Streda}(1977)}]{SmrckaStreda1977}%
  \BibitemOpen
  \bibfield  {author} {\bibinfo {author} {\bibfnamefont {L.}~\bibnamefont
  {Smrcka}}\ and\ \bibinfo {author} {\bibfnamefont {P.}~\bibnamefont
  {Streda}},\ }\bibfield  {title} {\bibinfo {title} {Transport coefficients in
  strong magnetic fields},\ }\href
  {https://doi.org/10.1088/0022-3719/10/12/021} {\bibfield  {journal} {\bibinfo
   {journal} {Journal of Physics C: Solid State Physics}\ }\textbf {\bibinfo
  {volume} {10}},\ \bibinfo {pages} {2153} (\bibinfo {year}
  {1977})}\BibitemShut {NoStop}%
\bibitem [{\citenamefont {Van Mu\~noz}\ \emph {et~al.}(2025)\citenamefont {Van
  Mu\~noz}, \citenamefont {Kaye}, \citenamefont {Barnett},\ and\ \citenamefont
  {Beck}}]{Munoz2025}%
  \BibitemOpen
  \bibfield  {author} {\bibinfo {author} {\bibfnamefont {L.}~\bibnamefont {Van
  Mu\~noz}}, \bibinfo {author} {\bibfnamefont {J.}~\bibnamefont {Kaye}},
  \bibinfo {author} {\bibfnamefont {A.}~\bibnamefont {Barnett}},\ and\ \bibinfo
  {author} {\bibfnamefont {S.}~\bibnamefont {Beck}},\ }\bibfield  {title}
  {\bibinfo {title} {High-order and adaptive optical conductivity calculations
  using wannier interpolation},\ }\href
  {https://doi.org/10.1103/PhysRevB.111.195162} {\bibfield  {journal} {\bibinfo
   {journal} {Phys. Rev. B}\ }\textbf {\bibinfo {volume} {111}},\ \bibinfo
  {pages} {195162} (\bibinfo {year} {2025})}\BibitemShut {NoStop}%
\bibitem [{\citenamefont {Rademaker}(2022)}]{Rademaker2022}%
  \BibitemOpen
  \bibfield  {author} {\bibinfo {author} {\bibfnamefont {L.}~\bibnamefont
  {Rademaker}},\ }\bibfield  {title} {\bibinfo {title} {Spin-orbit coupling in
  transition metal dichalcogenide heterobilayer flat bands},\ }\href
  {https://doi.org/10.1103/PhysRevB.105.195428} {\bibfield  {journal} {\bibinfo
   {journal} {Phys. Rev. B}\ }\textbf {\bibinfo {volume} {105}},\ \bibinfo
  {pages} {195428} (\bibinfo {year} {2022})}\BibitemShut {NoStop}%
\bibitem [{Note1()}]{Note1}%
  \BibitemOpen
  \bibinfo {note} {The interlayer hybridization leads to spin loop currents
  \protect \textit {both} in the $f$ and the $c$ layer. However, as a result of
  the bandwidth mismatch, they are much weaker in the latter, compared to the
  kinetic energy. The reason is that the kinetic energy due to an $f$-$f$ bond
  is $-4 t_f \chi ^{ff} \cos \phi ^{f}$, i.e.\ it is minimized for $\phi ^{f} =
  0$, and likewise for $c$. Since $|t_f| \ll |t_c|$, a phase mismatch between
  hopping amplitude and bond expectation value is energetically much more
  costly in the $c$ band than in the $f$ band. As a result, such phase
  mismatches and therefore spin loop currents are most prominent in the $f$
  band}\BibitemShut {NoStop}%
\bibitem [{\citenamefont {Wang}\ and\ \citenamefont {Zhang}(2012)}]{Wang2012}%
  \BibitemOpen
  \bibfield  {author} {\bibinfo {author} {\bibfnamefont {Z.}~\bibnamefont
  {Wang}}\ and\ \bibinfo {author} {\bibfnamefont {S.-C.}\ \bibnamefont
  {Zhang}},\ }\bibfield  {title} {\bibinfo {title} {Simplified topological
  invariants for interacting insulators},\ }\href
  {https://doi.org/10.1103/PhysRevX.2.031008} {\bibfield  {journal} {\bibinfo
  {journal} {Phys. Rev. X}\ }\textbf {\bibinfo {volume} {2}},\ \bibinfo {pages}
  {031008} (\bibinfo {year} {2012})}\BibitemShut {NoStop}%
\bibitem [{\citenamefont {Lau}\ \emph {et~al.}(2025)\citenamefont {Lau},
  \citenamefont {Gleis}, \citenamefont {Kaplan}, \citenamefont {Chandra},\ and\
  \citenamefont {Coleman}}]{Lau2025}%
  \BibitemOpen
  \bibfield  {author} {\bibinfo {author} {\bibfnamefont {L.~L.~H.}\
  \bibnamefont {Lau}}, \bibinfo {author} {\bibfnamefont {A.}~\bibnamefont
  {Gleis}}, \bibinfo {author} {\bibfnamefont {D.}~\bibnamefont {Kaplan}},
  \bibinfo {author} {\bibfnamefont {P.}~\bibnamefont {Chandra}},\ and\ \bibinfo
  {author} {\bibfnamefont {P.}~\bibnamefont {Coleman}},\ }\bibfield  {title}
  {\bibinfo {title} {Oscillate and renormalize: Fast phonons reshape the kondo
  effect in flat-band systems},\ }\bibfield  {journal} {\bibinfo  {journal}
  {Physical Review B}\ }\textbf {\bibinfo {volume} {111}},\ \href
  {https://doi.org/10.1103/xxyt-4bql} {10.1103/xxyt-4bql} (\bibinfo {year}
  {2025})\BibitemShut {NoStop}%
\bibitem [{\citenamefont {Si}\ \emph {et~al.}(2001)\citenamefont {Si},
  \citenamefont {Rabello}, \citenamefont {Ingersent},\ and\ \citenamefont
  {Smith}}]{Si2001}%
  \BibitemOpen
  \bibfield  {author} {\bibinfo {author} {\bibfnamefont {Q.}~\bibnamefont
  {Si}}, \bibinfo {author} {\bibfnamefont {S.}~\bibnamefont {Rabello}},
  \bibinfo {author} {\bibfnamefont {K.}~\bibnamefont {Ingersent}},\ and\
  \bibinfo {author} {\bibfnamefont {J.~L.}\ \bibnamefont {Smith}},\ }\bibfield
  {title} {\bibinfo {title} {Locally critical quantum phase transitions in
  strongly correlated metals},\ }\href {https://doi.org/10.1038/35101507}
  {\bibfield  {journal} {\bibinfo  {journal} {Nature}\ }\textbf {\bibinfo
  {volume} {413}},\ \bibinfo {pages} {804} (\bibinfo {year}
  {2001})}\BibitemShut {NoStop}%
\bibitem [{\citenamefont {Coleman}\ \emph {et~al.}(2001)\citenamefont
  {Coleman}, \citenamefont {Pépin}, \citenamefont {Si},\ and\ \citenamefont
  {Ramazashvili}}]{Coleman2001}%
  \BibitemOpen
  \bibfield  {author} {\bibinfo {author} {\bibfnamefont {P.}~\bibnamefont
  {Coleman}}, \bibinfo {author} {\bibfnamefont {C.}~\bibnamefont {Pépin}},
  \bibinfo {author} {\bibfnamefont {Q.}~\bibnamefont {Si}},\ and\ \bibinfo
  {author} {\bibfnamefont {R.}~\bibnamefont {Ramazashvili}},\ }\bibfield
  {title} {\bibinfo {title} {How do fermi liquids get heavy and die?},\ }\href
  {https://doi.org/10.1088/0953-8984/13/35/202} {\bibfield  {journal} {\bibinfo
   {journal} {Journal of Physics: Condensed Matter}\ }\textbf {\bibinfo
  {volume} {13}},\ \bibinfo {pages} {R723} (\bibinfo {year}
  {2001})}\BibitemShut {NoStop}%
\bibitem [{\citenamefont {Senthil}\ \emph {et~al.}(2003)\citenamefont
  {Senthil}, \citenamefont {Sachdev},\ and\ \citenamefont
  {Vojta}}]{Senthil2003}%
  \BibitemOpen
  \bibfield  {author} {\bibinfo {author} {\bibfnamefont {T.}~\bibnamefont
  {Senthil}}, \bibinfo {author} {\bibfnamefont {S.}~\bibnamefont {Sachdev}},\
  and\ \bibinfo {author} {\bibfnamefont {M.}~\bibnamefont {Vojta}},\ }\bibfield
   {title} {\bibinfo {title} {Fractionalized fermi liquids},\ }\href
  {https://doi.org/10.1103/physrevlett.90.216403} {\bibfield  {journal}
  {\bibinfo  {journal} {Physical Review Letters}\ }\textbf {\bibinfo {volume}
  {90}},\ \bibinfo {pages} {216403} (\bibinfo {year} {2003})}\BibitemShut
  {NoStop}%
\bibitem [{\citenamefont {Senthil}\ \emph {et~al.}(2004)\citenamefont
  {Senthil}, \citenamefont {Vojta},\ and\ \citenamefont
  {Sachdev}}]{Senthil2004}%
  \BibitemOpen
  \bibfield  {author} {\bibinfo {author} {\bibfnamefont {T.}~\bibnamefont
  {Senthil}}, \bibinfo {author} {\bibfnamefont {M.}~\bibnamefont {Vojta}},\
  and\ \bibinfo {author} {\bibfnamefont {S.}~\bibnamefont {Sachdev}},\
  }\bibfield  {title} {\bibinfo {title} {Weak magnetism and non-fermi liquids
  near heavy-fermion critical points},\ }\href
  {https://doi.org/10.1103/physrevb.69.035111} {\bibfield  {journal} {\bibinfo
  {journal} {Physical Review B}\ }\textbf {\bibinfo {volume} {69}},\ \bibinfo
  {pages} {035111} (\bibinfo {year} {2004})}\BibitemShut {NoStop}%
\bibitem [{\citenamefont {Gleis}\ \emph {et~al.}(2024)\citenamefont {Gleis},
  \citenamefont {Lee}, \citenamefont {Kotliar},\ and\ \citenamefont {von
  Delft}}]{Gleis2024}%
  \BibitemOpen
  \bibfield  {author} {\bibinfo {author} {\bibfnamefont {A.}~\bibnamefont
  {Gleis}}, \bibinfo {author} {\bibfnamefont {S.-S.~B.}\ \bibnamefont {Lee}},
  \bibinfo {author} {\bibfnamefont {G.}~\bibnamefont {Kotliar}},\ and\ \bibinfo
  {author} {\bibfnamefont {J.}~\bibnamefont {von Delft}},\ }\bibfield  {title}
  {\bibinfo {title} {Emergent properties of the periodic anderson model: A
  high-resolution, real-frequency study of heavy-fermion quantum criticality},\
  }\href {https://doi.org/10.1103/physrevx.14.041036} {\bibfield  {journal}
  {\bibinfo  {journal} {Physical Review X}\ }\textbf {\bibinfo {volume} {14}},\
  \bibinfo {pages} {041036} (\bibinfo {year} {2024})}\BibitemShut {NoStop}%
\bibitem [{\citenamefont {Gleis}\ \emph {et~al.}(2025)\citenamefont {Gleis},
  \citenamefont {Lee}, \citenamefont {Kotliar},\ and\ \citenamefont {von
  Delft}}]{Gleis2025}%
  \BibitemOpen
  \bibfield  {author} {\bibinfo {author} {\bibfnamefont {A.}~\bibnamefont
  {Gleis}}, \bibinfo {author} {\bibfnamefont {S.-S.~B.}\ \bibnamefont {Lee}},
  \bibinfo {author} {\bibfnamefont {G.}~\bibnamefont {Kotliar}},\ and\ \bibinfo
  {author} {\bibfnamefont {J.}~\bibnamefont {von Delft}},\ }\bibfield  {title}
  {\bibinfo {title} {Dynamical scaling and planckian dissipation due to
  heavy-fermion quantum criticality},\ }\href
  {https://doi.org/10.1103/physrevlett.134.106501} {\bibfield  {journal}
  {\bibinfo  {journal} {Physical Review Letters}\ }\textbf {\bibinfo {volume}
  {134}},\ \bibinfo {pages} {106501} (\bibinfo {year} {2025})}\BibitemShut
  {NoStop}%
\end{thebibliography}%
%

%



\clearpage

\thispagestyle{empty}

\setcounter{equation}{0}
\setcounter{figure}{0}
\setcounter{page}{1}

\renewcommand{\theequation}{S\arabic{equation}}
\renewcommand{\thefigure}{S\arabic{figure}}
\renewcommand{\thepage}{S\arabic{page}}

\setcounter{secnumdepth}{2} 
\renewcommand{\thefigure}{S\arabic{figure}}
\setcounter{figure}{0}
\setcounter{section}{0}
\setcounter{equation}{0}
\renewcommand{\thesection}{S-\Roman{section}}
\renewcommand{\theequation}{S\arabic{equation}}
%

%
\title{Supplemental Material for
``\maintitle''}

\date{\today}
\maketitle

In this supplemental material, we provide technical details on the dynamical mean-field theory plus Hartree-Fock calculations in Sec.~\ref{sec:sup_DMFT_plus_HF}
and details on the numerical evaluation of the conductivity tensor in Sec.~\ref{sec:sup_conductivity}.

\section{Dynamical mean-field theory plus Hartree Fock}
\label{sec:sup_DMFT_plus_HF}

In this section, we briefly summarize the dynamical mean-field theory~(DMFT) plus Hartree-Fock~(HF) scheme used in this work.
Since DMFT and HF are well-known methods, the discussion below only focuses on technical details specific to this work.

The self-consistency loop is organized as follows.
For a given self-energy, we compute the local lattice Green's functions, extract the DMFT hybridization function from those,
and first update the DMFT contribution to $\Sigma^f_{\mr{loc}}(\omega)$ by solving the corresponding effective Anderson impurity model with the numerical renormalization group~(NRG).
Using the resulting lattice Green's functions, we then recompute the expectation values entering the HF self-energies and update the HF contributions.
The chemical potential is updated during self-consistency to fix the filling $\nu$.
This procedure is iterated to self-consistency, and Broyden mixing~\cite{Zitko2009} is used to accelerate convergence.
Typically, between 30 and 50 iterations are required to reach a self-consistent solution.

The hybridization function of the effective Anderson impurity model is
\begin{align}
\Delta^{f}(\omega) = \omega + \mu - \frac{\Delta}{2} - \Sigma^{f}_{\mr{loc}}(\omega) - [G^{f}_{\mr{loc}}(\omega)]^{-1} \, .
\end{align}
Here,
\begin{align}
G^{f}_{\mr{loc}}(\omega) = \int_{\mr{BZ}} \frac{d^2 k}{\mc{V}_{\mr{BZ}}} G^{f}_{\vec{k}\sigma}(\omega) \, ,
\end{align}
is the local $f$-band Green's function, obtained from the momentum-resolved lattice Green's function $G^{f}_{\vec{k}\sigma}(\omega)$ by integration over the Brillouin zone of volume $\mc{V}_{\mr{BZ}}$.
Since time-reversal symmetry is preserved in all calculations discussed in this work, both $G^{f}_{\mr{loc}}(\omega)$ and $\Delta^{f}(\omega)$ are independent of $\sigma$.
The corresponding impurity problem, therefore, exhibits SU(2) spin symmetry.

We solve the resulting effective Anderson impurity model using NRG, employing the QSpace-based MuNRG package.
Throughout this work, we use a logarithmic discretization parameter $\Lambda = 2$ and $n_z = 4$ $z$-shifts to improve spectral resolution.
During the iterative diagonalization, we keep up to $4000$ low-energy SU(2) multiplets.
Dynamical correlation functions of the discretized impurity model are evaluated with the standard full-density-matrix NRG procedure.
The impurity self-energy is computed using the symmetric improved estimator of Ref.~\cite{Kugler2022}.

A minor technical complication is that the DMFT hybridization function of a Kondo insulator typically contains an isolated pole close to the real-frequency axis in the vicinity of $\omega = 0$.
To more reliably capture this pole, we slightly modify our discretization procedure.
Instead of directly logarithmically discretizing $\Delta^{f}(\omega)$, we rewrite it as
\begin{align}
\Delta^{f}(\omega) &= \frac{t_{0}^2}{\omega - \epsilon_0 - \Delta_1(\omega)} \, ,
\end{align}
with
\begin{subequations}
\begin{align}
t_0^2 &= -\frac{1}{\pi} \mr{Im} \int_{-\infty}^{\infty} \mr{d}\omega \Delta^{f}(\omega)
\\
\epsilon_0 &= -\frac{1}{t_0^2 \pi} \mr{Im} \int_{-\infty}^{\infty} \mr{d}\omega \,  \omega \, \Delta^{f}(\omega)  \, .
\end{align}
\end{subequations}
In this representation, a pole of $\Delta^{f}(\omega)$ captured by a near-zero of $\omega - \epsilon_0 - \Delta_1(\omega)$, while $\Delta_1(\omega)$ itself is considerably smoother.
We therefore apply the logarithmic discretization to $\Delta_1(\omega)$ rather than directly to $\Delta^{f}(\omega)$.

\section{Numerical evaluation of the conductivity tensor}
\label{sec:sup_conductivity}

In this section, we describe how we evaluate the dc conductivity tensor on the real-frequency axis. We use the symmetrized decomposition of Ref.~\cite{BonbienManchon2020}, 
which separates the conductivity into a Fermi-surface and a Fermi-sea contribution~\cite{Bastin1971,SmrckaStreda1977}. The spin-resolved conductivity is
\begin{equation}
\sigma^{\alpha\beta}_{\sigma}
=
\sigma_{\mr{surf},\sigma}^{\alpha\beta}
+
\sigma_{\mr{sea},\sigma}^{\alpha\beta},
\label{eq:sig_split}
\end{equation}
see Eqs.~\eqref{eq:sigmaI} and~\eqref{eq:sigmasea} for explicit formulas. 
The full conductivity is obtained by summing over $\sigma$. Throughout this section,
\begin{equation}
\int_{\boldsymbol{k}} \equiv \int_{\mathrm{BZ}} \frac{\mathrm{d}^2 k}{(2\pi)^2},
\qquad
f(\omega)=\frac{1}{e^{\omega/k_{\mr{B}}T}+1}.
\label{eq:kmeasure}
\end{equation}

The retarded Green's function is given by (cf.\ Eqs.~\eqref{eq:SEf_HF} and~\eqref{eq:Ginv} of the main text)
\begin{align}
\label{eq:Gr_def}
\bG^{\mr{r}}_{\vec{k}\sigma}(\omega) &= \left[(\omega + \mu)\mathds{1} - \mathbf{h}_{\vec{k}\sigma} - \bSigma^{\mr{r}}_{\bk\sigma}(\omega) \right]^{-1} \, ,
\\
\mathbf{h}_{\vec{k}\sigma} &=
\begin{pmatrix}
\epsilon_{f\bk} - \frac{\Delta}{2} & V_{\bk} \\
V^{\ast}_{\bk} & \epsilon_{c\bk\sigma} + \frac{\Delta}{2}
\end{pmatrix} \, ,
\end{align}
where $\mathds{1}$ is the $2 \times 2$ identity matrix.
We use the superscript $\mr{r}$ in this section to distinguish $\bG^\mr{r}$ from the advanced Green's function,
\begin{align}
\bG^{\mathrm{a}}_{\vec{k}\sigma}(\omega) = \bigl(\bG^{\mathrm{r}}_{\vec{k}\sigma}(\omega)\bigr)^\dagger .
\end{align}
The current vertices are given by
\begin{align}
\mathbf{j}^{\alpha}_{\vec{k}\sigma} &= - e \bv^{\alpha}_{\bk\sigma}
\\
\bv^{\alpha}_{\bk\sigma} &= -\frac{\partial [\bG^{\mr{r}}_{\vec{k}\sigma}(\omega)]^{-1}}{\partial k^{\alpha}} = \frac{\partial \mathbf{h}_{\vec{k}\sigma}}{\partial k^{\alpha}} + \frac{\partial \bSigma^{\mr{r}}_{\bk\sigma}(\omega)}{\partial k^{\alpha}} \, ,
\end{align}
where $\alpha \in \{x, y\}$ and $e$ is the elementary charge.
Because the momentum-dependent part of the self-energy is frequency independent, the velocities $\bv^{\alpha}_{\bk\sigma}$ and current vertices $\mathbf{j}^{\alpha}_{\vec{k}\sigma}$ are also frequency-independent.
Both $\bv^{\alpha}_{\bk\sigma}$ and $\mathbf{j}^{\alpha}_{\vec{k}\sigma}$ are renormalized from their non-interacting values only due to the non-local Fock contributions to the self-energy.

The Fermi-surface term is~\cite[Eq.~(9)]{BonbienManchon2020}
\begin{equation}
\begin{aligned}
\sigma_{\mr{surf},\sigma}^{\alpha\beta}
&=
\frac{\hbar}{4\pi}
\int_{\vec{k}}
\int_{-\infty}^{\infty}\mathrm{d}\omega\,
\partial_\omega f(\omega)
\\
&\times
\operatorname{tr}\!\Bigl[
\mathbf{j}^{\alpha}_{\vec{k}\sigma}
\bigl(
\bG^{\mr{r}}_{\vec{k}\sigma}(\omega)
- \bG^{\mr{a}}_{\vec{k}\sigma}(\omega)
\bigr)
\mathbf{j}^{\beta}_{\vec{k}\sigma}
\\
&\qquad\qquad\times
\bigl(
\bG^{\mr{r}}_{\vec{k}\sigma}(\omega)
- \bG^{\mr{a}}_{\vec{k}\sigma}(\omega)
\bigr)
\Bigr].
\end{aligned}
\label{eq:sigmaI}
\end{equation}
while the Fermi-sea term is~\cite[Eq.~(10)]{BonbienManchon2020}
\begin{equation}
\begin{aligned}
\sigma_{\mr{sea},\sigma}^{\alpha\beta}
&=
-\frac{\hbar}{4\pi}
\int_{\vec{k}}
\int_{-\infty}^{\infty}\mathrm{d}\omega\,
f(\omega)
\\
&\times
\operatorname{tr}\!\Bigl[
\Bigl\{
\mathbf{j}^{\alpha}_{\vec{k}\sigma}
\bigl(
\partial_\omega \bG^{\mr{r}}_{\vec{k}\sigma}(\omega)
+
\partial_\omega \bG^{\mr{a}}_{\vec{k}\sigma}(\omega)
\bigr)
\mathbf{j}^{\beta}_{\vec{k}\sigma}
\\
&\qquad
- \mathbf{j}^{\beta}_{\vec{k}\sigma}
\bigl(
\partial_\omega \bG^{\mr{r}}_{\vec{k}\sigma}(\omega)
+
\partial_\omega \bG^{\mr{a}}_{\vec{k}\sigma}(\omega)
\bigr)
\mathbf{j}^{\alpha}_{\vec{k}\sigma}
\Bigr\}
\\
&\qquad\qquad\times
\bigl(
\bG^{\mr{r}}_{\vec{k}\sigma}(\omega)
- \bG^{\mr{a}}_{\vec{k}\sigma}(\omega)
\bigr)
\Bigr].
\end{aligned}
\label{eq:sigmasea}
\end{equation}
Here and below, $\int_{\vec{k}}$ denotes the Brillouin-zone integral introduced in Eq.~\eqref{eq:kmeasure}.

The numerical evaluation of Eqs.~\eqref{eq:sigmaI} and \eqref{eq:sigmasea} is challenging because the integrands develop sharp structures when the imaginary part of the self-energy is small.
We use a modified form of the iterated adaptive integration method used in Ref.~\cite{Munoz2025}.
In contrast to Ref.~\cite{Munoz2025}, we evaluate the frequency integral first.
For that, we use the fact that the self-energy is computed on a predefined logarithmic frequency grid using NRG; self-energy values in between grid points are obtained with linear interpolation.
Due to that, $[\bG^{\mr{r,a}}(\omega)]^{-1}$ is a linear function on intervals between grid points, which we can use to evaluate the frequency integrals exactly on each of these intervals.
The remainder of this section is devoted to this exact evaluation of the frequency integrals.
The remaining momentum integrals are evaluated with an iterative adaptive integrator, using MATLAB's \texttt{integral2} with the option \texttt{'Method','iterated'}.

\subsection{Piecewise-linear representation of $\bG^{-1}$}

Let $\{\omega_i\}$ be our logarithmic real-frequency grid. 
On each interval
\begin{equation}
I_i=[\omega_i,\omega_{i+1}],
\qquad
\Delta \omega_i=\omega_{i+1}-\omega_i,
\label{eq:interval}
\end{equation}
we know the retarded self-energy at the endpoints and interpolate linearly in between,
\begin{equation}
\begin{aligned}
\bSigma^{\mr{r}}_{\vec{k}\sigma}(\omega)
&\approx
\bSigma^{\mr{r}}_{\vec{k}\sigma,i}
+
(\omega-\omega_i)\,\mathbf{S}_{i},
\\
\mathbf{S}_{i}
&=
\frac{
\bSigma^{\mr{r}}_{\vec{k}\sigma,i+1}
- \bSigma^{\mr{r}}_{\vec{k}\sigma,i}
}{\Delta \omega_i}.
\end{aligned}
\label{eq:sigma_linear}
\end{equation}
Here $\bSigma^{\mr{r}}_{\vec{k}\sigma,i}\equiv \bSigma^{\mr{r}}_{\vec{k}\sigma}(\omega_i)$. In the present calculations, only the local DMFT part of the self-energy depends on frequency. Hence $\mathbf{S}_{i}$ is independent of $\vec{k}$ and $\sigma$. The derivation below does not rely on this simplification.

Equation~\eqref{eq:sigma_linear} implies that on $I_i$ the inverse Green's function is linear in $\omega$,
\begin{equation}
\bigl[\bG^{\mr{r}}_{\vec{k}\sigma}(\omega)\bigr]^{-1}
=
\mathbf{A}_{i,\vec{k}\sigma}
+
\omega\,\mathbf{B}_{i,\vec{k}\sigma},
\label{eq:Ginv_affine}
\end{equation}
with
\begin{equation}
\begin{aligned}
\mathbf{B}_{i,\vec{k}\sigma}
&=
\mathds{1}-\mathbf{S}_{i},
\\
\mathbf{A}_{i,\vec{k}\sigma}
&=
\mu\,\mathds{1}
-
\mathbf{h}_{\vec{k}\sigma}
-
\bSigma^{\mr{r}}_{\vec{k}\sigma,i}
+
\omega_i \mathbf{S}_{i}.
\end{aligned}
\label{eq:AB_def}
\end{equation}
Since the matrix dimension is $2\times 2$, the inverse can be written analytically as
\begin{equation}
\bG^{\mr{r}}_{\vec{k}\sigma}(\omega)
=
\frac{\operatorname{adj}\!\bigl(\mathbf{A}_{i,\vec{k}\sigma}+\omega \mathbf{B}_{i,\vec{k}\sigma}\bigr)}
{\det\!\bigl(\mathbf{A}_{i,\vec{k}\sigma}+\omega \mathbf{B}_{i,\vec{k}\sigma}\bigr)}
\equiv
\frac{\mathbf{N}_{i,\vec{k}\sigma}(\omega)}{p_{i,\vec{k}\sigma}(\omega)}.
\label{eq:Gr_adj}
\end{equation}
For $2\times 2$ matrices the adjugate is linear in its argument, so
\begin{equation}
\begin{aligned}
\mathbf{N}_{i,\vec{k}\sigma}(\omega)
&=
\mathbf{N}_{0,i,\vec{k}\sigma}
+
\omega\,\mathbf{N}_{1,i,\vec{k}\sigma},
\\
\mathbf{N}_{0,i,\vec{k}\sigma}
&=
\operatorname{adj}\!\bigl(\mathbf{A}_{i,\vec{k}\sigma}\bigr),
\\
\mathbf{N}_{1,i,\vec{k}\sigma}
&=
\operatorname{adj}\!\bigl(\mathbf{B}_{i,\vec{k}\sigma}\bigr),
\end{aligned}
\label{eq:N_linear}
\end{equation}
while the determinant is quadratic,
\begin{equation}
p_{i,\vec{k}\sigma}(\omega)
=
\alpha_{2,i,\vec{k}\sigma}\omega^2
+
\alpha_{1,i,\vec{k}\sigma}\omega
+
\alpha_{0,i,\vec{k}\sigma},
\label{eq:p_quad}
\end{equation}
with
\begin{equation}
\begin{aligned}
\alpha_{2,i,\vec{k}\sigma}
&=
\det\!\bigl(\mathbf{B}_{i,\vec{k}\sigma}\bigr),
\\
\alpha_{1,i,\vec{k}\sigma}
&=
\operatorname{tr}\!\Bigl[
\mathbf{N}_{1,i,\vec{k}\sigma}\mathbf{A}_{i,\vec{k}\sigma}
\Bigr],
\\
\alpha_{0,i,\vec{k}\sigma}
&=
\det\!\bigl(\mathbf{A}_{i,\vec{k}\sigma}\bigr).
\end{aligned}
\label{eq:alpha_def}
\end{equation}
Factoring the determinant gives
\begin{equation}
p_{i,\vec{k}\sigma}(\omega)
=
\alpha_{2,i,\vec{k}\sigma}
\bigl(\omega-r_{1,i,\vec{k}\sigma}\bigr)
\bigl(\omega-r_{2,i,\vec{k}\sigma}\bigr),
\label{eq:p_factor}
\end{equation}
This defines the two complex poles $r_{1,i,\vec{k}\sigma}$ and $r_{2,i,\vec{k}\sigma}$ associated with the interval. On the real axis, the advanced Green's function and self-energy follow from $\bG^{\mr{a}}_{\vec{k}\sigma}(\omega)=\bigl(\bG^{\mr{r}}_{\vec{k}\sigma}(\omega)\bigr)^\dagger$ and $\bSigma^{\mr{a}}_{\vec{k}\sigma}(\omega)=\bigl(\bSigma^{\mr{r}}_{\vec{k}\sigma}(\omega)\bigr)^\dagger$. Accordingly, the unstarred quantities $\mathbf{A}_{i,\vec{k}\sigma}$, $\mathbf{B}_{i,\vec{k}\sigma}$, $\mathbf{N}_{i,\vec{k}\sigma}$, $p_{i,\vec{k}\sigma}$, $\alpha_{m,i,\vec{k}\sigma}$, and $r_{j,i,\vec{k}\sigma}$ below refer to the retarded branch, while advanced quantities are written explicitly through Hermitian and complex conjugation.

For later use, we define
\begin{align}
L_i(z) &= \log \left(\frac{\omega_{i+1}-z}{\omega_i-z}\right)
\label{eq:Ldef}
\\
R_i^{(1)}(z)
&=
\frac{1}{\omega_{i+1}-z}
-
\frac{1}{\omega_i-z},
\\
R_i^{(2)}(z)
&=
\frac{1}{(\omega_{i+1}-z)^2}
-
\frac{1}{(\omega_i-z)^2}
\label{eq:Rdef}
\end{align}
for complex $z$.

In the remainder of this section, we suppress the momentum and spin indices to avoid clutter.

\subsection{Fermi-surface contribution}

On a given interval, we write the Fermi-surface integrand in terms of the two kernels
\begin{align}
\Phi^{\mathrm{rr}}_{\alpha\beta}(\omega)
&=
\operatorname{tr}\!\Bigl[
\mathbf{j}^{\alpha}\bG^{\mr{r}}(\omega)\mathbf{j}^{\beta}\bG^{\mr{r}}(\omega)
\Bigr],
\label{eq:Phi_rr}\\
\Phi^{\mathrm{ar}}_{\alpha\beta}(\omega)
&=
\operatorname{tr}\!\Bigl[
\bG^{\mr{a}}(\omega)\mathbf{j}^{\alpha}\bG^{\mr{r}}(\omega)\mathbf{j}^{\beta}
\Bigr].
\label{eq:Phi_ar}
\end{align}
The trace in Eq.~\eqref{eq:sigmaI} then becomes
\begin{equation}
\operatorname{tr}\!\Bigl[
\mathbf{j}^{\alpha}\bigl(\bG^{\mr{r}}-\bG^{\mr{a}}\bigr)
\mathbf{j}^{\beta}\bigl(\bG^{\mr{r}}-\bG^{\mr{a}}\bigr)
\Bigr]
=
2\,\operatorname{Re}\Phi^{\mathrm{rr}}_{\alpha\beta}
-
\Phi^{\mathrm{ar}}_{\alpha\beta}
-
\Phi^{\mathrm{ar}}_{\beta\alpha}.
\label{eq:surface_split}
\end{equation}

Because $\mathbf{N}(\omega)$ is linear and $p(\omega)$ is quadratic,
\begin{equation}
\Phi^{\mathrm{rr}}_{\alpha\beta}(\omega)
=
\frac{q_{\alpha\beta}(\omega)}{p(\omega)^2},
\qquad
q_{\alpha\beta}(\omega)
=
q_{2,\alpha\beta}\omega^2+q_{1,\alpha\beta}\omega+q_{0,\alpha\beta},
\label{eq:qpoly}
\end{equation}
with
\begin{align}
q_{0,\alpha\beta}
&=
\operatorname{tr}\!\bigl[
\mathbf{j}^{\alpha}\mathbf{N}_0\,\mathbf{j}^{\beta}\mathbf{N}_0
\bigr],
\nonumber\\
q_{1,\alpha\beta}
&=
\operatorname{tr}\!\bigl[
\mathbf{j}^{\alpha}\mathbf{N}_0\,\mathbf{j}^{\beta}\mathbf{N}_1
+
\mathbf{j}^{\alpha}\mathbf{N}_1\,\mathbf{j}^{\beta}\mathbf{N}_0
\bigr],
\nonumber\\
q_{2,\alpha\beta}
&=
\operatorname{tr}\!\bigl[
\mathbf{j}^{\alpha}\mathbf{N}_1\,\mathbf{j}^{\beta}\mathbf{N}_1
\bigr].
\label{eq:qcoeffs}
\end{align}
Its partial-fraction decomposition is
\begin{equation}
\Phi^{\mathrm{rr}}_{\alpha\beta}(\omega)
=
\sum_{j=1}^{2}
\left[
\frac{A_{j,\alpha\beta}}{(\omega-r_j)^2}
+
\frac{B_{j,\alpha\beta}}{\omega-r_j}
\right],
\label{eq:RR_pf}
\end{equation}
with
\begin{align}
A_{j,\alpha\beta}
&=
\frac{q_{\alpha\beta}(r_j)}
{\alpha_2^{\,2}(r_j-r_{3-j})^2},
\label{eq:RR_A}\\
B_{j,\alpha\beta}
&=
\frac{
q'_{\alpha\beta}(r_j)(r_j-r_{3-j})-2q_{\alpha\beta}(r_j)
}
{\alpha_2^{\,2}(r_j-r_{3-j})^3}.
\label{eq:RR_B}
\end{align}

Similarly,
\begin{equation}
\begin{aligned}
\Phi^{\mathrm{ar}}_{\alpha\beta}(\omega)
&=
\frac{u_{\alpha\beta}(\omega)}{p(\omega)\,p(\omega)^*},
\\
u_{\alpha\beta}(\omega)
&=
u_{2,\alpha\beta}\omega^2+u_{1,\alpha\beta}\omega+u_{0,\alpha\beta},
\end{aligned}
\label{eq:upoly}
\end{equation}
with
\begin{align}
u_{0,\alpha\beta}
&=
\operatorname{tr}\!\bigl[
\mathbf{N}_0^\dagger \mathbf{j}^{\alpha}\mathbf{N}_0\mathbf{j}^{\beta}
\bigr],
\nonumber\\
u_{1,\alpha\beta}
&=
\operatorname{tr}\!\bigl[
\mathbf{N}_0^\dagger \mathbf{j}^{\alpha}\mathbf{N}_1\mathbf{j}^{\beta}
+
\mathbf{N}_1^\dagger \mathbf{j}^{\alpha}\mathbf{N}_0\mathbf{j}^{\beta}
\bigr],
\nonumber\\
u_{2,\alpha\beta}
&=
\operatorname{tr}\!\bigl[
\mathbf{N}_1^\dagger \mathbf{j}^{\alpha}\mathbf{N}_1\mathbf{j}^{\beta}
\bigr].
\label{eq:ucoeffs}
\end{align}
We write the interval integral as
\begin{equation}
\int_{I_i}\mathrm{d}\omega\, w(\omega)\,\Phi^{\mathrm{ar}}_{\alpha\beta}(\omega)
=
2\,\operatorname{Re}
\sum_{j=1}^{2}
C_{j,\alpha\beta}
\int_{I_i}\mathrm{d}\omega\,
\frac{w(\omega)}{\omega-r_j},
\label{eq:AR_interval}
\end{equation}
the residues are
\begin{equation}
C_{j,\alpha\beta}
=
\frac{u_{\alpha\beta}(r_j)}
{|\alpha_2|^2(r_j-r_{3-j})(r_j-r_1^*)(r_j-r_2^*)}.
\label{eq:AR_res}
\end{equation}

Finally, on $I_i$ we approximate
\begin{equation}
\begin{aligned}
\partial_\omega f(\omega) &= f'(\omega)
\approx
w_{0,i}^{(\mathrm{surf})}
+
w_{1,i}^{(\mathrm{surf})}\omega,
\\
w_{1,i}^{(\mathrm{surf})}
&=
\frac{f'(\omega_{i+1})-f'(\omega_i)}{\Delta \omega_i},
\\
w_{0,i}^{(\mathrm{surf})}
&=
f'(\omega_i)-w_{1,i}^{(\mathrm{surf})}\omega_i.
\end{aligned}
\label{eq:weight_surface}
\end{equation}
Here $f'(\omega)=-\bigl[4k_{\mr{B}}T\cosh^2\!\bigl(\omega/(2k_{\mr{B}}T)\bigr)\bigr]^{-1}$.
The corresponding interval integrals are
\begin{align}
I^{\mathrm{rr}}_{0,\alpha\beta}(i)
&=
\sum_{j=1}^{2}
\left[
-A_{j,\alpha\beta}\,R_i^{(1)}(r_j)
+
B_{j,\alpha\beta}\,L_i(r_j)
\right],
\label{eq:I0_rr}\\
I^{\mathrm{rr}}_{1,\alpha\beta}(i)
&=
\sum_{j=1}^{2}
\Bigl[
A_{j,\alpha\beta}\Bigl(L_i(r_j)-r_j R_i^{(1)}(r_j)\Bigr)
\notag\\
&+
B_{j,\alpha\beta}\Bigl(\Delta \omega_i+r_j L_i(r_j)\Bigr)
\Bigr],
\label{eq:I1_rr} \\
I^{\mathrm{ar}}_{0,\alpha\beta}(i)
&=
2\,\operatorname{Re}
\sum_{j=1}^{2} C_{j,\alpha\beta}\,L_i(r_j),
\label{eq:I0_ar} \\
I^{\mathrm{ar}}_{1,\alpha\beta}(i)
&=
2\,\operatorname{Re}
\sum_{j=1}^{2}
C_{j,\alpha\beta}
\bigl(\Delta \omega_i+r_jL_i(r_j)\bigr).
\label{eq:I1_ar}
\end{align}
The contribution of interval $I_i$ to the surface term is
\begin{equation}
\begin{aligned}
\mathcal{S}^{\mathrm{surf}}_{i,\alpha\beta}
&=
w_{0,i}^{(\mathrm{surf})}
\Bigl[
2\,\operatorname{Re}I^{\mathrm{rr}}_{0,\alpha\beta}(i)
-
I^{\mathrm{ar}}_{0,\alpha\beta}(i)
-
I^{\mathrm{ar}}_{0,\beta\alpha}(i)
\Bigr]
\\
&+
w_{1,i}^{(\mathrm{surf})}
\Bigl[
2\,\operatorname{Re}I^{\mathrm{rr}}_{1,\alpha\beta}(i)
-
I^{\mathrm{ar}}_{1,\alpha\beta}(i)
-
I^{\mathrm{ar}}_{1,\beta\alpha}(i)
\Bigr].
\end{aligned}
\label{eq:S_surface_interval}
\end{equation}
Summing Eq.~\eqref{eq:S_surface_interval} over all intervals yields the frequency integral in Eq.~\eqref{eq:sigmaI}.

\subsection{Fermi-sea contribution}

For the sea term, we first note that, on each interval,
\begin{equation}
\begin{aligned}
\partial_\omega \bG^{\mr{r}}(\omega)
&=
-\,\bG^{\mr{r}}(\omega)\,\mathbf{B}\,\bG^{\mr{r}}(\omega)
\\
&=
\frac{\mathbf{N}_1 p - \mathbf{N} p'}{p^2},
\\
\partial_\omega \bG^{\mr{a}}(\omega)
&=
\frac{\mathbf{N}_1^\dagger p^* - \mathbf{N}^\dagger p'^*}{(p^*)^2},
\end{aligned}
\label{eq:dG_exact}
\end{equation}
with $\mathbf{N}=\mathbf{N}(\omega)$, $p=p(\omega)$, and $p' = \partial_{\omega} p(\omega)$.
We then define
\begin{align}
\Psi^{\mathrm{rr}}_{\alpha\beta}(\omega)
&=
\operatorname{tr}\!\bigl[
\mathbf{j}^{\alpha}(\partial_\omega \bG^{\mr{r}})\mathbf{j}^{\beta}\bG^{\mr{r}}
\bigr]
=
\frac{R^{\mathrm{rr}}_{\alpha\beta}(\omega)}{p(\omega)^3},
\label{eq:Psi_rr}\\
\Psi^{\mathrm{ra}}_{\alpha\beta}(\omega)
&=
\operatorname{tr}\!\bigl[
\mathbf{j}^{\alpha}(\partial_\omega \bG^{\mr{r}})\mathbf{j}^{\beta}\bG^{\mr{a}}
\bigr]
=
\frac{R^{\mathrm{ra}}_{\alpha\beta}(\omega)}{p(\omega)^2\,p(\omega)^*},
\label{eq:Psi_ra}\\
\Psi^{\mathrm{ar}}_{\alpha\beta}(\omega)
&=
\operatorname{tr}\!\bigl[
\mathbf{j}^{\alpha}(\partial_\omega \bG^{\mr{a}})\mathbf{j}^{\beta}\bG^{\mr{r}}
\bigr]
=
\frac{R^{\mathrm{ar}}_{\alpha\beta}(\omega)}{p(\omega)\,(p(\omega)^*)^2},
\label{eq:Psi_ar}\\
\Psi^{\mathrm{aa}}_{\alpha\beta}(\omega)
&=
\operatorname{tr}\!\bigl[
\mathbf{j}^{\alpha}(\partial_\omega \bG^{\mr{a}})\mathbf{j}^{\beta}\bG^{\mr{a}}
\bigr]
=
\frac{R^{\mathrm{aa}}_{\alpha\beta}(\omega)}{(p(\omega)^*)^3}.
\label{eq:Psi_aa}
\end{align}
The numerators are cubic polynomials that follow directly from Eq.~\eqref{eq:dG_exact},
\begin{align}
R^{\mathrm{rr}}_{\alpha\beta}(\omega)
&=
\operatorname{tr}\!\bigl[
\mathbf{j}^{\alpha}\bigl(\mathbf{N}_1 p - \mathbf{N} p'\bigr)\mathbf{j}^{\beta}\mathbf{N}
\bigr],
\label{eq:Rrr_num}\\
R^{\mathrm{ra}}_{\alpha\beta}(\omega)
&=
\operatorname{tr}\!\bigl[
\mathbf{j}^{\alpha}\bigl(\mathbf{N}_1 p - \mathbf{N} p'\bigr)\mathbf{j}^{\beta}\mathbf{N}^\dagger
\bigr],
\label{eq:Rra_num}\\
R^{\mathrm{ar}}_{\alpha\beta}(\omega)
&=
\operatorname{tr}\!\bigl[
\mathbf{j}^{\alpha}\bigl(\mathbf{N}_1^\dagger p^* - \mathbf{N}^\dagger p'^*\bigr)\mathbf{j}^{\beta}\mathbf{N}
\bigr],
\label{eq:Rar_num}\\
R^{\mathrm{aa}}_{\alpha\beta}(\omega)
&=
\operatorname{tr}\!\bigl[
\mathbf{j}^{\alpha}\bigl(\mathbf{N}_1^\dagger p^* - \mathbf{N}^\dagger p'^*\bigr)\mathbf{j}^{\beta}\mathbf{N}^\dagger
\bigr].
\label{eq:Raa_num}
\end{align}
The sea integrand can then be written as
\begin{equation}
\begin{aligned}
\mathcal{K}^{\mathrm{sea}}_{\alpha\beta}(\omega)
={}&
\bigl[
\Psi^{\mathrm{rr}}_{\alpha\beta}
-
\Psi^{\mathrm{ra}}_{\alpha\beta}
+
\Psi^{\mathrm{ar}}_{\alpha\beta}
-
\Psi^{\mathrm{aa}}_{\alpha\beta}
\bigr]
\\
&-
\bigl[
\Psi^{\mathrm{rr}}_{\beta\alpha}
-
\Psi^{\mathrm{ra}}_{\beta\alpha}
+
\Psi^{\mathrm{ar}}_{\beta\alpha}
-
\Psi^{\mathrm{aa}}_{\beta\alpha}
\bigr].
\end{aligned}
\label{eq:Ksea_split}
\end{equation}

The corresponding pole structure is
\begin{itemize}
\item $\Psi^{\mathrm{rr}}$ has triple poles at $r_1$ and $r_2$;
\item $\Psi^{\mathrm{aa}}$ has triple poles at $r_1^*$ and $r_2^*$;
\item $\Psi^{\mathrm{ra}}$ has double poles at $r_1,r_2$ and simple poles at $r_1^*,r_2^*$;
\item $\Psi^{\mathrm{ar}}$ has simple poles at $r_1,r_2$ and double poles at $r_1^*,r_2^*$.
\end{itemize}

For $\Psi^{\mathrm{rr}}$ we use
\begin{equation}
\Psi^{\mathrm{rr}}_{\alpha\beta}(\omega)
=
\sum_{j=1}^{2}
\left[
\frac{A^{(3)}_{j,\alpha\beta}}{(\omega-r_j)^3}
+
\frac{B^{(3)}_{j,\alpha\beta}}{(\omega-r_j)^2}
+
\frac{C^{(3)}_{j,\alpha\beta}}{\omega-r_j}
\right],
\label{eq:rr3_pf}
\end{equation}
where
\begin{align}
A^{(3)}_{j,\alpha\beta}
&=
\frac{R^{\mathrm{rr}}_{\alpha\beta}(r_j)}
{\alpha_2^{\,3}(r_j-r_{3-j})^3},
\label{eq:rr3A}\\
B^{(3)}_{j,\alpha\beta}
&=
\frac{
\bigl(R^{\mathrm{rr}}_{\alpha\beta}\bigr)'(r_j)(r_j-r_{3-j})
-3R^{\mathrm{rr}}_{\alpha\beta}(r_j)
}
{\alpha_2^{\,3}(r_j-r_{3-j})^4},
\label{eq:rr3B}\\
C^{(3)}_{j,\alpha\beta}
&=
\frac{
\bigl(R^{\mathrm{rr}}_{\alpha\beta}\bigr)''(r_j)(r_j-r_{3-j})^2
} 
{2\,\alpha_2^{\,3}(r_j-r_{3-j})^5}.
\notag\\
&\quad
-\frac{
6\bigl(R^{\mathrm{rr}}_{\alpha\beta}\bigr)'(r_j)(r_j-r_{3-j})
-12R^{\mathrm{rr}}_{\alpha\beta}(r_j)
}
{2\,\alpha_2^{\,3}(r_j-r_{3-j})^5}.
\label{eq:rr3C}
\end{align}
The coefficients for $\Psi^{\mathrm{aa}}$ follow from the substitutions $r_j\to r_j^*$ and $\alpha_2\to\alpha_2^*$, together with $R^{\mathrm{rr}}\to R^{\mathrm{aa}}$.

For $\Psi^{\mathrm{ra}}$ we write
\begin{equation}
\begin{aligned}
\Psi^{\mathrm{ra}}_{\alpha\beta}(\omega)
={}&
\sum_{j=1}^{2}
\left[
\frac{a_{j,\alpha\beta}}{(\omega-r_j)^2}
+
\frac{b_{j,\alpha\beta}}{\omega-r_j}
\right]
\\
&+
\sum_{j=1}^{2}
\frac{c_{j,\alpha\beta}^*}{\omega-r_j^*},
\end{aligned}
\label{eq:ra_pf}
\end{equation}
with
\begin{align}
a_{j,\alpha\beta}
&=
\frac{R^{\mathrm{ra}}_{\alpha\beta}(r_j)}
{\alpha_2^{\,2}\alpha_2^*
(r_j-r_{3-j})^2
(r_j-r_1^*)(r_j-r_2^*)},
\label{eq:ra_a}\\
b_{j,\alpha\beta}
&=
\frac{
\bigl(R^{\mathrm{ra}}_{\alpha\beta}\bigr)'(r_j)
}
{\alpha_2^{\,2}\alpha_2^*
(r_j-r_{3-j})^2
(r_j-r_1^*)(r_j-r_2^*)},
\notag\\
&\quad
-\frac{
R^{\mathrm{ra}}_{\alpha\beta}(r_j)
\left[
\frac{2}{r_j-r_{3-j}}
+\frac{1}{r_j-r_1^*}
+\frac{1}{r_j-r_2^*}
\right]
}
{\alpha_2^{\,2}\alpha_2^*
(r_j-r_{3-j})^2
(r_j-r_1^*)(r_j-r_2^*)},
\label{eq:ra_b}\\
c_{j,\alpha\beta}^*
&=
\frac{
R^{\mathrm{ra}}_{\alpha\beta}(r_j^*)
}
{
p(r_j^*)^2\,p'^*(r_j^*)
}.
\label{eq:ra_c}
\end{align}
For $\Psi^{\mathrm{ar}}$ we analogously use
\begin{equation}
\begin{aligned}
\Psi^{\mathrm{ar}}_{\alpha\beta}(\omega)
={}&
\sum_{j=1}^{2}
\frac{d_{j,\alpha\beta}}{\omega-r_j}
\\
&+
\sum_{j=1}^{2}
\left[
\frac{\hat{a}_{j,\alpha\beta}^*}{(\omega-r_j^*)^2}
+
\frac{\hat{b}_{j,\alpha\beta}^*}{\omega-r_j^*}
\right],
\end{aligned}
\label{eq:ar_pf}
\end{equation}
with
\begin{align}
d_{j,\alpha\beta}
&=
\frac{
R^{\mathrm{ar}}_{\alpha\beta}(r_j)
}
{
p'(r_j)\,\bigl[p^*(r_j)\bigr]^2
},
\label{eq:ar_d}\\
\hat{a}_{j,\alpha\beta}^*
&=
\frac{
R^{\mathrm{ar}}_{\alpha\beta}(r_j^*)
}
{
\alpha_2\alpha_2^{*2}
(r_j^*-r_{3-j}^*)^2
(r_j^*-r_1)(r_j^*-r_2)
},
\label{eq:ar_a}\\
\hat{b}_{j,\alpha\beta}^*
&=
\frac{
\bigl(R^{\mathrm{ar}}_{\alpha\beta}\bigr)'(r_j^*)
}
{
\alpha_2\alpha_2^{*2}
(r_j^*-r_{3-j}^*)^2
(r_j^*-r_1)(r_j^*-r_2)
}.
\notag\\
&\quad
-\frac{
R^{\mathrm{ar}}_{\alpha\beta}(r_j^*)
\left[
\frac{2}{r_j^*-r_{3-j}^*}
+\frac{1}{r_j^*-r_1}
+\frac{1}{r_j^*-r_2}
\right]
}
{
\alpha_2\alpha_2^{*2}
(r_j^*-r_{3-j}^*)^2
(r_j^*-r_1)(r_j^*-r_2)
}.
\label{eq:ar_b}
\end{align}

The required integrals are
\begin{align}
\int \frac{w_{0}+w_{1}\omega}{(\omega-r)^3}\,\mathrm{d}\omega
&=
-\frac{w_0}{2(\omega-r)^2}
\notag\\
&\quad
+
w_1\left[
-\frac{1}{\omega-r}
-\frac{r}{2(\omega-r)^2}
\right],
\label{eq:prim_triple}\\
\int \frac{w_{0}+w_{1}\omega}{(\omega-r)^2}\,\mathrm{d}\omega
&=
-w_0\frac{1}{\omega-r}
\notag\\
&\quad
+
w_1\left[
\log(\omega-r)-r\frac{1}{\omega-r}
\right],
\label{eq:prim_double2}\\
\int \frac{w_{0}+w_{1}\omega}{\omega-r}\,\mathrm{d}\omega
&=
w_0\log(\omega-r)
\notag\\
&\quad
+
w_1\left[
(\omega-r)+r\log(\omega-r)
\right],
\label{eq:prim_simple2}
\end{align}
which are evaluated between $\omega_i$ and $\omega_{i+1}$.
On $I_i$, we further approximate
\begin{align}
f(\omega)
&\approx
w_{0,i}^{(\mathrm{sea})}+w_{1,i}^{(\mathrm{sea})}\omega \, ,
\label{eq:weight_sea}
\end{align}
with
\begin{equation}
\begin{aligned}
w_{1,i}^{(\mathrm{sea})}
&=
\frac{f(\omega_{i+1})-f(\omega_i)}{\Delta \omega_i},
\\
w_{0,i}^{(\mathrm{sea})}
&=
f(\omega_i)-w_{1,i}^{(\mathrm{sea})}\omega_i.
\end{aligned}
\label{eq:weight_sea_coeffs}
\end{equation}
The contribution of interval $I_i$ to the sea term is
\begin{equation}
\mathcal{S}^{\mathrm{sea}}_{i,\alpha\beta}
=
w_{0,i}^{(\mathrm{sea})}\,
\mathcal{I}^{\mathrm{sea}}_{0,i,\alpha\beta}
+
w_{1,i}^{(\mathrm{sea})}\,
\mathcal{I}^{\mathrm{sea}}_{1,i,\alpha\beta},
\label{eq:sea_interval}
\end{equation}
where
\begin{align}
\nonumber
\mathcal{I}^{\mathrm{sea}}_{m,i,\alpha\beta}
&=
\bigl[
I^{\mathrm{rr}}_{m,i,\alpha\beta}
-
I^{\mathrm{ra}}_{m,i,\alpha\beta}
+
I^{\mathrm{ar}}_{m,i,\alpha\beta}
-
I^{\mathrm{aa}}_{m,i,\alpha\beta}
\bigr]
\\ 
&-
\bigl[
I^{\mathrm{rr}}_{m,i,\beta\alpha}
-
I^{\mathrm{ra}}_{m,i,\beta\alpha}
+
I^{\mathrm{ar}}_{m,i,\beta\alpha}
-
I^{\mathrm{aa}}_{m,i,\beta\alpha}
\bigr]
\\ \nonumber
&=
2 \mr{Re} \, \bigl[
I^{\mathrm{rr}}_{m,i,\alpha\beta}
-
I^{\mathrm{ra}}_{m,i,\alpha\beta}
+
I^{\mathrm{ar}}_{m,i,\alpha\beta}
-
I^{\mathrm{aa}}_{m,i,\alpha\beta}
\bigr] \, .
\label{eq:sea_interval_moments}
\end{align}
for $m=0,1$.
Summing Eq.~\eqref{eq:sea_interval} over all intervals yields the frequency integral in Eq.~\eqref{eq:sigmasea}.

\clearpage

\end{document}